\documentclass{aa}

\usepackage{natbib}
\bibpunct{(}{)}{;}{a}{}{,} 
\usepackage{graphicx}

\usepackage{txfonts}

\usepackage{breqn}
\usepackage{tabularx}
\usepackage{siunitx}
\usepackage{booktabs}
\usepackage{color}
\usepackage{multirow}
\usepackage[switch]{lineno} 
\linenumbers

\usepackage{hyperref}
\hypersetup{
    colorlinks=true,
    linkcolor=blue,
    citecolor=blue,
    urlcolor=blue,
    filecolor=blue,
    pdftitle={RR Lyrae stars as distance indicators I. Sample selection and characterisation},
    pdfauthor={Manuel Sánchez Benavente}
}

\usepackage{orcidlink}

\definecolor{lightseagreen}{rgb}{0.13, 0.7, 0.67}
\definecolor{lightskyblue}{rgb}{0.53, 0.81, 0.98}
\definecolor{lightslategray}{rgb}{0.47, 0.53, 0.6}
\definecolor{pastelviolet}{rgb}{0.8, 0.6, 0.79}
\definecolor{plum(web)}{rgb}{0.8, 0.6, 0.8}
\definecolor{skobeloff}{rgb}{0.0, 0.48, 0.45}
\definecolor{ruddypink}{rgb}{0.88, 0.56, 0.59}
\definecolor{color1}{rgb}{0.22,0.45,0.70}
\definecolor{colortitle}{rgb}{0.8, 0.33, 0.0}
\definecolor{colortitle2}{rgb}{0.00, 0.18, 0.66}
\definecolor{colortitle3}{rgb}{1.00, 0.00, 1.00}
\definecolor{colortitle4}{rgb}{0.59, 0.29, 0.00}
\definecolor{orange}{rgb}{1.0, 0.5, 0.0}
\definecolor{teal}{rgb}{0.0, 0.5, 0.5}
\definecolor{davygrey}{rgb}{0.33, 0.33, 0.33}
\definecolor{cadet}{rgb}{0.33, 0.41, 0.47} 
\definecolor{blush}{rgb}{0.87, 0.36, 0.51}
\definecolor{asparagus}{rgb}{0.53, 0.66, 0.42}
\definecolor{deeppink}{rgb}{1.0, 0.08, 0.58}

\newcommand{\gloess}{GLOESS}
\newcommand{\nodata}{~...~}
\newcommand{\feh}{\mathrm{\left[Fe/H\right]}}

\begin{document}
   \title{RR Lyrae stars as distance indicators}

   \subtitle{I. Sample selection and characterisation}

   \author{M. Sánchez-Benavente \inst{1,2}
          \corrauth{msb@iac.es} \orcidlink{0000-0003-2693-279X}
    \and
          M. Monelli \inst{3,4,1,2}
    \and
          M. Dall'Ora \inst{5}
    \and
          V. F. Braga \inst{4}
    \and
          J. P. Mullen \inst{9,10}
    \and
          M. Marengo \inst{8}
    \and
          N. Matsunaga \inst{12}
    \and
          G. Bono \inst{6,4}
    \and
          G. Fiorentino \inst{4}
    \and
          C. Gallart \inst{1,2}
    \and
          M. Marconi \inst{5}
    \and
          I. Plaks \inst{11} \orcidlink{0009-0003-3216-3402}
    \and
          A. Capodagli \inst{8}
    \and
          G. Ceci \inst{7,6,4}
    \and 
          C. E. Martínez-Vázquez \inst{13}
    \and
          M. Tantalo \inst{4}
          }     

   \institute{IAC - Instituto de Astrofísica de Canarias, Calle Vía Lactea s/n, E-38205 La Laguna, Tenerife, Spain
         \and
             Departmento de Astrofísica, Universidad de La Laguna, E-38206 La Laguna, Tenerife, Spain 
         \and 
             INAF - Osservatorio Astronomico d'Abruzzo, via Mentore Maggini s.n.c., 64100 Teramo, Italy
         \and 
             INAF - Osservatorio Astronomico di Roma, via Frascati 33, 00078 Monte Porzio Catone, Italy
        \and 
             INAF – Osservatorio Astronomico di Capodimonte, Salita Moiariello 16, 80131 Naples, Italy 
         \and
             Dipartimento di Fisica, Università di Roma Tor Vergata, via della Ricerca Scientifica 1, 00133, Rome, Italy
        \and
             Dipartimento di Fisica, Sapienza Università di Roma, P.le A. Moro 5, 00185, Roma, Italy
        \and
             Florida State University, Department of Physics, 77 Chieftain Way, Tallahassee, FL 32306, USA 
         \and
            Department of Physics, University of North Florida, 1 UNF Dr., Jacksonville, FL 32224, USA
         \and
            Department of Physics and Astronomy, Vanderbilt University, VU Station 1807, Nashville, TN 37235, USA
         \and
            Department of Physics and Astronomy, Iowa State University, Ames, IA 50011, USA
         \and
            Department of Astronomy, School of Science, The University of Tokyo, 7-3-1 Hongo, Bunkyo-ku, Tokyo 113-0033, Japan
        \and
            NSF NOIRLab, 670 North A’ ohoku Place, Hilo, HI 96720, USA
            }

   \date{Received July, 24 2026; accepted October, 10 2026}

  \abstract
   {RR Lyrae (RRL) stars are well-established Population II distance indicators, with period–metallicity–luminosity (PLZ) relations linking pulsational properties to absolute luminosity. Gaia parallaxes now allow these relations to be anchored to geometric distances with unprecedented precision, offering an independent route to the extragalactic distance scale.} 
   {We present the first paper in a series aimed at deriving new empirical PLZ and metallicity–luminosity relations for RRL in the optical and near-infrared. We describe the selection and characterisation of a sample of bright nearby field RRL, with distances anchored to reliable Gaia DR3 parallaxes.}
   {We conducted a multi-band photometric campaign of 172 RRL: 101 in Johnson–Cousins UBVRI with the IAC80 telescope and 130 in JHKs with the IRSF telescope, with 59 stars in common. The iron abundances come from the Spectroscopic Rome RRL Catalog catalogue, mostly from high-resolution spectroscopy. The light curves were modelled with GLOESS (optical) and near-infrared templates, and the mean magnitudes, amplitudes, and Fourier parameters were derived via a bootstrap Monte Carlo method. The linear relations were fitted with a Bayesian Markov Chain Monte Carlo method.}
   {The sample spans -3.4 to 0.2 dex in $\feh$. We found tight relations between the Fourier parameter $\phi_{31}$ in each optical and Gaia band and its V-band value for RRab stars (scatter below 0.06). The near-infrared amplitude ratios with respect to V or G increase continuously with period, without the break reported in globular clusters. We provide a homogeneous calibration of the colour at minimum light across all filter combinations from $U-B$ to $H-K_s$, revealing significant period and metallicity dependences, and we use it to derive reddening estimates for 15 stars outside current 3D extinction maps.}
   {This sample and its characterisation provide the foundation for the PLZ and metallicity–luminosity calibrations to be presented in Paper II.}


   \maketitle
   \nolinenumbers

\section{Introduction}
\label{sec:introduction}

\begin{figure*}[th]
\centering
\includegraphics{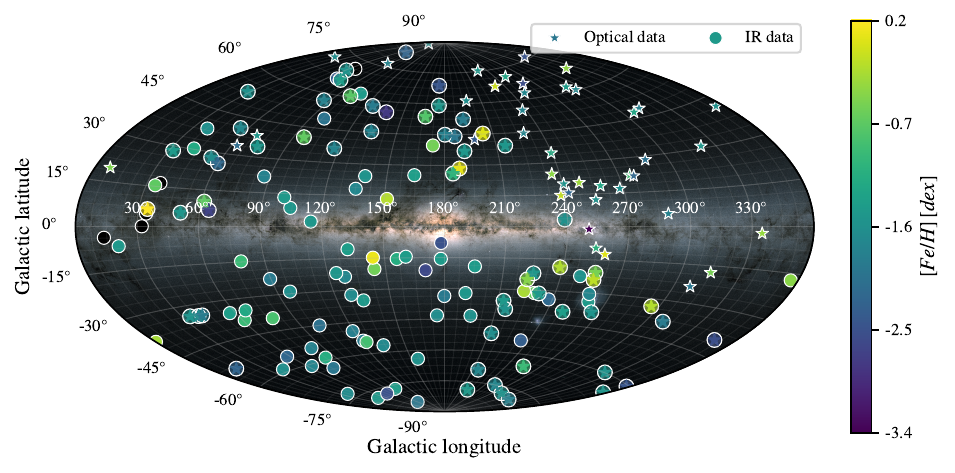}
  \caption{Location of all the targets in Galactic coordinates, colour-coded by their metallicity. Stars that were observed in the Johnson-Cousins filters are shown as stars, those observed in the NIR are shown as circles, and those that were observed in both systems share both symbols.}
     \label{plot:skymap}
\end{figure*}

RR~Lyrae (RRL) stars are old ($\gtrsim 10$~Gyr), low-mass Population~II variables residing on the intersection of the Cepheid instability strip and the horizontal branch \citep{catelan2015}. Ubiquitous across the Milky Way and Local Group, their light-curve structures closely reflect their physical properties, making them distance indicators, metallicity probes, and Galactic population tracers. RRLs are classified by their pulsation mode: fundamental-mode (RRab) stars exhibit asymmetric saw-tooth-shaped light curves and longer periods ($0.4-1.25$ d: \citealt{bono1997b, braga2020}), while first-overtone (RRc) stars show more symmetric and sinusoidal shorter-period ($0.2-0.5$ d) variations. Double-mode (RRd) stars pulsate in both modes, allowing a period fundamentalisation via their characteristic period ratio $P_{\rm 1O}/P_{\rm F} \approx 0.746$ \citep{iben1971,braga2016}. These modes correlate directly with temperature and colour across the instability strip, making multi-band light curves sensitive indicators of both stellar composition and environment.
While the light curves of many RRLs are stable, up to $\sim 50\%$ of RRab stars exhibit the Blazhko effect \citep{skarka2013}, which is a quasi-periodic modulation in the amplitude and phase that was first reported by \citet{blazhko1907} and \citet{shapley1916}. Because these modulation periods can range from days to years, the effect is cannot always be detected within the baseline of a single observing campaign. On much longer timescales, evolution off the zero-age horizontal branch (ZAHB)also induces gradual secular period drifts that can be detected by analysing observed-minus-calculated (O$-$C) diagrams of maximum-light timing \citep[e.g.][]{leborgne2007}.

In variable stars, the shape of the light curve is directly related to some of their physical parameters, as first shown by \cite{simon1981} for Cepheid stars. \cite{jurcsik1996} later demonstrated that, at optical wavelengths, the same principle applies to RRab stars, where the light-curve shape is linked to metallicity. The usual way to quantify the light-curve morphology is through Fourier decomposition, which provides a compact quantitative description of their shapes. By fitting a Fourier series to the phased light curve, amplitude ratios and phase differences are obtained that capture subtle features beyond global quantities, such as the mean magnitude and amplitude. Further works extended this analysis to different filters and environments \citep{smolec2005, nemec2013, martinez-vazquez2016, ngeow2016, iorio2021, mullen2021} using larger and more modern datasets. The same method was applied to RRc stars later than to RRab \citep{morgan2007}, because of the added technical difficulty and the lower number of sources, and it has since been expanded in the same way as for RRab stars \citep[see ][ and references therein]{mullen2022}. More recently, machine-learning techniques have also been employed to study these correlations in large samples \citep{dekany2022, muraveva2025}. The ability to determine metallicities photometrically, without the need for spectroscopic follow-up, is a crucial diagnostic for mapping the chemical enrichment of old stellar populations in the Milky Way and in external galaxies. This unique opportunity becomes even more relevant in combination with long-term, multi-band variability surveys such as the Optical Gravitational Lensing Experiment \citep[OGLE, ][]{soszynski2019}, Gaia, and the upcoming Vera C. Rubin Observatory Legacy Survey of Space and Time \citep{ivezic2019}.

A second diagnostic tool provided by the light curve is the colour of the star at minimum light. \citet{sturch1966} first recognised that during this broad phase (typically between phases 0.5 and 0.8), the stellar atmosphere of an RRL reaches a thermally stable state near minimum temperature. Consequently, the intrinsic colour at minimum light is largely insensitive to the pulsation amplitude, the period, or the metallicity, at least when filters that are combined are sufficiently red to minimise atmospheric variations along the pulsation cycle. This makes the colour at minimum light a useful proxy for interstellar reddening: by comparing the observed colour with the expected intrinsic value, the line-of-sight extinction to each individual star can be estimated without requiring a separate reddening map. The technique was calibrated in the Johnson-Cousins system by several authors using samples of field and cluster RRL \citep{blanco1992, guldenschuh2005, kunder2010}, that provided specific calibrations for $B-V$, $V-R$, and $V-I$ colours. The method was applied at broader wavelengths, for example in the DECam system by \citet{vivas2017} and \citet{saha2019}, to construct high-resolution 2D reddening maps of complex regions such as Baade's Window. However, a single self-consistent calibration covering the full optical-to-near-infrared (NIR) wavelength range from $U$ through $K_s$ is still missing from the literature. Our dataset, which simultaneously covers eight photometric bands for a large and chemically diverse sample of nearby field RRL, provides an excellent opportunity to fill this gap.

\begin{figure}[h]
\centering
\includegraphics[width=\columnwidth]{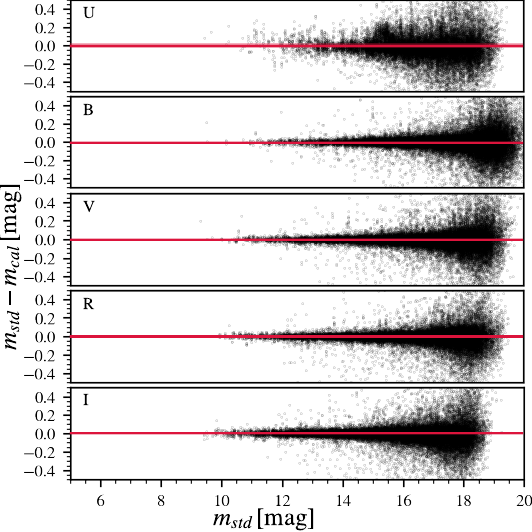}
  \caption{Difference between the tabulated standard magnitude and the calibrated magnitude obtained from applying Equation~\ref{eq:photcal} to the instrumental magnitudes of standard stars. The red line depicts the mean value of that difference for stars with a magnitude brighter than 15. The differences, presented in Table~\ref{table:photcal}, are compatible with zero.}
  \label{plot:zeropoint_calibration}
\end{figure}

The RRL distance determinations traditionally follow three different paths: i) optical luminosity-metallicity ($LZ$), ii) NIR period-luminosity-metallicity ($PLZ$), and iii) reddening-free period-Wesenheit-metallicity ($PWZ$) relations. The classic optical $M_V - \feh$ relation \citep{sandage1981,sandage1990,bono2003} remains a simple consistency check, but includes intrinsic evolutionary scatter and high extinction and distance sensitivity, despite recent Gaia-based updates \citep{muraveva2018, garofalo2022}. Conversely, NIR $PLZ$ relations \citep{longmore1986} dramatically reduce extinction effects, yield lower intrinsic scatter, and feature smaller amplitudes that simplify mean-magnitude determinations. This framework was extensively calibrated theoretically \citep{bono2001, bono2003, catelan2004, marconi2015} and empirically \citep{sollima2006, bhardwaj2023, zgirski2023}. The same formalism was extended to the mid-infrared, where extinction corrections become negligible and the intrinsic scatter reaches its minimum \citep{neeley2017, mullen2023}. To bypass star-by-star reddening corrections entirely, the period-Wesenheit framework uses a pseudo-magnitude defined as $W = X - R_X(Y-Z)$ \citep{madore1982}, where $R_X$ is the ratio of total to selective extinction in the adopted filter combination. This proved to be a powerful tool for heavily obscured fields \citep{soszynski2003, braga2015}. The Wesenheit index inherently mimics a period-luminosity-colour relation, allowing reddening-free distances to be estimated directly from a light curve. Its metallicity coefficient is generally comparable to that of the PLZ relations, with one notable exception: the optical $W(V,B-V)$ index, for which the metallicity term is found to be consistent with zero \citep{marconi2015}, allowing individual stellar distances to be estimated from this specific pseudo-magnitude without an external metallicity measurement, in contrast to the general case, where an independent abundance estimate is still required \citep{bono2008, braga2015}. Comprehensive theoretical models \citep{marconi2015, marconi2022, neeley2017} and Gaia-anchored empirical studies \citep{neeley2019, garofalo2022, mullen2023} now map $PWZ$ calibrations across nearly the entire optical, NIR, and mid-infrared passband regime, with the primary caveat of possible systematics, since they rely on the assumption of a universal extinction law.

Despite these advances, existing empirical PLZ and PWZ calibrations in the Johnson-Cousins optical bands and in the NIR still have three limitations: heterogeneous literature samples observed at different epochs and with different instruments, often with too few NIR phase points to recover reliable intensity-mean magnitudes; inconsistent metallicity scales mixing high- and low-resolution spectroscopy; and, before Gaia DR3, distance anchors with substantial uncertainties, with parallax errors remaining a major source of scatter in small samples (Benedict et al. 2002; Muraveva et al. 2018; Neeley et al. 2019). The last Gaia release has changed this landscape fundamentally: the DR3 (Lindegren et al. 2021; Eyer et al. 2023) parallaxes now typically reach fractional uncertainty $\lesssim5\%$ for nearby halo field stars, enabling empirical PLZ and PWZ calibrations anchored directly to geometric distances. However, no existing Gaia-anchored calibration has simultaneously combined i) homogeneous, well-sampled multi-band light curves for a large sample of nearby field RRL, ii) geometric distance anchors of high quality, and iii) metallicities on a uniform scale derived predominantly from high-resolution spectroscopy. The dataset presented in this paper is designed to provide all three simultaneously.

This paper is the first in a series that will fulfil  precisely this programme. We have conducted an extensive photometric campaign of 172 bright, nearby Galactic field RRL in up to eight photometric bands (Johnson-Cousins $UBVRI$ plus NIR $JHK_s$), with distances anchored to precise trigonometric parallaxes from Gaia DR3 and metallicities drawn from a single homogeneous spectroscopic catalogue, the Spectroscopic Rome RRL Catalog (SR3C; Bono et al. 2026). The sample covers the broad metallicity range $-3.4 \lesssim \feh \lesssim 0.2$~dex, and the metal-poor extreme and the metal-rich tail well represented.

In this Paper I, we describe the sample selection, the observational strategy, the photometric reduction and calibration pipelines, and we comprehensively characterise of the pulsational properties of the sample. The empirical PLZ and LZ calibrations that are the primary goal of this programme will be presented in the companion Paper II. The paper is structured as follows. Section~\ref{sec:photometry} presents the instrumental setups, observing strategy, and photometry. Section~\ref{sec:project} describes the sample properties and the treatment of distances and extinctions. Section~\ref{sec:lc} details the construction of the best-fit light curves and the derivation of mean magnitudes and amplitudes. Section~\ref{sec:analysis} presents the analysis of the multi-wavelength light-curve properties, including the relations between the Fourier phase parameter $\phi_{31}$ in all available passbands and its $V$-band value, NIR/optical amplitude ratios, and the first homogeneous calibration of the intrinsic colour at minimum light for all available filter combinations, from $U-B$ through $H-K_s$. Section~\ref{sec:conclusions} summarises our conclusions. 

\begin{figure}[th]
\centering
\includegraphics[width=\columnwidth]{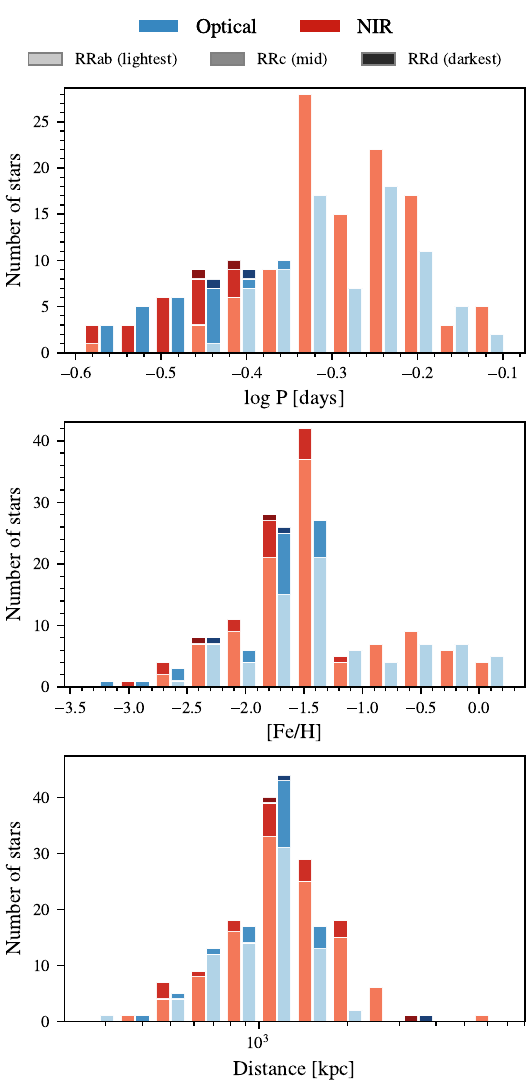}
  \caption{Upper panel: Marginal distribution of $\log(P)$ for the stars in both samples. The Johnson-Cousins sample is shown in blue and the NIR sample in red. RRab, RRc, and RRd are presented with increasing darkness of the base colour.
   Central panel: Marginal distribution of $\feh$ for our targets, following the same colour code.
  Lower panel: Marginal distribution of distances for our targets, with the same colour coding. Most of the stars are contained within 750 and 1800 pc.}
   \label{plot:period_metal_distance}
\end{figure}

\section{Observations and photometry} \label{sec:photometry}

We observed a total of 172 RRL in the optical and NIR. Optical observations were conducted with the IAC80 telescope in Tenerife, Spain, covering 101 stars in the Johnson–Cousins $UBVRI$ system. The choice of this facility was motivated by its flexibility for long-term programs, its semi-automatic capabilities that minimize human intervention, and the availability of a broad filter set. Access to Johnson filters was especially important to tie our results to the large legacy datasets accumulated over decades \citep[e.g.][]{stetson2014b}, while also maintaining compatibility with Sloan $ugriz$ photometry and future surveys such as Rubin/LSST. In the NIR, we used the InfraRed Survey Facility (IRSF) in Sutherland, South Africa, observing 130 stars in $JHK_s$ passbands. The IRSF was one of the few facilities suitable for bright halo stars at these wavelengths, given its small-aperture design and the availability of the required filters. In both wavelength regimes, the use of small telescopes was essential: our targets are bright and require only short exposures of a few seconds (longer in the bluest filters), making large telescopes inefficient. More importantly, the programme required extensive temporal coverage to build well-sampled light curves, which is best achieved at facilities where significant observing time can be allocated to a single long-term project. Our observing strategy consisted of repeated sequences of filters across multiple consecutive stars in each night, rather than densely monitoring a single target. This allowed us to collect phase points for many stars during a single observing night, steadily building full light-curve coverage over months of monitoring. Since the stars are widely distributed on the sky, multiple semesters were required to complete the survey, accounting also for weather and technical losses. Figure~\ref{plot:skymap} shows the distribution of the 172 target RRL in Galactic coordinates, colour-coded by metallicity.

\subsection{Optical dataset} \label{subsec:photometry_optical}

\begin{figure*}[ht]
\centering
\includegraphics[width=\columnwidth]{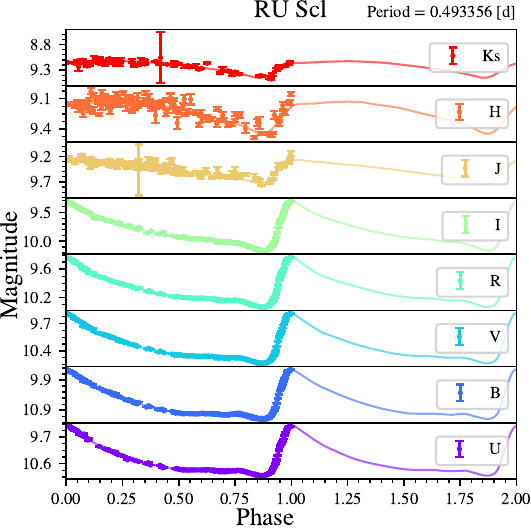}
\includegraphics[width=\columnwidth]{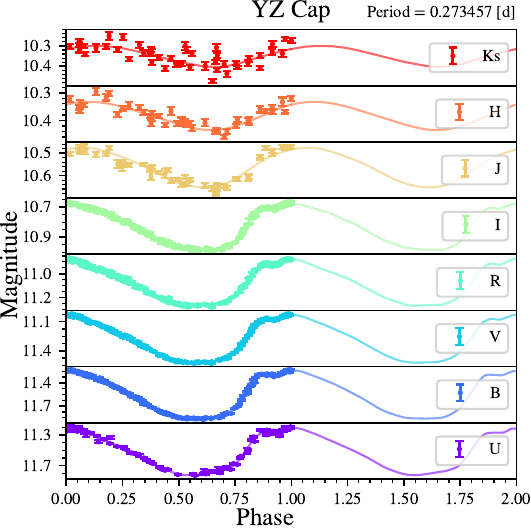}
  \caption{Phase-folded light curves for two typical stars from our sample: an RRab in the left panel, and an RRc in the right panel. For them, light curves for both sets of filters, Johnson-Cousins $UBVRI$ and NIR $JHK_s$ are available. Individual photometric measurements and their uncertainties are shown as the coloured points and error bars. The solid coloured line shows the model for the light curve obtained with the procedure described in Section~\ref{subsec:lcfit}: \gloess{} for the optical passbands, and templates for the NIR. Object name and period are shown in the title.}
     \label{plot:lightcurves}
\end{figure*}

\paragraph{Instrumentation}

Optical data were acquired with IAC80, an $0.82$m $f/11.3$ Cassegrain telescope with a German equatorial mount located at the Teide Observatory (OT) in Tenerife, Spain. Two instruments were in use during our observing period, CAMELOT and CAMELOT2. While both were used, the majority of the data were collected with CAMELOT2, which was installed in mid 2019. The first instrument consisted of a $2048\times2048$ pixel E2V Technologies CCD42-40 back illuminated detector, with a pixel scale of 0.304 arcsec/pixel, corresponding to a $10.4\times10.4$ arcmin field of view. CAMELOT2 is a $4096\times4112$ pixel Spectral Instruments back-illuminated CCD, with a pixel scale of $0.336$ arcsec/pixel, corresponding to a theoretical $23\times23$ arcmin field of view. Both detectors made use of the same twelve-position filter wheel, originally designed for the former. This wheel generates a circular vignetting pattern in CAMELOT2 images, leading to an effective field of view of $16.7$ arcmin diameter. A huge collection of filters\footnote{\url{https://research.iac.es/OOCC/iac-managed-telescopes/iac80/iac80-filters/}} can be installed in the filter wheel. For this particular project, the Johnson-Bessell \citep{bessell1990} $UBVRI$ set of filters was used.

\paragraph{Observations}

Observations were carried out between February 2017 and April 2026. In this period, a total of 679 nights were allocated to our projects, with 531 nights on which data were successfully obtained, with some of them being service nights that partially covered one of our programmes. Of these, 338 nights were used to obtain data in the $UBVRI$ filters.

For each target, we typically obtained a sequence of $UBVRI$ exposures, with exposure times that varied from a few seconds for the brighter targets up to $60$s for filter $U$ in most targets. These short exposures meant that guiding was not required, so targets could be changed after each sequence and images could be obtained directly upon pointing, enabling a more efficient observing cadence. During a typical night, between three and eight different fields were successively observed (although in 25$\%$ of the nights more than eight fields were visited), according to their visibility on the sky and to the gaps that needed to be filled in the light curves. Almost one fifth of the nights included the observation of standard fields, but only a fraction of these yielded a robust photometric solution. For every night, a set of semi-automatic custom scripts generated the targets to be observed. The whole campaign resulted in a total of $114{,}180$ individual science images for the selected targets. A summary of how these observations were distributed among the years of the project is available in Table \ref{table:obs_dates}.

A custom pipeline constructed with \texttt{Gnuastro}\footnote{\url{https://ascl.net/1801.009}} \citep{gnuastro2015}, \texttt{SExtractor} \citep{sextractor1996}, \texttt{astrometry} \citep{astrometry2010} and \texttt{SCAMP} \citep{scamp2006} was used for the reduction of the raw images, based on standard procedures, including bias subtraction and flat fielding using dome and sky flats.

\paragraph{Photometry}

From the reduced images, we performed aperture photometry to obtain instrumental magnitudes ($m_U$, $m_B$, $m_V$, $m_R$, $m_I$) using SExtractor. Standard magnitudes ($U$, $B$, $V$, $R$, $I$) and colours ($U-B$, $B-V$, $V-R$, $V-I$) for the standard fields observed were gathered from the large database of homogeneous photometry from P. B. Stetson, built up over the past decades \citep{stetson2000}. 

The photometric calibrations were computed employing a custom pipeline based on Python's \texttt{NumPy} \citep{numpy2020}, \texttt{Astropy} \citep{astropy2013, astropy2018, astropy2022}, \texttt{Scipy} \citep{scipy2020}, and \texttt{pandas} \citep{pandas2022}. We defined the following relations between the instrumental and the standardised magnitude:

\begin{equation} \label{eq:photcal}
X - m_X = \alpha_X + \beta_X \left(z-1\right)+\gamma_X (\Delta_X)
\end{equation}

where $X$ is the label for each of the five filters calibrated, $\alpha_X$ is the zero-point, $\beta_X$ is the extinction coefficient, $z$ is the airmass, $\gamma_X$ is the colour coefficient, and $\Delta_X$ is the standard colour associated with each filter, that is, $\{U:U-B,\,B:B-V,\,V:B-V,\,R:V-R,\,I:V-I\}$. These coefficients were computed for each photometric night in which standard fields were observed by performing a weighted least-squares fit, and are unique for each of those nights. We observed a median of seven standard fields per photometric night, with a median of 200 stars per field that were used for the fit. The mean values of the coefficients for all the nights, along with their standard deviations, are presented in Table \ref{table:photcal}. As a sanity check, for each photometric night we calibrated the frames containing the standard stars with the resulting calibration, and computed the difference among the measured calibrated magnitude and the tabulated one. We report the average values of that residuals in the last column of Table~\ref{table:photcal}, all of which are negligible for stars brighter than $m=15$. The results of this global operation are shown in Figure~\ref{plot:zeropoint_calibration}.

Each RRL was observed in at least one photometric night. To calibrate the instrumental magnitudes from non-photometric nights, we tied them to a reference frame from a photometric night for that field. We calculated the magnitude offset between each reduced frame and the reference frame by finding the common stars, subtracting their instrumental magnitudes, and averaging using a mean weighted by photometric errors. This average was only computed, however, after removing from the sample of common stars those that are classified as variables by the Gaia DR3 general catalogue \citep{eyer2023} and rejecting stars whose individual magnitude offset was more than $2\cdot1.4826\cdot MAD$ (median absolute deviation) away from the weighted mean. This offset was then applied to the instrumental magnitudes of the target reduced frame, thus absorbing extinction, colour, or other atmospheric and instrumental terms between the two images. Following this procedure, any image taken on a photometric night can be used as a reference. As the last step before the photometric calibration, all the observation dates were transformed into Barycentric Julian Dates.

After photometrically tying all the frames of a given star and filter to a common reference, we performed the photometric calibration itself. We then applied the coefficients from Equation~\ref{eq:photcal} obtained for the photometric night of the reference frame to the whole new dataset. As a first step, only $\alpha_X$ and $\beta_X$ terms were applied to the instrumental magnitude, as the calibrated colour was unknown. After this, a new colour is computed taking, for each point in filter $X$, the closest point in observing time for filter $Y$, and a solution is found using a fixed-point iteration method. In an ideal case, the whole system could be treated as a system of linear equations and be solved through a least-squares method, but in our case, as the number of points for each filter can be different, empty terms in the equations would bias the results. The typical mean errors of the calibrated magnitudes are $0.02$ mag in U, $0.007$ mag in B, $0.004$ mag in V and R, and $0.003$ mag in I.

\begin{table}
\caption{Number of images per year of observing campaign.}
\label{table:obs_dates}
\centering
\sisetup{
  table-number-alignment = center,
  table-text-alignment = center
}
\begin{tabular}{ccc}
\toprule
\toprule
{Year} & {Optical} & {NIR} \\
\midrule
\midrule
2014 & \nodata & 10160 \\
2015 & \nodata & 1075 \\
2016 & \nodata & 6815 \\
2017 & 5743 & 21720 \\
2018 & 5405 & 5060 \\
2019 & 9171 & 7100 \\
2020 & 13248 & \nodata \\
2021 & 11676 & \nodata \\
2022 & 16918 & \nodata \\
2023 & 24254 & 15075 \\
2024 & 12378 & \nodata \\
2025 & 9490 & 66800 \\
2026 & 5897 & \nodata \\
\bottomrule
\end{tabular}
\end{table}

\begin{table}
\caption{Coefficients of the photometric calibration.}  
\label{table:photcal}      
\centering
\resizebox{\columnwidth}{!}{%
\renewcommand{\arraystretch}{1.5}
{\setlength{\tabcolsep}{4pt}
\begin{tabular}{c
                S[table-format=-1.3] S[table-format=1.3]
                S[table-format=-1.3] S[table-format=1.3]
                S[table-format=-1.3] S[table-format=1.3]
                c}
\hline\hline
Filter & {$\alpha$} & {$\sigma_{\alpha}$} &
         {$\beta$} & {$\sigma_{\beta}$} &
         {$\gamma$} & {$\sigma_{\gamma}$} & {$m_{std,X}-m_{cal,X}$} \\
\hline
\hline
U & -6.549 & 0.444 & -0.390 & 0.097 & 0.156 & 0.041 & $(60.2\pm141.8)\cdot10^{-5}$\\
B & -4.034 & 0.409 & -0.205 & 0.033 & 0.119 & 0.014 & $(-58.7\pm368.9)\cdot10^{-5}$\\
V & -3.712 & 0.337 & -0.119 & 0.029 & -0.029 & 0.008 & $(-110.8\pm236.5)\cdot10^{-5}$ \\
R & -3.579 & 0.365 & -0.084 & 0.028 & -0.063 & 0.020 & $(-45.1\pm210.9)\cdot10^{-5}$ \\
I & -3.581 & 0.352 & -0.086 & 0.106 & 0.059 & 0.014 & $(2.4\pm229.6)\cdot10^{-5}$ \\
\hline
\end{tabular}}}
\tablefoot{Mean coefficients of the photometric calibration, and averaged difference between the tabulated standard magnitude and the measured calibrated magnitude in the calibration frames. Coefficients are defined in Equation~\ref{eq:photcal}. The zero-point is relative to 25.}
\end{table}

\subsection{NIR dataset} \label{subsec:photometry_NIR}

\paragraph{Instrumentation}

NIR data were acquired with the IRSF, an 1.4 $m$ $f/9.9$ Cassegrain telescope located at the South African Astronomical Observatory (SAAO) in Sutherland, South Africa. We employed the SIRIUS camera (Simultaneous Infrared Imager for Unbiased
Survey, \citep{nagashima1999, nagayama2003}, which consists of three $1024\times1024$ pixel HAWAII HgCdTe detectors with a field of view of $7.7\times7.7$ arcmin and pixel scale of $0.453$ arcsec/pixel. They operate simultaneously in the $JHK_s$ bands.

\paragraph{Observations}

Observations were carried out in three different periods: March 2014 to June 2019, June-July 2023 and November-December 2025. In these runs, a total of 163 nights were allocated to our projects, with 151 nights on which data for the RRL targets were successfully gathered.  

For each target, a series of dithered exposures was taken simultaneously in the $JHK_s$ bands. Typically, we acquired either five or ten dither positions with exposure times ranging from $2$s to $14$s, depending on target brightness. Several targets were pointed during each night. The whole campaign resulted in a total of $133{,}805$ exposures for the selected targets. A summary of how these observations were distributed in time is available in Table \ref{table:obs_dates}.

The pipeline constructed for the reduction of the raw optical images was adapted and used for this dataset, adding the extra steps of sky subtraction and dithered image combination. This process yielded $17,223$ final, combined science images from which we could extract the photometry.

\paragraph{Photometry}

From the reduced images, we performed aperture photometry to obtain instrumental magnitudes ($m_J$, $m_H$, $m_{K_{s}}$) using SExtractor. We adapted the pipeline used to calibrate the optical dataset to be used also with this sample, calibrating these magnitudes onto the 2MASS system by computing the magnitude offset between each reduced frame and 2MASS point sources in the field \citep{cutri2003} and calculating the average offset with a mean weighted by photometric errors. For this step, we first removed from the sample of common stars those that are classified as variables by the Gaia DR3 general catalogue and rejected stars whose individual magnitude offset was more than $2\cdot1.4826\cdot MAD$ away from the weighted mean. All observation dates were subsequently transformed into Barycentric Julian Dates. The typical mean errors of the calibrated magnitudes are $0.01$ mag in all three bands.

\section{Sample description and properties}\label{sec:project}

In this section, we describe the main properties of the RRL sample that we observed in the Johnson-Cousins \textit{U, B, V, R, I} and in the NIR \textit{J, H, Ks} passbands. We present their pulsational properties, $\feh$, and the general treatment followed concerning distances and extinction on each individual source. In Table~\ref{table:rrl_sample} we present the sample of 172 RRL along with these main properties.

\subsection{Optical sample} \label{subsec:opt_sample}

The optical sample includes 101 RRL stars: 77 fundamental mode pulsators (76.2\%), 22 first-overtone pulsators (21.8\%), and two double-mode pulsators (2.0\%), as determined by classifications in the literature and our own light curves. Among them, 33 stars (32.7\%) show the Blazhko effect, specifically 29 RRab (37.7\% of the fundamental mode subsample) and 4 RRc (18.2\% of the first-overtone subsample), according to the Blazhko Stars of the Galactic Field database \citep[\texttt{BlaSGalF}\footnote{\url{https://www.physics.muni.cz/~blasgalf}},][]{skarka2013}. Because Blazhko amplitude modulations can span from a few tenths to several hundredths of a magnitude and Blazhko periods can range from days to years, not all Blazhko stars necessarily display detectable modulation in our light curves. In some specific cases, even, the Blazhko effect has been reported to have turned off in the recent times (for example, RR Lyrae itself). Figure~\ref{plot:period_metal_distance} shows the period (upper panel), $\feh$ (centre), and distance (lower panel) marginal distributions for fundamental, first-overtone, and double-mode variables for the optical sample in blue colours. The sample covers typical pulsation periods and a wide range of iron abundances. 

Periods were initially adopted from the Gaia DR3 RRL catalogue \citep{clementini2023} and ASAS-SN \citep{shappee2014, jayasinghe2018}. These values were then refined using a Python based GUI that allows visual inspection of all photometric data and computation of optimised periods for our light curves. The GUI employs the Lomb–Scargle periodogram \citep{lomb1976, scargle1982} through its \texttt{astropy} implementation \citep{vanderplas2015}, computed over a user-defined frequency window as the core for period determination.

Notably, some stars not identified as Blazhko candidates exhibit period variations, occurring either between the Gaia DR3 epoch and our observations or within our own observing window. On the timescales probed by our data, these variations give rise to various distinct point sequences that cannot be reconciled with a single period. Such variations can be analysed using $O-C$ (observed minus calculated) diagrams of the times of maximum light. A valuable resource for this purpose is the GEOS (Groupe Européen d’Observation Stellaire) RRL database\footnote{\url{https://rr-lyr.irap.omp.eu/dbrr/index.php}} \citep{leborgne2007}, which currently contains approximately 160,000 maximum times for roughly 4,000 RRL stars. We inspected the $O-C$ diagrams for the nine stars displaying such behavior in our data, confirming either noisy or quadratic trends. These objects are flagged in the corresponding column of Table~\ref{table:rrl_sample}. For each of them we adopted the period that produces the cleanest phased light curve, minimising the number of points displaced by an imperfect period.

\subsection{NIR sample} \label{subsec:nir_sample}

The NIR sample of RRL targeted a total of 130 stars. 109 RRab (83.8\%), 19 RRc (14.6\%), and two RRd (1.5\%), based on literature classifications and our own analysis. Among them, 48 exhibit the Blazhko effect (36.2\%), divided into 42 RRab (38.5\% of all fundamental pulsators) and six RRc (31.6\% of first-overtone pulsators), according to the \texttt{BlaSGalF} database cited in Section~\ref{subsec:opt_sample}, and the ASAS-SN light curve for V535 Mon. Additionally, it is worth mentioning that a total of 59 stars are present in the NIR and optical samples (47 RRab, ten RRc, and two RRd). Figure~\ref{plot:period_metal_distance} shows the period (upper panel), $\feh$ (centre) and distance (lower panel) marginal distributions for RRab, RRc, and RRd for the NIR sample in red colours. Note that the current sample covers the typical range in periods and a very broad range in iron abundances.

Periods were determined using the same approach as described in Section~\ref{subsec:opt_sample}. For stars in common with the optical sample, we adopted the periods derived from the optical data unless clear evidence of period variation between the two observing campaigns was present. In that case, the procedures outlined in Section~\ref{sec:lc} were applied independently, and the period that led to the cleanest light curve for each set of filters was adopted.

\subsection{Distances and extinction} \label{subsec:reddening}
For the whole sample, parallaxes and their uncertainties were obtained from the third data release of the Gaia space mission (Gaia DR3, \citealt{eyer2023}), and corrected by the zero-point offset proposed by \citet{lindegren2021} employing the public Python package\footnote{\url{https://gitlab.com/icc-ub/public/gaiadr3_zeropoint}}. The distribution of distances for the optical and NIR samples is shown in the lower panel of Figure~\ref{plot:period_metal_distance}.
To ensure homogeneous reddening corrections, we employed the 3D extinction map by \citet{edenhofer2024}. They provide a dimensionless measurement of extinction as defined in \citet{zhang2023}, which can be transformed back into extinction in any filter contained within the limits of the extinction curve\footnote{\url{https://zenodo.org/records/7811871}} presented in that same source. The main map covers distances from 69 $pc$ to 1.25 $kpc$ with a resolution of 14$'$. An additional map covers up to 2 $kpc$ by using less data in its construction. By employing the \texttt{dustmaps} Python package \citep{green2018}, we computed $E(B-V)$ and its uncertainty from the extinction curve and then applied the empirical reddening law provided by \citet{cardelli1989} to calculate the extinction in the different filters, as the $U$ filter lies outside the extinction curve from \citet{zhang2023}. From our whole sample, 93 RRL lie within the limits of the first map, and an additional 64 stars can be found in the second map. The treatment followed for the remaining 15 stars that were outside both maps is described in Section~\ref{subsec:colour_min}.

Since the release of Gaia DR3, a number of studies have claimed that the \citet{lindegren2021} correction overcorrects for the parallax offset \citep[e.g.][]{gilligan2021,riess2021a,cruzreyes2023,khan2023}. As an alternative check, we also computed distances from parallaxes corrected by an additional offset of $-13 \,\mu\mathrm{as}$ \citep[the median of the literature values proposed for the magnitude range spanned by our stars]{groenewegen2024}, to test the sensitivity of our reddening estimates to the assumed parallax zero-point. This consideration does not impact the reddening estimation: the median variation between $E(B-V)$ (computed with and without the counter-correction) is 0.4$\%$, with only five stars showing up to a 2$\%$ difference. In every single case, the one $\sigma$ uncertainty for each $E(B-V)$ is at least a factor of two bigger than this difference. 

\subsection{Iron abundances} \label{subsec:feh}

 Iron abundances ($\feh$) for the full sample are drawn from SR3C (Spectroscopic Rome RRL Catalog; \citealt{bono2026}). Among the 172 RRL stars, 145 have abundances derived from high-resolution (HR) spectra, 22 from low-resolution (LR) spectra, and no measurement is available for the remaining 5, all of which belong to the NIR sample. The marginal distribution of $\feh$ for the optical and NIR samples is shown in the central panel of Figure~\ref{plot:period_metal_distance}. The five NIR stars without a $\feh$ estimate are necessarily excluded. A breakdown by subsample is provided in Table~\ref{table:fe_sources}. For our sample of stars, HR measurements of $\feh$ were estimated from individual Fe{\sc i} and Fe{\sc ii} lines and have been gathered from the following sources: i) \citet{dorazi2025a}, ii) \citet{dorazi2024}, iii) \citet{crestani2021a, crestani2021b}, iv) \citet{gilligan2021}, v) GALAH DR3. The metallicity scale adopted throughout is the same as used by \citet{magurno2018, magurno2019} for cluster RRL variables. $\feh$ from GALAH DR3 \citep{buder2021} were brought onto the same scale via a linear analytical relation \citep[Equation 1 of ][]{bono2026}. On the other hand, $\feh$ abundances based on LR spectra ($\Delta S$ method) come from the following sources: i) \citet{fabrizio2021}, ii) \citet{zinn2020}. Full details on the spectroscopic treatment, including proprietary data, the total sample composition, and literature measurements, are given in \citet{bono2026} and references therein.

\begin{table}[h]
\caption{Source for the iron abundances for the whole sample of stars.}
\label{table:fe_sources}
\begin{center}
\resizebox{\columnwidth}{!}{%
\begin{tabular}{l ccc ccc ccc}
\hline\hline
Dataset & \multicolumn{3}{c}{RRab} & \multicolumn{3}{c}{RRc} & \multicolumn{3}{c}{RRd} \\
\cmidrule(lr){2-4} \cmidrule(lr){5-7} \cmidrule(lr){8-10} 
 & HR & LR & X & HR & LR & X & HR & LR & X\\
\hline
Optical & 66 & 11 & \nodata & 18 & 4 & \nodata & 1 & 1 & \nodata \\
NIR     & 97 & 9 & 3 & 15 & 2 & 2 & 1 & 1 & \nodata \\
Common  & 43 & 4  & \nodata & 10 & 1 & \nodata & 1 & 1 & \nodata \\
\hline
Whole    & 120 & 16 & 3 & 24 & 5 & 2 & 1 & 1 & \nodata \\
\hline
\end{tabular}
}
\end{center}
\tablefoot{HR stands for high-resolution spectra, LR for low-resolution, and the X for stars with no source for the iron abundance. The row “Whole” gives the combined, non duplicated sample of 172 RRL (Optical + NIR - Common).}
\end{table}

\section{Light-curve properties} \label{sec:lc}

\subsection{Light-curve quality criteria} \label{subsec:quality_crit}
We constructed the light curves for each individual star for all the filters available by phasing the photometry time series with the period contained in Table \ref{table:rrl_sample}. Some examples are shown in Figure \ref{plot:lightcurves}. Hereafter, we employed the following parameters to asses the quality of a light curve:

i) Phase coverage ($\Phi_c$), defined as the fraction of bins occupied by at least a phase point in the phase range $[0,1)$ in a grid of 50 bins, that is, every $0.02$ in phase.

ii) Maximum gap in phase in the light curve ($\Delta_{\Phi}$).

iii) Number of points of the light curve ($n_p$).

iv) A uniformity index constructed from the Kuiper test, by comparing the distribution of points in phase with the distribution that would arise from the same number of point distributed uniformly ($U_K$).

In the next subsection we define the methods used to generate light-curve models from the data: the \gloess{} algorithm and template fitting. The values chosen for these four parameters are different depending on whether the light curves are constructed through the \gloess{} algorithm (for the $UBVRI/$Gaia filters) or from templates, as their sparsity differs considerably. For the former, we selected light curves with $\Phi_c>0.6$, $\Delta_{\Phi}<0.15$, and $U_K<0.4$, as the \gloess{} algorithm does not behave well if gaps are too large, or if phase is not well covered. For the latter, the values were $\phi_c>0.3$, $\Delta_{\Phi}<0.4$, and $U_K<0.5$, as fitting a template makes it so that those parameters do not need to be so tight. Whenever the light-curve parameters such as amplitude, magnitude at the minimum, or Fourier parameters are involved in any kind of fitting procedure, we also selected $n_p>10$ as a limiting last threshold that, in practice, only affects the NIR data.

\subsection{Construction of the uniformly phase-sampled light curves} \label{subsec:lcfit}

In order to compute the different parameters that can be derived from the light curves, such as amplitudes, mean magnitudes, instantaneous magnitudes, or combinations of these parameters, we needed to create evenly phase-sampled light-curve models from the non-uniformly phase-sampled photometry. For this purpose we decided to follow two different approaches, depending on the degree of sparseness of the number of points available:

i) For the Johnson-Cousins \textit{U, B, V, R, I}  bands, where the light curves are well sampled, we build a smoothed model using \gloess{} \citep[Gaussian-windowed LOcal regrESSion, ][]{persson2004}, a variant of the LOESS algorithm \citep[LOcal regrESSion, ][]{cleveland1979,cleveland1988}. A phase grid uniformly spaced between zero and one is generated and extended by two full periods on either side (phase -2 to 3) to avoid edge effects. For each point in this grid, a second degree polynomial is fitted locally to the photometric data. Unlike standard LOESS, every observed point contributes to every fit, weighted by a Gaussian function of its phase distance to the grid point, $e^{-\Delta_\phi^2/\sigma_\phi^2}$, further convolved with each point's photometric uncertainty. The width $\sigma_\phi$ of this Gaussian window is implemented in two ways: (a) a single fixed value for the whole light curve (the fixed-window version), or (b) a value that varies with the local density of points (the adaptive-window version), set at each phase to the distance to the k-th nearest neighbour (k = square root of the number of points in the light curve), and then smoothed with a Gaussian filter. The adaptive-window version captures fine structure in densely sampled regions and improves the fit near rapidly changing features such as the rising and descending branches of RRab stars; however, when the rising or descending branch is under-sampled relative to the rest of the curve, the same adaptivity can over smooth and flatten its shape. We therefore compute both versions for every star and filter and adopt whichever gives the lower RMS residual as the final light-curve model.

ii) For the NIR \textit{J, H Ks}, there is much more variety between the different stars in terms of phase coverage and number of phase points. For that reason, we decided to use the templates provided by \cite{braga2019a}. We prepared a custom version of \verb|gatspy| \citep{vanderplas2015} in order for those templates to be available in the pool of the \verb|RRLyraeTemplateModeler| class. A $\chi^2$ minimisation between the observed photometry and the \citet{braga2019a} templates is performed, solving for amplitude, phase shift, and mean magnitude simultaneously. Employing the results from Section~\ref{subsec:amplitude_ratio} (where we compute the ratios between the amplitudes in the different passbands and the Johnson $V$ and Gaia $G$ bands), caps on the minimum and maximum amplitudes were established when fitting the templates. This was done by limiting the amplitudes to the values contained within the intrinsic scatter of the relations derived in Section~\ref{subsec:amplitude_ratio}. 

iii) To obtain Gaia band G$_{BP}$, G, and G$_{RP}$ light-curve parameters analogous to those derived in i) and ii) above, we used the \gloess{} algorithm and the SDSS template library of \citet{sesar2010} to fit the light curves extracted from Gaia data \citep{clementini2023}, using the same $\chi^2$ minimisation approach for the template fitting. The SOS Cep\&RRL pipeline \citep[the Gaia DPAC “Specific Object Study” pipeline that characterises RRL and Cepheid light curves,][]{clementini2023} already provides these parameters, but some stars of our sample were not covered by it, and some parameters (period, amplitude, mean magnitude, etc.) were missing. Only the stars that did not fulfil the minimum quality criteria from Section~\ref{subsec:quality_crit} fell back to template fitting, as this method yields more realistic results than SOS for poorly sampled stars.

We applied the corresponding approach independently to every filter of each star in an iterative way to perform a sigma-clipping over the data. Distances between the phased photometry and the uniform light-curve model can be computed either as the vertical distance ($\Delta_{mag}$) at a fixed phase, or as the perpendicular distance to the curve in the (phase, magnitude) plane, and are to be used as the residual over which the 3$\sigma$ outliers are rejected. The two differ most where the light curve is steep, such as the rising branch of RRab stars. There, a small phase uncertainty translates into a large vertical $\Delta_{mag}$ even for an otherwise well behaved point, which can make the vertical distance criterion reject good points (or fail to reject bad ones). The perpendicular distance instead measures the geometric separation from the model, correctly accounting for the local slope of the curve, and is therefore more robust in these steep regions. The results of this whole procedure are shown in Figure \ref{plot:lightcurves}, where the best fit is represented as a solid line over the phased photometry.

\subsection{Mean magnitude and amplitude} \label{subsec:meanmag}

The mean magnitude can be estimated by taking the uniform light curve (either \gloess{} or template constructed), transforming it to flux, computing the mean as the integral of the curve over the phase range [0,1], and then converting back to magnitude. The amplitude is determined as the difference between the minimum and maximum magnitudes of the uniform light curve. To ensure accuracy, we sampled each uniform light curve at every $0.001$ in phase, that is, $1000$ points over the full cycle.

While a single fit to each light curve yields point estimates of the mean magnitude and amplitude, it does not capture the uncertainties arising from finite sampling and photometric errors. To address this, we adopted a bootstrap Monte Carlo approach. Specifically, for each star we generated $1000$ realisations of the light-curve data by randomly resampling the observed magnitudes, errors, and phases (with replacement), and then perturbing the magnitudes by adding Gaussian noise scaled by their photometric errors. For every resampled and perturbed dataset, we reconstructed the uniform light curve using the procedure depicted in the previous section, and measured its mean magnitude and amplitude.

This approach produces an empirical distribution of light curves consistent with the observations, from which average values and realistic uncertainties can be derived for every parameter under study. The final values reported in Table \ref{table:magnitude} correspond to the mean and standard deviation of these distributions.
Importantly, this method naturally incorporates the intrinsic scatter present in the data, such as the modulation associated with Blazhko stars, into the resulting uncertainty estimates.

\subsection{Comparison with the literature}

\begin{figure}[t]
\centering
\includegraphics[width=\columnwidth]{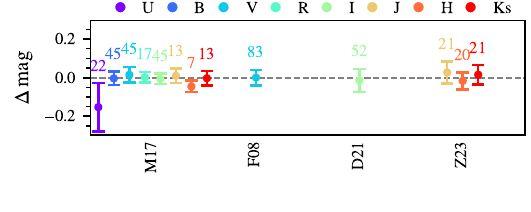}
  \caption{Comparison between our intensity-averaged mean magnitudes and literature values, grouped by comparison sample along the x-axis. Points show the median magnitude offset (this work minus literature) for each filter, with error bars showing the standard deviation of the offset distribution. Colours indicate the photometric filter, from $U$ (purple) to $K_s$ (red), as shown in the legend. The number above each point gives the number of stars in common with that literature sample. The dashed grey line marks zero offset.}
     \label{plot:comparison_literature}
\end{figure}

We compared the intensity averaged magnitudes of our photometry with those in the literature by subtracting the literature mean magnitude from our own measured mean magnitude. As the most comprehensive analysis to present date, we choose \citet[M17]{monson2017} as the main source for comparison, given their extensive coverage of 55 Galactic field RRL in a variable number of filters (Johnson $UBV$, Kron-Cousins $R_{C}I_{C}$, 2MASS $JHK_s$ and Spitzer $[3.6]$ and $[4.5]$ bandpasses). Their main study was performed in the $BVI$ filters, with the remaining passbands coming from a variety of literature sources available at the time.
Figure~\ref{plot:comparison_literature} summarises the offsets and their dispersion for every filter and comparison sample discussed below. Across all bands, our photometry agrees with M17 to within the combined uncertainties, with the largest offset found in U ($\Delta U=-0.153\pm0.125$), likely reflecting the smaller number of stars in common (22) and the larger intrinsic scatter of that band.

We further compared our V-band photometry with \citet[F08]{feast2008}, who provide mean magnitudes for 142 RRL derived from Hipparcos photometry transformed into the Johnson V system using a colour correction common to each pulsation type, rather than the star's own observed colour; our I-band photometry with \citet[D21]{dekany2021}, who recalibrate mean magnitudes from 129 light curves gathered mainly from ASAS \citep{pojmanski1998} and M17; and our NIR $JHK_s$ photometry with \citet[Z23]{zgirski2023}, who provide $JHK_s$ photometry for 28 Galactic RRL. In all three cases the offsets are compatible with zero within the uncertainties (Figure~\ref{plot:comparison_literature}), confirming the consistency of our photometric calibration across the optical and NIR regimes.

\section{Analysis of the multi-wavelength light curves} \label{sec:analysis}
The comprehensive set of light curves, amplitudes, and mean magnitudes derived in Section~\ref{sec:lc} allows for a detailed investigation of the pulsational behaviour of our RRL sample across multiple wavelength ranges. In this section, we use these parameters to explore the physical properties of the stars and the relationships between their light-curve morphologies in the optical and NIR.

We begin by defining the quality criteria adopted for the light curves analysed in this section (Section~\ref{subsec:quality_crit}). Next, we investigate the relations between the $\phi_{31}$ Fourier parameter determined for a given photometric band and its value for the V band, examining how light-curve shape changes across filters (Section~\ref{subsec:fourier_relation}). We subsequently perform an analogous study of the pulsation amplitudes, focusing on amplitude ratios and their dependence on wavelength, in particular with respect to the V and G bands (Section~\ref{subsec:amplitude_ratio}). Finally, we recalibrate the colour at minimum light as a reddening indicator using our multi-band dataset and independent reddening estimates from 3D reddening maps, extending this relation to a wide range of filter combinations (Section~\ref{subsec:colour_min}). Every single relation is derived by applying the unified linear fit procedure described in Appendix~\ref{sec:linear_fit}.

\subsection{Relation between Fourier parameter $\phi_{31}$ determined for different passbands} \label{subsec:fourier_relation}

\begin{figure}
\centering
\includegraphics[width=\columnwidth]{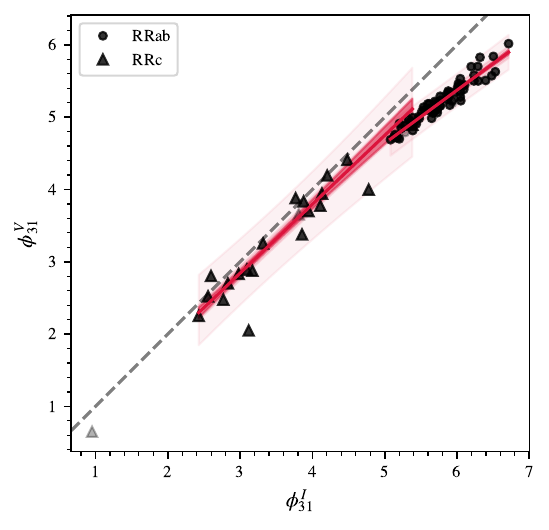}
  \caption{Fourier parameter $\phi_{31}$ in the $I$ vs. $V$ band for both RRab (circles) and RRc (triangles) stars. Grey points come from light curves not fulfilling the minimum requirements established in Section~\ref{subsec:quality_crit} and therefore were not included in the fit. The red line and the shaded areas are described in Section~\ref{subsec:fourier_relation}, and show the linear relation and its uncertainty, alongside the intrinsic scatter of the relation itself. The dashed grey line marks the 1:1 relation. Fit parameters are presented in Table~\ref{table:phi31_relation}.}
     \label{plot:phi31_relations}
\end{figure}

Following \cite{mullen2021, mullen2022}, we determined the Fourier parameters of our optical light curves by fitting the mentioned Fourier series (Equation~\ref{eq:fourier_series}) to the smooth evenly sampled uniform light-curve models obtained through the \gloess{} algorithm (Section \ref{subsec:lcfit}), rather than to the data themselves. Fourier parameters obtained for the NIR light curves are also determined by modelling their light curves using the \gloess{} algorithm, as obtaining those parameters from the templates that are fitted to NIR data (see Section~\ref{subsec:lcfit}) would have no physical meaning, due to their fixed shape. The general representation of the stellar magnitude as a function of phase phi, $m(\Phi)$, is

\begin{equation}
\label{eq:fourier_series}
m(\Phi) = A_0 + \sum_{i=1}^{N} \left[ a_i \cos(2\pi i \Phi) + b_i \sin(2\pi i \Phi) \right],
\end{equation}
where $m(\Phi)$ is the stellar magnitude modelled using the Fourier series as a function of phase $\Phi$ (ranging from 0 to 1), and $a_i$ and $b_i$ are the $i$th-order Fourier coefficients for the cosine and the sine terms, respectively. From this general representation, the equivalent sine-only or cosine-only expansion can be written,
\begin{equation}
\label{eq:fourier_series_sin}
m(\Phi) = A_0 + \sum_{i=1}^{N} A_i \sin\!\big(2\pi i \Phi + \phi_i^s\big).
\end{equation}
Here, $A_0$ is the zeroth-order term that approximates the mean magnitude of the star, and $A_i$ and $\phi_i^s$ are the $i$th-order Fourier amplitudes and phases, respectively. The cosine-only expansion is completely equivalent, with phases defined differently, as detailed in Equation~\ref{eq:fourier_relation}.

From these amplitudes and phases, the conventional Fourier parameters used to characterise the light-curve morphology can be computed:
\begin{align}
\label{eq:fourier_parameters}
R_{ij} &= \frac{A_i}{A_j}, \\
\phi_{ij} &= \phi_i - i \phi_j,
\end{align}
where $R_{ij}$ are amplitude ratios (e.g., $R_{21}$ and $R_{31}$) and $\phi_{ij}$ are epoch-independent phase differences (e.g., $\phi_{21}$ and $\phi_{31}$).

We chose to fit the general formulation in Equation~\ref{eq:fourier_series} rather than the sine-only expression of Equation~\ref{eq:fourier_series_sin} because the former is fully linear in the coefficients, ensuring larger numerical stability. The two formulations are strictly equivalent, and the transformation between them is straightforward,
\begin{align}
\label{eq:fourier_relation}
A_i &= \sqrt{a_i^2 + b_i^2}, \\
\phi_i^s &= \arctan2(a_i, b_i), &\phi_i^c = \arctan2(-b_i, a_i), 
\end{align}
where $\phi_i^s$ and $\phi_i^c$ are the $i$th-order Fourier phases in the sine-only and cosine-only representations, respectively. In practice, either formulation can be adopted; the choice of the general series simply provides a more robust fitting framework while preserving access to the conventional sine or cosine-based Fourier parameters.

Uncertainties and final values of the Fourier parameters were derived through the same bootstrap Monte Carlo approach described in Section \ref{subsec:meanmag}. For each of the $1000$ resampled and perturbed datasets, we reconstructed the light curve with \gloess{} and fitted a fifth-order Fourier series, following \cite{mullen2021}. This procedure yields, for every star and filter, an ensemble of smooth light curves and their corresponding Fourier decompositions. From the resulting distributions of Fourier amplitudes and phases, we computed circular means and standard deviations, which we adopt as the final parameter values and their uncertainties. In this way, the reported Fourier parameters consistently incorporate the effects of photometric errors, sampling, and phase coverage. 

The phase difference $\phi_{31}$, is the one that is mainly used in the literature to constrain the relation between $\feh$ and the light-curve shape. The recent calibrations of the $\feh$-period-$\phi_{31}$ relation in the $V$ band provided by \citet{mullen2021,mullen2022} with a large sample of RRL stars with spectroscopic $\feh$ yields a robust model from which $\feh$ estimates can be obtained. For these two reasons, we investigated the relation between $\phi_{31}$ value in the $V$ band ($\phi_{31}^{V}$) and the corresponding counterpart in each of the other passbands ($\phi_{31}^{\lambda}$). The full set of determined $\phi_{31}$ is presented in Table \ref{table:phi31}. Taking $\phi_{31}^{\lambda}$ as the independent variable and $\phi_{31}^V$ as the dependent variable, we performed a linear fit following the prescription detailed in Appendix~\ref{sec:linear_fit}, only selecting the light curves that met the quality threshold established in Section~\ref{subsec:quality_crit} for the \gloess{} constructed light curves, as $\phi_{31}$ comes directly from them. We also removed stars with Blazkho behavior from the sample, as their Fourier parameters are less constrained. The whole set of results of the linear fit for RRab and RRc are shown in Table~\ref{table:phi31_relation}, with a representative example of the fit for the $VI$ relation presented in Figure~\ref{plot:phi31_relations}. Results of the linear fit on Table~\ref{table:phi31_relation} are presented in the form $\phi_{31}^{\lambda}-\phi_{31,0}^{\lambda}=\alpha+\beta \left(\phi_{31}^{V}-\phi_{31,0}^{V} \right)$, with $\phi_{31,0}=\overline{\phi_{31}}$ (a line over a variable denotes the mean value of that variable). Throughout Section~\ref{sec:analysis}, the solid line in each figure is the median of the posterior relation from the MCMC fit of Appendix~\ref{sec:linear_fit}, that is, at each $x$, the median of $\alpha+\beta x$ over all samples after burn-in. The darker shaded band is the $1\sigma$ credible interval on this fitted relation itself (uncertainties on $\alpha$ and $\beta$ determination); the two lighter bands are the  $1\sigma$ and  $3\sigma$ prediction intervals, which additionally include the fitted intrinsic scatter  $\sigma_{int}$ and show where new individual data points are expected to fall. RRab show tight relations across all filter pairs, yielding slope coefficients greater than one for filters bluer than $V$, and less than one for redder filters. In contrast, RRc slope coefficients remain below one in all cases, though they approach unity in the better sampled $BRI$ bands. Additionally, intrinsic scatter is much larger for RRc than for RRab. When examining the $V$-band to NIR $\phi_{31}$ relation for RRc, our NIR light curves are sufficiently noisy to produce uncertainties ($\Delta\phi^{NIR}_{31}$) exceeding 0.6 radians in all cases, and 1.2 radians for half the sample. Because extracting Fourier phase parameters from nearly sinusoidal RRc light curves is highly sensitive to noise, particularly for high-order parameters such as $\phi_3$, attempting a reliable fit between $\phi^V_{31}$ and $\phi^{NIR}_{31}$ is impractical.

In Appendix~\ref{sec:phi31_comparison} we compare the  $\phi_{31}^I-\phi_{31}^V$ relation derived here with similar relations widely used in the literature. For the RRab, we find the best agreement with the theoretical models of \citet{dorfi1999}, while the empirical relations of \citet{deb2010} and \citet{skowron2016} lie systematically below ours. For the RRc, all the compared linear fits are consistent within uncertainties.

\subsection{NIR/optical amplitude ratios} \label{subsec:amplitude_ratio}

\begin{figure}
\centering
\includegraphics[width=\columnwidth]{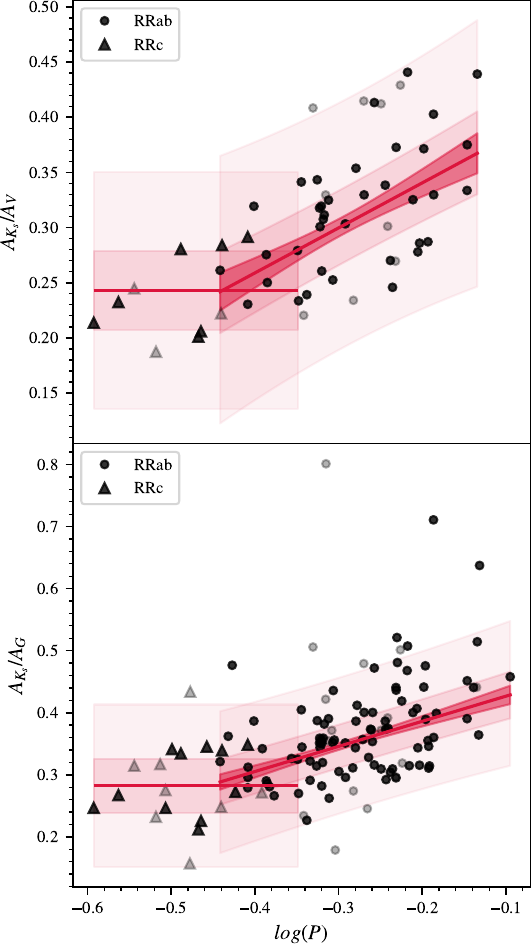}
  \caption{Upper panel: Amplitude ratio $A_{K_s}/A_V$ as a function of the logarithm of period for RRab (circles) and RRc (triangles) stars. Grey points come from light curves not fulfilling the minimum requirements established in Section~\ref{subsec:quality_crit} and therefore were not included in the fit. For RRab, the red line and the shaded areas show the linear relation adopted in Section~\ref{subsec:amplitude_ratio} and its uncertainty, alongside the intrinsic scatter of the relation itself; their general computation is described in Section~\ref{subsec:fourier_relation}. For RRc, they show the weighted mean and weighted standard deviation.
  Lower panel: Same as i), but for the amplitude ratio $A_{K_s}/A_G$. 
  The resulting parameters of the whole set of fits are presented in Table~\ref{table:ampl_ratio_relation}.}
     \label{plot:amplitude_period_relation}
\end{figure}

We investigate the amplitude ratios in the optical and in the NIR passbands, as they are a useful tool for the template fitting. Amplitudes for individual stars in every filter available can be found in Table~\ref{table:amplitude}. We took the amplitude ratio between a given filter and the Johnson-Cousins $V$ or Gaia $G$ filters as the $y$ variable, and the $\log_{10}P$ of the RRL as the $x$ variable. Performing the linear fit as detailed in Appendix~\ref{sec:linear_fit}, only selecting the light curves that met the quality thresholds established in Section~\ref{subsec:quality_crit} and also restricting the sample to the non-Blazhko stars, we obtained the results shown in Table~\ref{table:ampl_ratio_relation}. We note that the fit was only performed for the RRab stars, where a trend is clearly appreciable, and not for the RRc stars, for which we present a weighted mean and standard deviation as the measure of these ratios. The whole set of results is presented in the form $r^{\lambda}-r_{0}^{\lambda}=\alpha+\beta \left(\log_{10}(P)-L_0 \right)$, where $r^{\lambda}$ is the ratio between the amplitude in a given filter $\lambda$ and the reference filter ($V$ or $G$), $r^{\lambda}_0=\overline{r^{\lambda}}$, and $L_0=\overline{\log_{10}(P)}$. We show an example of the fit for the $A_V/A_{K_s}$ relation in Figure~\ref{plot:amplitude_period_relation}. 

Most of the works regarding these relations have been addressed by employing globular cluster (GC) data, thus with samples of stars with a small spread in $\feh$. In the case of RRab stars, and for the optical bands, \cite{stetson2014a}, \cite{braga2016} and \cite{jurcsik2018} find constant or quasi-constant ratios for $A_B/A_V$, $A_B/A_R$ and $A_I/A_V$. \citet{braga2024} show a small positive trend for the ratio $A_I/A_V$ for the RRL in the bulge, a constant ratio for the LMC, and a big scatter in both samples. Our results show that for the $A_R/A_V$ and $A_I/A_V$ the dependence of the ratio with period exhibits a positive but almost negligible trend, confirming their results. This is not the case with $A_B/A_V$, where we find a clear sign of a decreasing relation. That trend is even stronger with the bluer $U$ filter. For the $A_G/A_V$ a positive trend also appears, consistent with the fact that the Gaia $G$ passband includes a large portion of the Johnson-Cousins $RI$ passbands. For the NIR bands, \cite{braga2018} and \cite{bhardwaj2020,bhardwaj2021} find an increasing trend with period, but not in a continuous way: they define a break that splits RRab into two groups with short and long periods, defining constant ratios for each. This split changes its position depending on the cluster, and \cite{bhardwaj2020,bhardwaj2021} suggest this shift is related to the Oosterhoff type of the cluster. However, as we show in Figure~\ref{plot:amplitude_period_relation} for the $K_s$ band (results are similar for the other two NIR bands), we do not find this shift but a continuous increase in the values of the ratio for every $A_{JHK_s}/A_V$ or $A_{JHK_s}/A_G$, as our sample is heterogeneous in $\feh$. \cite{jurcsik2018} also find this increasing linear trend in the $A_{K_s}/A_I$ ratio for bulge stars. When one puts these two results together, it suggests that this behaviour in the GCs is just a selection bias due to the absence of metal-intermediate stars \citep{fabrizio2019}. In the case of RRc stars, the constant amplitude ratio is always smaller than the one for RRab, as noted by previous works.

\subsection{RRL colours at minimum light} \label{subsec:colour_min}

We computed the colour at minimum light ($C^{min}$) for every single RRL and filter combination using the three ways already in use in the literature, that is, employing the magnitude at phase $\phi=0.65$ \citep{vivas2017}, the magnitude at the minimum of the light curve \citep{kunder2008}, and the average magnitude over phase $\phi=0.5-0.8$ \citep{guldenschuh2005}. For the final fit, we decided to use the latter, even though \cite{kunder2008} found that the magnitude at the light-curve minimum is less strongly correlated with the RRL amplitude, for two main reasons: i) measuring the magnitude at a specific phase point is a non-trivial task unless the light curve is very well populated, and ii) averaging over phase can dilute the errors introduced when templates are used to fit a light curve. Each of that measurements was obtained alongside with its uncertainty following the same process described in Section~\ref{subsec:meanmag} for the magnitude and the amplitude.

To calibrate $C^{min}$ as a extinction proxy, we investigated its dependence on period and on $\feh$. For every possible colour combination, we took this average colour at minimum light as the independent variable and computed two different fits: the first, comprising $\log_{10}P$ and $\feh$ as dependent variables; the second, only with $\log_{10}P$. For each of the two fits, we checked whether the dependence on $\feh$ had a statistical significance over a period-only dependence by computing the Bayesian Information Criterion (BIC) for the two models. We performed the linear fit as detailed in Appendix~\ref{sec:linear_fit}, only selecting the light curves that met the quality thresholds established in Section~\ref{subsec:quality_crit}. The fit was carried out for RRab and RRc stars. For RRc stars, a weighted mean and linear fit yield similar numbers for the BIC, meaning that none of the models is favoured. However, the GC data used for validation in Section~\ref{sec:colour_min_validation} suggests an increasing trend of the RRc colour at minimum with period, so we kept that model as the preferred one.  The sample of stars that fulfil the thresholds comprises 77 RRab/22 RRc stars in the optical, 93 RRab/16 RRc in the NIR (88 RRab/13 RRc in the combinations including the $K_s$ filter), and 45 RRab/10 RRc in the optical-NIR relations (42 RRab/9 RRc in the combinations including the $K_s$ filter).

\begin{figure}
\centering
\includegraphics[width=\columnwidth]{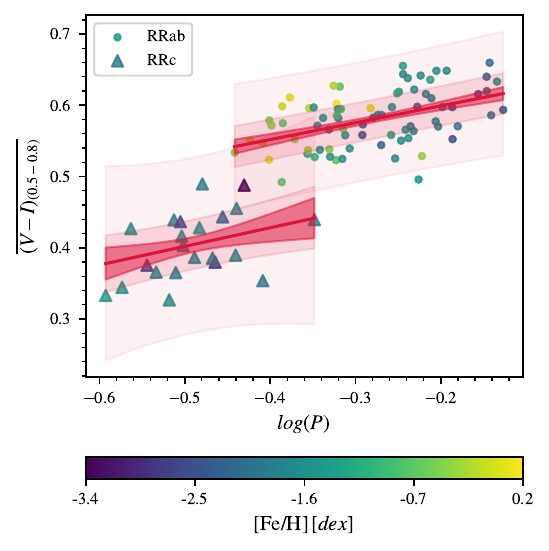}
  \caption{Colour at light-curve minimum as a function of the logarithm of the period, as defined in Section~\ref{subsec:colour_min}, for RRab (circles) and RRc (triangles) stars. The red line and shaded areas show the linear relation adopted in Section~\ref{subsec:colour_min} and its uncertainty, alongside the intrinsic scatter of the relation itself; their general computation is described in Section~\ref{subsec:fourier_relation}. The points are colour-coded by $\feh$.
  The resulting parameters of the whole set of fits are presented in Table~\ref{table:colour_min_relation}. }
     \label{plot:colour_min_relation}
\end{figure}

We present the whole set of results for every possible filter combination in Table~\ref{table:colour_min_relation}, following the equation $C^{min}-C^{min}_{0}=\alpha+\beta \left( \log_{10}(P)-L_0 \right)+\gamma \left( \feh-\feh_0 \right)$, where $C^{min}_{0}=\overline{C^{min}}$, $L_0=\overline{\log_{10}(P)}$, and $\feh_0 = \overline{\feh}$. When the difference in BIC between the fit with one independent variable (period) or two (period and $\feh$) shows positive values ($\gtrsim2$), we report an existing dependence on $\feh$. An example of the fit for the $V-I$ colour is shown in Figure~\ref{plot:colour_min_relation}. 

Regarding RRab, \cite{guldenschuh2005} reported based on a sample of only 16 RRab that for the $V-R$ colour, $C_{0}^{min}=0.579\pm0.024$ and no dependence on period or $\feh$. As shown in Figure~\ref{plot:colour_min_relation}, we find a significant trend with period for the same filter combination, with the $y$-axis pivot ($C^{min}_{0}=0.581$) and intrinsic scatter ($0.026$) being compatible with their result, not finding a dependence on $\feh$. For the $V-R$ combination, \cite{kunder2010} found a value of $C^{min}_{0}=0.27\pm0.02$ employing a sample of 31 RRab stars, with no dependence on period or $\feh$. We find a similar result, with a negligible dependence of colour on period, $0.062\pm0.052$, almost compatible with zero within $1\sigma$, with a $C^{min}_{0}=0.297$ and an intrinsic scatter of $0.023$, and no dependence with $\feh$. For the $B-V$ combination, \citet{sturch1966} found $0.38\lesssim(B-V)_{min}\lesssim0.46$, compatible with our result of an almost zero value for the slope of the relation with period ($\beta=-0.031\pm0.065$) and a mean value of $0.422$ with an intrinsic scatter of $0.028$. However, this changes when a possible dependence on $\feh$ is included, which in this case is small but significant. Every combination with the two bluest bands $U$ and $B$ yields a statistically significant dependence on $\feh$. 

Regarding RRc, we find an increasing trend of the colour at minimum with period for almost every possible filter combination but for $U-B$, $BVRI$ combinations with $J$, and $H-K_s$. However, uncertainties in the slope coefficient for those combinations are comparable to the coefficient itself. As noted earlier in this section, the data support the linear and constant value (weighted mean) models, resulting in weak constraints on the slope for some filter combinations, and making the use of $C^{min}_0$ an equally valid approach. Remarkably, we do find that a dependence with $\feh$ is strongly disfavoured for every filter combination.

Finally, there were reasons to expect the $V-K_s$ colour to be the most precise for RRLs, as it is less sensitive to metallicity and atmospheric shocks and should theoretically trace temperature variations with less scatter \citep{cacciari2000}. Bluer colours are more sensitive to metallicity, micro/macro turbulence and affected by non-linear phenomena (shocks) that cause the appearance of peculiar spectral features. This means that bluer colours at minimum would show a larger scatter when compared with colours that are significantly less affected by the quoted phenomena. However, the measured intrinsic scatter for $V-K_s$ remains comparable to combinations including the $VRI$ bands. This is likely because the current scatter is dominated by photometric errors and by uncertainties in the reddening map rather than by intrinsic physical properties.

The performance of these relations is validated against a sample of GC with independent literature reddening estimates; the full procedure is presented in Appendix~\ref{sec:colour_min_validation}. All filter combinations reproduce the literature $E(B-V)$ values well. Relations involving NIR passbands are tested against a smaller sample due to the scarcity of NIR light curves in GCs. For the $UB$ combinations specifically, including the $\feh$ term measurably reduces the residuals with respect to the literature values.

We employed these results to compute $E(B-V)$ for the subsample of stars that could not get an estimate from the \cite{edenhofer2024} reddening map, as described in Section~\ref{subsec:reddening}. A total of 15 stars lay outside the coverage of the mentioned map: 12 RRab, two RRc, and one RRd. All of the RRab and RRc stars are only present in the NIR subsample, so we employed the relation for the $J-H$ colour, given its smaller intrinsic scatter. We computed their colour excess by subtracting the predicted colour from the observed colour and transformed it into colour excess $E(B-V)$ by employing the \cite{cardelli1989} reddening law. These stars' $E(B-V)$ are marked with a dagger in Table~\ref{table:rrl_sample}. Finally, for the RRd, CU Com, we took the mean and standard deviation of the RRc sample in the $V-R$ colour (the one with smaller scatter in the optical) and repeated that same procedure, as the first-overtone mode is dominant in RRd stars.

\section{Conclusions} \label{sec:conclusions}

We present the first paper in a series in which we derive new empirical PLZ, LZ, and PWZ calibrations for RRL stars. This work provides the foundational dataset, anchored simultaneously to homogeneous, multi-band photometry, Gaia DR3 geometric distances, and a uniform, predominantly HR spectroscopic metallicity scale. As discussed in the Introduction, previous Gaia-anchored calibrations have lacked this simultaneous synthesis; the sample characterised here was specifically designed to address this gap and support the derivations in Paper II.

The dataset itself is one of the main products of this work. It comprises 172 bright, nearby Galactic field RRL stars observed across up to 11 photometric bands: 101 stars with Johnson-Cousins $UBVRI$ photometry from the IAC80 telescope (77 RRab, 22 RRc, and two RRd), 130 with $JHK_s$ NIR photometry from the IRSF (109 RRab, 19 RRc, and two RRd), with 59 targets in common, and all but one with Gaia $G$, $BP$, and $RP$ light curves. Iron abundances, adopted from the SR3C catalogue \citep{bono2026} for 167 stars (145 of them from HR spectroscopy), span a wide range ($-3.4 \lesssim \feh \lesssim 0.2$ dex). This continuous coverage of the metal-poor and metal-rich tails is an improvement over GC samples, which are limited by narrow metallicity spreads and therefore offer only a restricted diagnostic power. Our photometric calibration, tied to the P. B. Stetson standard stars \citep{stetson2000} for the Johnson-Cousins $UBVRI$ and to 2MASS \citep{cutri2003} for the NIR $JHK_s$, reaches an internal photometric precision below 0.01 mag in $BVRIJHK_s$ and 0.02 mag in $U$, showing statistical consistency with independent literature photometry within the uncertainties. The distances rely on Gaia DR3 parallaxes corrected for the \citet{lindegren2021} zero-point offset. We verified that adopting the alternative zero-point offset proposed in different sources of the literature \citep{groenewegen2024} changes our reddening estimates by a median of only $0.4\%$, which is well within their uncertainties. This confirms that our results are robust against this systematic effect. 

Beyond the catalogue construction, the multi-wavelength nature of this dataset yields three primary constraints on the pulsational morphology of field RRL stars. First, the Fourier phase parameter $\phi_{31}$ is tightly correlated with its $V$-band value for RRab stars throughout the entire optical range, from $U$ to $I$ and in all three Gaia bands, with an intrinsic scatter below $0.07$. These relations are sufficiently precise to transform $\phi_{31}$ based $\text{[Fe/H]}$ estimates \citep[e.g.][]{mullen2021,mullen2022} between passbands, which is useful for combining or cross-calibrating heterogeneous survey datasets (OGLE, ASAS-SN, and Gaia). As expected from lower amplitudes and noisier light curves, the corresponding NIR relations exhibit larger scatter. Second, the NIR amplitude ratios $A_{JHK_s}/A_V$ and $A_{JHK_s}/A_G$ increase continuously with period, with no sign of the period break reported in several GC studies \citep[e.g.][]{braga2018,bhardwaj2020,bhardwaj2021}. Because our field sample spans a much broader metallicity range than any single cluster, this suggests that the reported break is a selection effect driven by the absence of metal intermediate stars in GC, and is not an intrinsic pulsational feature \citep{fabrizio2019}. Third, we provide the first internally consistent calibration of the intrinsic colour at minimum light from $U-B$ to $H-K_s$. While recovering mean values consistent with earlier determinations \citep{sturch1966,guldenschuh2005,kunder2010}, our sample reveals statistically significant period dependences, and for combinations involving the bluer $UB$ bands, metallicity dependences that have not been detected previously. We validated the relations against 44 globular clusters, leading to a good agreement with literature reddening estimates \citep{harris2010,schlafly2011}, including independent confirmation of the metallicity dependence noted above for the $UB$ combinations. These relations were subsequently applied to derive $E(B-V)$ for the 15 stars from our sample that are located outside the boundaries of the \citet{edenhofer2024} 3D extinction map.

Certain limitations must be considered when these relations are applied more broadly. The RRc and RRd subsamples are relatively small, particularly where the optical and NIR datasets intersect (ten RRc and two RRd stars). This means that relations derived for these subclasses, especially in the optical-NIR combinations, carry larger uncertainties and should be treated as indicative only. Finally, our reddening determinations depend on the assumption of a universal \citet{cardelli1989} extinction law when transforming between the \citet{zhang2023} extinction curve and individual filters, which does not account for potential line-of-sight variations.

In summary, we established the observational and methodological framework for the remainder of this series. We provide a chemically diverse field sample with homogeneous multi-band light curves, precise geometric distances, and uniform high-resolution metallicities, that will be used in Paper II to derive empirical PLZ, PWZ, and LZ relations. Methodologically, the transformation relations between passbands (Section~\ref{subsec:fourier_relation}), the amplitude ratio trends (Section~\ref{subsec:amplitude_ratio}), and the colour at minimum calibrations (Section~\ref{subsec:colour_min}), supported by a Bayesian fitting framework that rigorously handles measurement errors, intrinsic scatter, and outliers, are reusable tools that can be applied to template fitting and reddening estimation for the deeper samples anticipated from upcoming surveys.

\section*{Data availability}
Tables \ref{table:rrl_sample}, \ref{table:magnitude}, \ref{table:amplitude}, and \ref{table:phi31} are only available in electronic form at the CDS via anonymous ftp to cdsarc.u-strasbg.fr (130.79.128.5) or via \url{http://cdsweb.u-strasbg.fr/cgi-bin/qcat?J/A+A/}.

\begin{acknowledgements}
M.S.B. and M.Mo. warmly thank Alejandro Oscoz Abad and José Acosta Pulido, the telescope operators, and the support astronomers of the Telescope Operation Group of the IAC, for the continuous support that allowed us to realize long term observations with the IAC80.
We are thankful to the IRSF team, in particular Takahiro Nagayama and Takayoshi Kusune, for their long-term supports for our NIR observations. We also acknowledge Takumi Hanaue, Jill Neeley, and Christina Gilligan, who contributed to collect a part of the NIR data used in this work.
This paper made use of the IAC HTCondor facility (https://research.cs.wisc.edu/htcondor/), partly financed by the Ministry of Economy and Competitiveness with FEDER funds, code IACA13-3E-2493.
M.S.B and M.Mo. acknowledge support from Spanish Ministry of Science, Innovation and Universities (MICIU) through the Spanish State Research Agency under the grants "RR Lyrae stars, a lighthouse to distant galaxies and early galaxy evolution" and the European Regional Development Fund (ERDF) with reference PID2021-127042OB-I00. 
CG, M.S.B and M.Mo. acknowledge support from the Agencia Estatal de Investigación del Ministerio de Ciencia e Innovación (AEI-MCINN) under grant "At the forefront of Galactic Archaeology: evolution of the luminous and dark matter components of the Milky Way and Local Group dwarf galaxies in the Gaia era" with reference PID2023-150319NB-C21/C22/10.13039/501100011033. 
M.Mo. acknowledges support from the project ``ASTRA: Across Space and Time: Relics \& Archaeology'' (P.I. M. Marconi) funded by INAF-Instituto Nazionale di Astrofisica.
M.Mo. and G.F. acknowledge the INAF projects "Participation in LSST - Large Synoptic Survey Telescope" (LSST inkind contribution ITA-INA-S22, PI: G. Fiorentino), OB.FU. 1.05.03.06 and "MINI-GRANTS (2023) DI RSN2" (PI: G. Fiorentino), OB.FU. 1.05.23.04.02.
This research was supported by the International Space Science Institute (ISSI) in Bern/Beijing through ISSI/ISSI-BJ International Team project ID \#24-603 - "EXPANDING Universe" (EXploiting Precision AstroNomical Distance INdicators in the Gaia Universe).
M.Marconi was supported by the grants Project PRIN MUR 2022 (code 2022ARWP9C) "Early Formation and Evolution of Bulge and HalO (EFEBHO)", PI: Marconi, M.,  funded by European Union – Next Generation EU, and the Large grant INAF 2023 MOVIE (PI: M. Marconi). M.Marengo is supported by the National Science Foundation under grant No. AST-2407965.
Supported by the European Union (ERC-2024-AdG "ChronoGal: Chronology of our Galaxy from Gaia CMD-fitting", Grant Agreement 101201412). Views and opinions expressed are however those of the author(s) only and do not necessarily reflect those of the European Union or the European Research Council. Neither the European Union nor the granting authority can be held responsible for them.
This article is partly based on observations made with the IAC80 telescope operated on the island of Tenerife by the Instituto de Astrof\'isica de Canarias in the Spanish Observatorio del Teide. 
The IRSF project is a collaboration between Nagoya University and the South African Astronomical Observatory (SAAO) supported by the Grants-in-Aid for Scientific Research on Priority Areas (A) (No. 10147207 and No. 10147214), Optical \& Near-Infrared Astronomy Inter-University Cooperation Program, the Ministry of Education, Culture, Sports, Science and Technology (MEXT) of Japan and NRF of South Africa.
\end{acknowledgements}

\bibliographystyle{aa} 
\bibliography{bibliography}

@ARTICLE{blazhko1907,
       author = {{Bla{\v{z}}ko}, S.},
        title = "{Mitteilung {\"u}ber ver{\"a}nderliche Sterne}",
      journal = {Astronomische Nachrichten},
         year = 1907,
        month = aug,
       volume = {175},
        pages = {325},
          doi = {10.1002/asna.19071752002},
       adsurl = {https://ui.adsabs.harvard.edu/abs/1907AN....175..325B}
}

@ARTICLE{shapley1916,
       author = {{Shapley}, H.},
        title = "{On the changes in the spectrum, period, and lightcurve of the Cepheid variable RR Lyrae.}",
      journal = {\apj},
         year = 1916,
        month = apr,
       volume = {43},
        pages = {217-233},
          doi = {10.1086/142246},
       adsurl = {https://ui.adsabs.harvard.edu/abs/1916ApJ....43..217S}
}

@ARTICLE{sturch1966,
       author = {{Sturch}, C.},
        title = "{Intrinsic UBV colors of RR LYR stars.}",
      journal = {\apj},
         year = 1966,
        month = mar,
       volume = {143},
        pages = {774},
          doi = {10.1086/148557},
       adsurl = {https://ui.adsabs.harvard.edu/abs/1966ApJ...143..774S}
}

@ARTICLE{iben1971,
       author = {{Iben}, Jr., I. and {Huchra}, J.},
        title = "{Comments on the Instability Strip for Halo Population Variables}",
      journal = {\aap},
         year = 1971,
        month = sep,
       volume = {14},
        pages = {293},
       adsurl = {https://ui.adsabs.harvard.edu/abs/1971A&A....14..293I}
}

@ARTICLE{lomb1976,
       author = {{Lomb}, N.~R.},
        title = "{Least-Squares Frequency Analysis of Unequally Spaced Data}",
      journal = {\apss},
         year = 1976,
        month = feb,
       volume = {39},
       number = {2},
        pages = {447-462},
          doi = {10.1007/BF00648343},
       adsurl = {https://ui.adsabs.harvard.edu/abs/1976Ap&SS..39..447L}
}

@article{cleveland1979,
    author = {William S. Cleveland},
    title = {Robust Locally Weighted Regression and Smoothing Scatterplots},
    journal = {Journal of the American Statistical Association},
    volume = {74},
    number = {368},
    pages = {829--836},
    year = {1979},
    publisher = {ASA Website},
    doi = {10.1080/01621459.1979.10481038},
    URL = {https://www.tandfonline.com/doi/abs/10.1080/01621459.1979.10481038},
    eprint = {https://www.tandfonline.com/doi/pdf/10.1080/01621459.1979.10481038}
}

@ARTICLE{simon1981,
       author = {{Simon}, N.~R. and {Lee}, A.~S.},
        title = "{The structural properties of cepheid light curves.}",
      journal = {\apj},
         year = 1981,
        month = aug,
       volume = {248},
        pages = {291-297},
          doi = {10.1086/159153},
       adsurl = {https://ui.adsabs.harvard.edu/abs/1981ApJ...248..291S}
}

@ARTICLE{sandage1981,
       author = {{Sandage}, A.},
        title = "{The Oosterhoff period groups and the age of globular clusters. II. Properties of RR Lyrae stars in six clusters: the P-L-A relation.}",
      journal = {\apj},
         year = 1981,
        month = aug,
       volume = {248},
        pages = {161-176},
          doi = {10.1086/159140},
       adsurl = {https://ui.adsabs.harvard.edu/abs/1981ApJ...248..161S}
}

@ARTICLE{madore1982,
       author = {{Madore}, B.~F.},
        title = "{The period-luminosity relation. IV. Intrinsic relations and reddenings for the Large Magellanic Cloud Cepheids.}",
      journal = {\apj},
         year = 1982,
        month = feb,
       volume = {253},
        pages = {575-579},
          doi = {10.1086/159659},
       adsurl = {https://ui.adsabs.harvard.edu/abs/1982ApJ...253..575M}
}

@ARTICLE{scargle1982,
       author = {{Scargle}, J.~D.},
        title = "{Studies in astronomical time series analysis. II. Statistical aspects of spectral analysis of unevenly spaced data.}",
      journal = {\apj},
         year = 1982,
        month = dec,
       volume = {263},
        pages = {835-853},
          doi = {10.1086/160554},
       adsurl = {https://ui.adsabs.harvard.edu/abs/1982ApJ...263..835S}
}

@ARTICLE{longmore1986,
       author = {{Longmore}, A.~J. and {Fernley}, J.~A. and {Jameson}, R.~F.},
        title = "{RR Lyrae stars in globular clusters : better distances from infrared measurements ?}",
      journal = {\mnras},
         year = 1986,
        month = may,
       volume = {220},
        pages = {279-287},
          doi = {10.1093/mnras/220.2.279},
       adsurl = {https://ui.adsabs.harvard.edu/abs/1986MNRAS.220..279L}
}

@article{cleveland1988,
    author = {William S. Cleveland and Susan J. Devlin},
    title = {Locally Weighted Regression: An Approach to Regression Analysis by Local Fitting},
    journal = {Journal of the American Statistical Association},
    volume = {83},
    number = {403},
    pages = {596--610},
    year = {1988},
    publisher = {ASA Website},
    doi = {10.1080/01621459.1988.10478639},
    URL = {https://www.tandfonline.com/doi/abs/10.1080/01621459.1988.10478639},
    eprint = {https://www.tandfonline.com/doi/pdf/10.1080/01621459.1988.10478639}
}

@ARTICLE{cardelli1989,
       author = {{Cardelli}, Jason A. and {Clayton}, Geoffrey C. and {Mathis}, John S.},
        title = "{The Relationship between Infrared, Optical, and Ultraviolet Extinction}",
      journal = {\apj},
         year = 1989,
        month = oct,
       volume = {345},
        pages = {245},
          doi = {10.1086/167900},
       adsurl = {https://ui.adsabs.harvard.edu/abs/1989ApJ...345..245C}
}

@ARTICLE{sandage1990,
       author = {{Sandage}, Allan and {Cacciari}, Carla},
        title = "{The Absolute Magnitudes of RR Lyrae Stars and the Age of the Galactic Globular Cluster System}",
      journal = {\apj},
         year = 1990,
        month = feb,
       volume = {350},
        pages = {645},
          doi = {10.1086/168417},
       adsurl = {https://ui.adsabs.harvard.edu/abs/1990ApJ...350..645S}
}

@ARTICLE{bessell1990,
       author = {{Bessell}, M.~S.},
        title = "{UBVRI passbands.}",
      journal = {\pasp},
         year = 1990,
        month = oct,
       volume = {102},
        pages = {1181-1199},
          doi = {10.1086/132749},
       adsurl = {https://ui.adsabs.harvard.edu/abs/1990PASP..102.1181B}
}

@ARTICLE{blanco1992,
       author = {{Blanco}, V.~M.},
        title = "{The reddening of Type AB RR Lyrae Stars}",
      journal = {\aj},
         year = 1992,
        month = aug,
       volume = {104},
        pages = {734},
          doi = {10.1086/116269},
       adsurl = {https://ui.adsabs.harvard.edu/abs/1992AJ....104..734B}
}

@ARTICLE{sextractor1996,
       author = {{Bertin}, E. and {Arnouts}, S.},
        title = "{SExtractor: Software for source extraction.}",
      journal = {\aaps},
         year = 1996,
        month = jun,
       volume = {117},
        pages = {393-404},
          doi = {10.1051/aas:1996164},
       adsurl = {https://ui.adsabs.harvard.edu/abs/1996A&AS..117..393B}
}

@ARTICLE{jurcsik1996,
       author = {{Jurcsik}, J. and {Kovacs}, G.},
        title = "{Determination of [Fe/H] from the light curves of RR Lyrae stars.}",
      journal = {\aap},
         year = 1996,
        month = aug,
       volume = {312},
        pages = {111-120},
       adsurl = {https://ui.adsabs.harvard.edu/abs/1996A&A...312..111J}
}

@ARTICLE{akritas1996,
       author = {{Akritas}, Michael G. and {Bershady}, Matthew A.},
        title = "{Linear Regression for Astronomical Data with Measurement Errors and Intrinsic Scatter}",
      journal = {\apj},
         year = 1996,
        month = oct,
       volume = {470},
        pages = {706},
          doi = {10.1086/177901},
archivePrefix = {arXiv},
       eprint = {astro-ph/9605002},
 primaryClass = {astro-ph},
       adsurl = {https://ui.adsabs.harvard.edu/abs/1996ApJ...470..706A}
}

@ARTICLE{harris1996,
       author = {{Harris}, William E.},
        title = "{A Catalog of Parameters for Globular Clusters in the Milky Way}",
      journal = {\aj},
         year = 1996,
        month = oct,
       volume = {112},
        pages = {1487},
          doi = {10.1086/118116},
       adsurl = {https://ui.adsabs.harvard.edu/abs/1996AJ....112.1487H}
}

@ARTICLE{pojmanski1998,
       author = {{Pojmanski}, G.},
        title = "{The All Sky Automated Survey}",
      journal = {\actaa},
         year = 1997,
        month = oct,
       volume = {47},
        pages = {467-481},
          doi = {10.48550/arXiv.astro-ph/9712146},
archivePrefix = {arXiv},
       eprint = {astro-ph/9712146},
 primaryClass = {astro-ph},
       adsurl = {https://ui.adsabs.harvard.edu/abs/1997AcA....47..467P}
}

@ARTICLE{dorfi1999,
       author = {{Dorfi}, E.~A. and {Feuchtinger}, M.~U.},
        title = "{Theoretical UBVI light curves of RR Lyrae stars}",
      journal = {\aap},
         year = 1999,
        month = aug,
       volume = {348},
        pages = {815-824},
       adsurl = {https://ui.adsabs.harvard.edu/abs/1999A&A...348..815D}
}

@INPROCEEDINGS{nagashima1999,
       author = {{Nagashima}, C. and {Nagayama}, T. and {Nakajima}, Y. and {Tamura}, M. and {Sugitani}, K. and {Nagata}, T. and {Hirao}, T. and {Nakaya}, H. and {Yanagisawa}, K. and {Sato}, S.},
        title = "{Development of SIRIUS --- A Simultaneous-Color InfraRed Imager for Unbiased Survey}",
    booktitle = {Star Formation 1999},
         year = 1999,
       editor = {{Nakamoto}, T.},
        month = dec,
        pages = {397-398},
       adsurl = {https://ui.adsabs.harvard.edu/abs/1999sf99.proc..397N}
}

@INPROCEEDINGS{cacciari2000,
       author = {{Cacciari}, C. and {Clementini}, G. and {Castelli}, F. and {Melandri}, F.},
        title = "{Revised Baade-Wesselink Analysis of RR Lyrae Stars}",
    booktitle = {IAU Colloquium 176: The Impact of Large-Scale Surveys on Pulsating Star Research},
         year = 2000,
       editor = {{Szabados}, L. and {Kurtz}, D.},
       series = {Astronomical Society of the Pacific Conference Series},
       volume = {203},
        month = jan,
        pages = {176-181},
          doi = {10.48550/arXiv.astro-ph/9909350},
archivePrefix = {arXiv},
       eprint = {astro-ph/9909350},
 primaryClass = {astro-ph},
       adsurl = {https://ui.adsabs.harvard.edu/abs/2000ASPC..203..176C}
}

@ARTICLE{stetson2000,
       author = {{Stetson}, Peter B.},
        title = "{Homogeneous Photometry for Star Clusters and Resolved Galaxies. II. Photometric Standard Stars}",
      journal = {\pasp},
         year = 2000,
        month = jul,
       volume = {112},
       number = {773},
        pages = {925-931},
          doi = {10.1086/316595},
archivePrefix = {arXiv},
       eprint = {astro-ph/0004144},
 primaryClass = {astro-ph},
       adsurl = {https://ui.adsabs.harvard.edu/abs/2000PASP..112..925S}
}

@ARTICLE{bono2001,
       author = {{Bono}, G. and {Caputo}, F. and {Castellani}, V. and {Marconi}, M. and {Storm}, J.},
        title = "{Theoretical insights into the RR Lyrae K-band period-luminosity relation}",
      journal = {\mnras},
         year = 2001,
        month = sep,
       volume = {326},
       number = {3},
        pages = {1183-1190},
          doi = {10.1046/j.1365-8711.2001.04655.x},
archivePrefix = {arXiv},
       eprint = {astro-ph/0105481},
 primaryClass = {astro-ph},
       adsurl = {https://ui.adsabs.harvard.edu/abs/2001MNRAS.326.1183B}
}

@INPROCEEDINGS{nagayama2003,
       author = {{Nagayama}, Takahiro and {Nagashima}, Chie and {Nakajima}, Yasushi and {Nagata}, Tetsuya and {Sato}, Shuji and {Nakaya}, Hidehiko and {Yamamuro}, Tomoyasu and {Sugitani}, Koji and {Tamura}, Motohide},
        title = "{SIRIUS: a near infrared simultaneous three-band camera}",
    booktitle = {Instrument Design and Performance for Optical/Infrared Ground-based Telescopes},
         year = 2003,
       editor = {{Iye}, Masanori and {Moorwood}, Alan F.~M.},
       series = {Society of Photo-Optical Instrumentation Engineers (SPIE) Conference Series},
       volume = {4841},
        month = mar,
        pages = {459-464},
          doi = {10.1117/12.460770},
       adsurl = {https://ui.adsabs.harvard.edu/abs/2003SPIE.4841..459N}
}

@dataset{cutri2003,
       author = {{Cutri}, R.~M. and {Skrutskie}, M.~F. and {van Dyk}, S. and {Beichman}, C.~A. and {Carpenter}, J.~M. and {Chester}, T. and {Cambresy}, L. and {Evans}, T. and {Fowler}, J. and {Gizis}, J. and {Howard}, E. and {Huchra}, J. and {Jarrett}, T. and {Kopan}, E.~L. and {Kirkpatrick}, J.~D. and {Light}, R.~M. and {Marsh}, K.~A. and {McCallon}, H. and {Schneider}, S. and {Stiening}, R. and {Sykes}, M. and {Weinberg}, M. and {Wheaton}, W.~A. and {Wheelock}, S. and {Zacarias}, N.},
        title = "{VizieR Online Data Catalog: 2MASS All-Sky Catalog of Point Sources (Cutri+ 2003)}",
 howpublished = {VizieR On-line Data Catalog: II/246.  Originally published in: University of Massachusetts and Infrared Processing and Analysis Center, (IPAC/California Institute of Technology) (2003)},
         year = 2003,
        month = jun,
          eid = {II/246},
       adsurl = {https://ui.adsabs.harvard.edu/abs/2003yCat.2246....0C}
}

@ARTICLE{soszynski2003,
       author = {{Soszy{\'n}ski}, I. and {Udalski}, A. and {Szyma{\'n}ski}, M. and {Kubiak}, M. and {Pietrzy{\'n}ski}, G. and {Wo{\'z}niak}, P. and {{\.Z}ebru{\'n}}, K. and {Szewczyk}, O. and {Wyrzykowski}, {\L}.},
        title = "{The Optical Gravitational Lensing Experiment. Catalog of RR Lyr Stars in the Large Magellanic Cloud}",
      journal = {\actaa},
         year = 2003,
        month = jun,
       volume = {53},
        pages = {93-116},
          doi = {10.48550/arXiv.astro-ph/0306041},
archivePrefix = {arXiv},
       eprint = {astro-ph/0306041},
 primaryClass = {astro-ph},
       adsurl = {https://ui.adsabs.harvard.edu/abs/2003AcA....53...93S}
}

@ARTICLE{bono2003,
       author = {{Bono}, G. and {Caputo}, F. and {Castellani}, V. and {Marconi}, M. and {Storm}, J. and {Degl'Innocenti}, S.},
        title = "{A pulsational approach to near-infrared and visual magnitudes of RR Lyr stars}",
      journal = {\mnras},
         year = 2003,
        month = oct,
       volume = {344},
       number = {4},
        pages = {1097-1106},
          doi = {10.1046/j.1365-8711.2003.06878.x},
archivePrefix = {arXiv},
       eprint = {astro-ph/0306142},
 primaryClass = {astro-ph},
       adsurl = {https://ui.adsabs.harvard.edu/abs/2003MNRAS.344.1097B}
}

@ARTICLE{catelan2004,
       author = {{Catelan}, M. and {Pritzl}, Barton J. and {Smith}, Horace A.},
        title = "{The RR Lyrae Period-Luminosity Relation. I. Theoretical Calibration}",
      journal = {\apjs},
         year = 2004,
        month = oct,
       volume = {154},
       number = {2},
        pages = {633-649},
          doi = {10.1086/422916},
archivePrefix = {arXiv},
       eprint = {astro-ph/0406067},
 primaryClass = {astro-ph},
       adsurl = {https://ui.adsabs.harvard.edu/abs/2004ApJS..154..633C}
}

@ARTICLE{persson2004,
       author = {{Persson}, S.~E. and {Madore}, Barry F. and {Krzemi{\'n}ski}, W. and {Freedman}, Wendy L. and {Roth}, M. and {Murphy}, D.~C.},
        title = "{New Cepheid Period-Luminosity Relations for the Large Magellanic Cloud: 92 Near-Infrared Light Curves}",
      journal = {\aj},
         year = 2004,
        month = nov,
       volume = {128},
       number = {5},
        pages = {2239-2264},
          doi = {10.1086/424934},
       adsurl = {https://ui.adsabs.harvard.edu/abs/2004AJ....128.2239P}
}

@ARTICLE{smolec2005,
       author = {{Smolec}, R.},
        title = "{Metallicity Dependence of the Blazhko Effect}",
      journal = {\actaa},
         year = 2005,
        month = mar,
       volume = {55},
        pages = {59-84},
          doi = {10.48550/arXiv.astro-ph/0503614},
archivePrefix = {arXiv},
       eprint = {astro-ph/0503614},
 primaryClass = {astro-ph},
       adsurl = {https://ui.adsabs.harvard.edu/abs/2005AcA....55...59S}
}

@ARTICLE{guldenschuh2005,
       author = {{Guldenschuh}, Katherine A. and {Layden}, Andrew C. and {Wan}, Yayun and {Whiting}, Alan and {van der Bliek}, Nicole and {Baca}, Pamela and {Carlin}, Jeffrey and {Freismuth}, Tucker and {Mora}, Marcelo and {Salyk}, Colette and {Vera}, Sergio and {Verdugo}, Miguel and {Young}, Amber},
        title = "{The Intrinsic Colors of RR Lyrae Variables: A Means to Determine Interstellar Reddening}",
      journal = {\pasp},
         year = 2005,
        month = jul,
       volume = {117},
       number = {833},
        pages = {721-725},
          doi = {10.1086/431178},
       adsurl = {https://ui.adsabs.harvard.edu/abs/2005PASP..117..721G}
}

@ARTICLE{sollima2006,
       author = {{Sollima}, A. and {Cacciari}, C. and {Valenti}, E.},
        title = "{The RR Lyrae period-K-luminosity relation for globular clusters: an observational approach}",
      journal = {\mnras},
         year = 2006,
        month = nov,
       volume = {372},
       number = {4},
        pages = {1675-1680},
          doi = {10.1111/j.1365-2966.2006.10962.x},
archivePrefix = {arXiv},
       eprint = {astro-ph/0608397},
 primaryClass = {astro-ph},
       adsurl = {https://ui.adsabs.harvard.edu/abs/2006MNRAS.372.1675S}
}

@INPROCEEDINGS{scamp2006,
       author = {{Bertin}, E.},
        title = "{Automatic Astrometric and Photometric Calibration with SCAMP}",
    booktitle = {Astronomical Data Analysis Software and Systems XV},
         year = 2006,
       editor = {{Gabriel}, C. and {Arviset}, C. and {Ponz}, D. and {Enrique}, S.},
       series = {Astronomical Society of the Pacific Conference Series},
       volume = {351},
        month = jul,
        pages = {112},
       adsurl = {https://ui.adsabs.harvard.edu/abs/2006ASPC..351..112B}
}

@ARTICLE{morgan2007,
       author = {{Morgan}, Siobahn M. and {Wahl}, Jennifer N. and {Wieckhorst}, Rachel M.},
        title = "{[Fe/H] relations for c-type RR Lyrae variables based upon Fourier coefficients}",
      journal = {\mnras},
         year = 2007,
        month = feb,
       volume = {374},
       number = {4},
        pages = {1421-1426},
          doi = {10.1111/j.1365-2966.2006.11247.x},
archivePrefix = {arXiv},
       eprint = {astro-ph/0506348},
 primaryClass = {astro-ph},
       adsurl = {https://ui.adsabs.harvard.edu/abs/2007MNRAS.374.1421M}
}

@ARTICLE{leborgne2007,
       author = {{Le Borgne}, J.~F. and {Paschke}, A. and {Vandenbroere}, J. and {Poretti}, E. and {Klotz}, A. and {Bo{\"e}r}, M. and {Damerdji}, Y. and {Martignoni}, M. and {Acerbi}, F.},
        title = "{Stellar evolution through the ages: period variations in galactic RRab stars as derived from the GEOS database and TAROT telescopes}",
      journal = {\aap},
         year = 2007,
        month = dec,
       volume = {476},
       number = {1},
        pages = {307-316},
          doi = {10.1051/0004-6361:20077957},
archivePrefix = {arXiv},
       eprint = {0710.1846},
 primaryClass = {astro-ph},
       adsurl = {https://ui.adsabs.harvard.edu/abs/2007A&A...476..307L}
}

@ARTICLE{feast2008,
       author = {{Feast}, Michael W. and {Laney}, Clifton D. and {Kinman}, Thomas D. and {van Leeuwen}, Floor and {Whitelock}, Patricia A.},
        title = "{The luminosities and distance scales of type II Cepheid and RR Lyrae variables}",
      journal = {\mnras},
         year = 2008,
        month = jun,
       volume = {386},
       number = {4},
        pages = {2115-2134},
          doi = {10.1111/j.1365-2966.2008.13181.x},
archivePrefix = {arXiv},
       eprint = {0803.0466},
 primaryClass = {astro-ph},
       adsurl = {https://ui.adsabs.harvard.edu/abs/2008MNRAS.386.2115F}
}

@ARTICLE{bono2008,
       author = {{Bono}, G. and {Caputo}, F. and {Fiorentino}, G. and {Marconi}, M. and {Musella}, I.},
        title = "{Cepheids in External Galaxies. I. The Maser-Host Galaxy NGC 4258 and the Metallicity Dependence of Period-Luminosity and Period-Wesenheit Relations}",
      journal = {\apj},
         year = 2008,
        month = sep,
       volume = {684},
       number = {1},
        pages = {102-117},
          doi = {10.1086/589965},
archivePrefix = {arXiv},
       eprint = {0805.1592},
 primaryClass = {astro-ph},
       adsurl = {https://ui.adsabs.harvard.edu/abs/2008ApJ...684..102B}
}

@ARTICLE{kunder2008,
       author = {{Kunder}, Andrea and {Chaboyer}, Brian},
        title = "{Metallicity Analysis of MACHO Galactic Bulge RR0 Lyrae Stars from their Light Curves}",
      journal = {\aj},
         year = 2008,
        month = dec,
       volume = {136},
       number = {6},
        pages = {2441-2452},
          doi = {10.1088/0004-6256/136/6/2441},
archivePrefix = {arXiv},
       eprint = {0809.1645},
 primaryClass = {astro-ph},
       adsurl = {https://ui.adsabs.harvard.edu/abs/2008AJ....136.2441K}
}

@ARTICLE{carretta2009,
       author = {{Carretta}, E. and {Bragaglia}, A. and {Gratton}, R. and {D'Orazi}, V. and {Lucatello}, S.},
        title = "{Intrinsic iron spread and a new metallicity scale for globular clusters}",
      journal = {\aap},
         year = 2009,
        month = dec,
       volume = {508},
       number = {2},
        pages = {695-706},
          doi = {10.1051/0004-6361/200913003},
archivePrefix = {arXiv},
       eprint = {0910.0675},
 primaryClass = {astro-ph.GA},
       adsurl = {https://ui.adsabs.harvard.edu/abs/2009A&A...508..695C}
}

@ARTICLE{goodman2010,
       author = {{Goodman}, Jonathan and {Weare}, Jonathan},
        title = "{Ensemble samplers with affine invariance}",
      journal = {Communications in Applied Mathematics and Computational Science},
         year = 2010,
        month = jan,
       volume = {5},
       number = {1},
        pages = {65-80},
          doi = {10.2140/camcos.2010.5.65},
       adsurl = {https://ui.adsabs.harvard.edu/abs/2010CAMCS...5...65G}
}

@ARTICLE{sesar2010,
       author = {{Sesar}, Branimir and {Ivezi{\'c}}, {\v{Z}}eljko and {Grammer}, Skyler H. and {Morgan}, Dylan P. and {Becker}, Andrew C. and {Juri{\'c}}, Mario and {De Lee}, Nathan and {Annis}, James and {Beers}, Timothy C. and {Fan}, Xiaohui and {Lupton}, Robert H. and {Gunn}, James E. and {Knapp}, Gillian R. and {Jiang}, Linhua and {Jester}, Sebastian and {Johnston}, David E. and {Lampeitl}, Hubert},
        title = "{Light Curve Templates and Galactic Distribution of RR Lyrae Stars from Sloan Digital Sky Survey Stripe 82}",
      journal = {\apj},
         year = 2010,
        month = jan,
       volume = {708},
       number = {1},
        pages = {717-741},
          doi = {10.1088/0004-637X/708/1/717},
archivePrefix = {arXiv},
       eprint = {0910.4611},
 primaryClass = {astro-ph.GA},
       adsurl = {https://ui.adsabs.harvard.edu/abs/2010ApJ...708..717S}
}

@ARTICLE{deb2010,
       author = {{Deb}, Sukanta and {Singh}, Harinder P.},
        title = "{Physical parameters of the Small Magellanic Cloud RR Lyrae stars and the distance scale}",
      journal = {\mnras},
         year = 2010,
        month = feb,
       volume = {402},
       number = {1},
        pages = {691-704},
          doi = {10.1111/j.1365-2966.2009.15927.x},
archivePrefix = {arXiv},
       eprint = {0910.4806},
 primaryClass = {astro-ph.SR},
       adsurl = {https://ui.adsabs.harvard.edu/abs/2010MNRAS.402..691D}
}

@ARTICLE{kunder2010,
       author = {{Kunder}, Andrea and {Chaboyer}, Brian and {Layden}, Andrew},
        title = "{The Determination of Reddening from Intrinsic VR Colors of RR Lyrae Stars}",
      journal = {\aj},
         year = 2010,
        month = feb,
       volume = {139},
       number = {2},
        pages = {415-424},
          doi = {10.1088/0004-6256/139/2/415},
archivePrefix = {arXiv},
       eprint = {0911.1770},
 primaryClass = {astro-ph.SR},
       adsurl = {https://ui.adsabs.harvard.edu/abs/2010AJ....139..415K}
}

@ARTICLE{astrometry2010,
       author = {{Lang}, Dustin and {Hogg}, David W. and {Mierle}, Keir and {Blanton}, Michael and {Roweis}, Sam},
        title = "{Astrometry.net: Blind Astrometric Calibration of Arbitrary Astronomical Images}",
      journal = {\aj},
         year = 2010,
        month = may,
       volume = {139},
       number = {5},
        pages = {1782-1800},
          doi = {10.1088/0004-6256/139/5/1782},
archivePrefix = {arXiv},
       eprint = {0910.2233},
 primaryClass = {astro-ph.IM},
       adsurl = {https://ui.adsabs.harvard.edu/abs/2010AJ....139.1782L}
}

@ARTICLE{hogg2010,
       author = {{Hogg}, David W. and {Bovy}, Jo and {Lang}, Dustin},
        title = "{Data analysis recipes: Fitting a model to data}",
      journal = {arXiv e-prints},
         year = 2010,
        month = aug,
          eid = {arXiv:1008.4686},
        pages = {arXiv:1008.4686},
          doi = {10.48550/arXiv.1008.4686},
archivePrefix = {arXiv},
       eprint = {1008.4686},
 primaryClass = {astro-ph.IM},
       adsurl = {https://ui.adsabs.harvard.edu/abs/2010arXiv1008.4686H}
}

@ARTICLE{harris2010,
       author = {{Harris}, William E.},
        title = "{A New Catalog of Globular Clusters in the Milky Way}",
      journal = {arXiv e-prints},
         year = 2010,
        month = dec,
          eid = {arXiv:1012.3224},
        pages = {arXiv:1012.3224},
          doi = {10.48550/arXiv.1012.3224},
archivePrefix = {arXiv},
       eprint = {1012.3224},
 primaryClass = {astro-ph.GA},
       adsurl = {https://ui.adsabs.harvard.edu/abs/2010arXiv1012.3224H}
}

@ARTICLE{schlafly2011,
       author = {{Schlafly}, Edward F. and {Finkbeiner}, Douglas P.},
        title = "{Measuring Reddening with Sloan Digital Sky Survey Stellar Spectra and Recalibrating SFD}",
      journal = {\apj},
         year = 2011,
        month = aug,
       volume = {737},
       number = {2},
          eid = {103},
        pages = {103},
          doi = {10.1088/0004-637X/737/2/103},
archivePrefix = {arXiv},
       eprint = {1012.4804},
 primaryClass = {astro-ph.GA},
       adsurl = {https://ui.adsabs.harvard.edu/abs/2011ApJ...737..103S}
}

@ARTICLE{skarka2013,
       author = {{Skarka}, M.},
        title = "{Known Galactic field Blazhko stars}",
      journal = {\aap},
         year = 2013,
        month = jan,
       volume = {549},
          eid = {A101},
        pages = {A101},
          doi = {10.1051/0004-6361/201220398},
archivePrefix = {arXiv},
       eprint = {1210.7120},
 primaryClass = {astro-ph.SR},
       adsurl = {https://ui.adsabs.harvard.edu/abs/2013A&A...549A.101S}
}

@ARTICLE{foreman-mackey2013,
       author = {{Foreman-Mackey}, Daniel and {Hogg}, David W. and {Lang}, Dustin and {Goodman}, Jonathan},
        title = "{emcee: The MCMC Hammer}",
      journal = {\pasp},
         year = 2013,
        month = mar,
       volume = {125},
       number = {925},
        pages = {306},
          doi = {10.1086/670067},
archivePrefix = {arXiv},
       eprint = {1202.3665},
 primaryClass = {astro-ph.IM},
       adsurl = {https://ui.adsabs.harvard.edu/abs/2013PASP..125..306F}
}

@ARTICLE{nemec2013,
       author = {{Nemec}, James M. and {Cohen}, Judith G. and {Ripepi}, Vincenzo and {Derekas}, Aliz and {Moskalik}, Pawel and {Sesar}, Branimir and {Chadid}, Merieme and {Bruntt}, Hans},
        title = "{Metal Abundances, Radial Velocities, and Other Physical Characteristics for the RR Lyrae Stars in The Kepler Field}",
      journal = {\apj},
         year = 2013,
        month = aug,
       volume = {773},
       number = {2},
          eid = {181},
        pages = {181},
          doi = {10.1088/0004-637X/773/2/181},
archivePrefix = {arXiv},
       eprint = {1307.5820},
 primaryClass = {astro-ph.SR},
       adsurl = {https://ui.adsabs.harvard.edu/abs/2013ApJ...773..181N}
}

@ARTICLE{astropy2013,
       author = {{Astropy Collaboration} and {Robitaille}, Thomas P. and {Tollerud}, Erik J. and {Greenfield}, Perry and {Droettboom}, Michael and {Bray}, Erik and {Aldcroft}, Tom and {Davis}, Matt and {Ginsburg}, Adam and {Price-Whelan}, Adrian M. and {Kerzendorf}, Wolfgang E. and {Conley}, Alexander and {Crighton}, Neil and {Barbary}, Kyle and {Muna}, Demitri and {Ferguson}, Henry and {Grollier}, Fr{\'e}d{\'e}ric and {Parikh}, Madhura M. and {Nair}, Prasanth H. and {Unther}, Hans M. and {Deil}, Christoph and {Woillez}, Julien and {Conseil}, Simon and {Kramer}, Roban and {Turner}, James E.~H. and {Singer}, Leo and {Fox}, Ryan and {Weaver}, Benjamin A. and {Zabalza}, Victor and {Edwards}, Zachary I. and {Azalee Bostroem}, K. and {Burke}, D.~J. and {Casey}, Andrew R. and {Crawford}, Steven M. and {Dencheva}, Nadia and {Ely}, Justin and {Jenness}, Tim and {Labrie}, Kathleen and {Lim}, Pey Lian and {Pierfederici}, Francesco and {Pontzen}, Andrew and {Ptak}, Andy and {Refsdal}, Brian and {Servillat}, Mathieu and {Streicher}, Ole},
        title = "{Astropy: A community Python package for astronomy}",
      journal = {\aap},
         year = 2013,
        month = oct,
       volume = {558},
          eid = {A33},
        pages = {A33},
          doi = {10.1051/0004-6361/201322068},
archivePrefix = {arXiv},
       eprint = {1307.6212},
 primaryClass = {astro-ph.IM},
       adsurl = {https://ui.adsabs.harvard.edu/abs/2013A&A...558A..33A}
}

@ARTICLE{shappee2014,
       author = {{Shappee}, B.~J. and {Prieto}, J.~L. and {Grupe}, D. and {Kochanek}, C.~S. and {Stanek}, K.~Z. and {De Rosa}, G. and {Mathur}, S. and {Zu}, Y. and {Peterson}, B.~M. and {Pogge}, R.~W. and {Komossa}, S. and {Im}, M. and {Jencson}, J. and {Holoien}, T.~W.-S. and {Basu}, U. and {Beacom}, J.~F. and {Szczygie{\l}}, D.~M. and {Brimacombe}, J. and {Adams}, S. and {Campillay}, A. and {Choi}, C. and {Contreras}, C. and {Dietrich}, M. and {Dubberley}, M. and {Elphick}, M. and {Foale}, S. and {Giustini}, M. and {Gonzalez}, C. and {Hawkins}, E. and {Howell}, D.~A. and {Hsiao}, E.~Y. and {Koss}, M. and {Leighly}, K.~M. and {Morrell}, N. and {Mudd}, D. and {Mullins}, D. and {Nugent}, J.~M. and {Parrent}, J. and {Phillips}, M.~M. and {Pojmanski}, G. and {Rosing}, W. and {Ross}, R. and {Sand}, D. and {Terndrup}, D.~M. and {Valenti}, S. and {Walker}, Z. and {Yoon}, Y.},
        title = "{The Man behind the Curtain: X-Rays Drive the UV through NIR Variability in the 2013 Active Galactic Nucleus Outburst in NGC 2617}",
      journal = {\apj},
         year = 2014,
        month = jun,
       volume = {788},
       number = {1},
          eid = {48},
        pages = {48},
          doi = {10.1088/0004-637X/788/1/48},
archivePrefix = {arXiv},
       eprint = {1310.2241},
 primaryClass = {astro-ph.HE},
       adsurl = {https://ui.adsabs.harvard.edu/abs/2014ApJ...788...48S}
}

@ARTICLE{stetson2014a,
       author = {{Stetson}, P.~B. and {Braga}, V.~F. and {Dall'Ora}, M. and {Bono}, G. and {Buonanno}, R. and {Ferraro}, I. and {Iannicola}, G. and {Marengo}, M. and {Neeley}, J.},
        title = "{Optical and Near-Infrared UBVRIJHK Photometry for the RR Lyrae Stars in the Nearby Globular Cluster M4 (NGC 6121)}",
      journal = {\pasp},
         year = 2014,
        month = jun,
       volume = {126},
       number = {940},
        pages = {521},
          doi = {10.1086/677195},
archivePrefix = {arXiv},
       eprint = {1406.7531},
 primaryClass = {astro-ph.SR},
       adsurl = {https://ui.adsabs.harvard.edu/abs/2014PASP..126..521S}
}

@ARTICLE{stetson2014b,
       author = {{Stetson}, Peter B. and {Fiorentino}, Giuliana and {Bono}, Giuseppe and {Bernard}, Edouard J. and {Monelli}, Matteo and {Iannicola}, Giacinto and {Gallart}, Carme and {Ferraro}, Ivan},
        title = "{Homogeneous Photometry VI: Variable Stars in the Leo I Dwarf Spheroidal Galaxy}",
      journal = {\pasp},
         year = 2014,
        month = jul,
       volume = {126},
       number = {941},
        pages = {616},
          doi = {10.1086/677352},
archivePrefix = {arXiv},
       eprint = {1406.6704},
 primaryClass = {astro-ph.GA},
       adsurl = {https://ui.adsabs.harvard.edu/abs/2014PASP..126..616S}
}

@ARTICLE{braga2015,
       author = {{Braga}, V.~F. and {Dall'Ora}, M. and {Bono}, G. and {Stetson}, P.~B. and {Ferraro}, I. and {Iannicola}, G. and {Marengo}, M. and {Neeley}, J. and {Persson}, S.~E. and {Buonanno}, R. and {Coppola}, G. and {Freedman}, W. and {Madore}, B.~F. and {Marconi}, M. and {Matsunaga}, N. and {Monson}, A. and {Rich}, J. and {Scowcroft}, V. and {Seibert}, M.},
        title = "{On the Distance of the Globular Cluster M4 (NGC 6121) Using RR Lyrae Stars. I. Optical and Near-infrared Period-Luminosity and Period-Wesenheit Relations}",
      journal = {\apj},
         year = 2015,
        month = feb,
       volume = {799},
       number = {2},
          eid = {165},
        pages = {165},
          doi = {10.1088/0004-637X/799/2/165},
archivePrefix = {arXiv},
       eprint = {1411.6826},
 primaryClass = {astro-ph.GA},
       adsurl = {https://ui.adsabs.harvard.edu/abs/2015ApJ...799..165B}
}

@ARTICLE{marconi2015,
       author = {{Marconi}, M. and {Coppola}, G. and {Bono}, G. and {Braga}, V. and {Pietrinferni}, A. and {Buonanno}, R. and {Castellani}, M. and {Musella}, I. and {Ripepi}, V. and {Stellingwerf}, R.~F.},
        title = "{On a New Theoretical Framework for RR Lyrae Stars. I. The Metallicity Dependence}",
      journal = {\apj},
         year = 2015,
        month = jul,
       volume = {808},
       number = {1},
          eid = {50},
        pages = {50},
          doi = {10.1088/0004-637X/808/1/50},
archivePrefix = {arXiv},
       eprint = {1505.02531},
 primaryClass = {astro-ph.SR},
       adsurl = {https://ui.adsabs.harvard.edu/abs/2015ApJ...808...50M}
}

@ARTICLE{gnuastro2015,
       author = {{Akhlaghi}, Mohammad and {Ichikawa}, Takashi},
        title = "{Noise-based Detection and Segmentation of Nebulous Objects}",
      journal = {\apjs},
         year = 2015,
        month = sep,
       volume = {220},
       number = {1},
          eid = {1},
        pages = {1},
          doi = {10.1088/0067-0049/220/1/1},
archivePrefix = {arXiv},
       eprint = {1505.01664},
 primaryClass = {astro-ph.IM},
       adsurl = {https://ui.adsabs.harvard.edu/abs/2015ApJS..220....1A}
}

@ARTICLE{vanderplas2015,
       author = {{VanderPlas}, Jacob T. and {Ivezi{\'c}}, {\v{Z}}eljko},
        title = "{Periodograms for Multiband Astronomical Time Series}",
      journal = {\apj},
         year = 2015,
        month = oct,
       volume = {812},
       number = {1},
          eid = {18},
        pages = {18},
          doi = {10.1088/0004-637X/812/1/18},
archivePrefix = {arXiv},
       eprint = {1502.01344},
 primaryClass = {astro-ph.IM},
       adsurl = {https://ui.adsabs.harvard.edu/abs/2015ApJ...812...18V}
}

@BOOK{catelan2015,
       author = {{Catelan}, M. and {Smith}, H.~A.},
        title = "{Pulsating Stars}",
         year = 2015,
    publisher = {Wiley-VCH Verlag GmbH \& Co. KGaA},
       adsurl = {https://ui.adsabs.harvard.edu/abs/2015pust.book.....C}
}

@ARTICLE{martinez-vazquez2016,
       author = {{Martinez-Vazquez}, C.~E. and {Monelli}, M. and {Bono}, G. and {Stetson}, P.~B. and {Gallart}, C. and {Bernard}, E.~J. and {Fiorentino}, G. and {Dall'Ora}, M.},
        title = "{A new Phi\_31-period-metallicity relation for RR Lyrae stars}",
      journal = {Commmunications of the Konkoly Observatory Hungary},
         year = 2016,
        month = may,
       volume = {105},
        pages = {53-56},
       adsurl = {https://ui.adsabs.harvard.edu/abs/2016CoKon.105...53M}
}

@ARTICLE{skowron2016,
       author = {{Skowron}, D.~M. and {Soszy{\'n}ski}, I. and {Udalski}, A. and {Szyma{\'n}ski}, M.~K. and {Pietrukowicz}, P. and {Skowron}, J. and {Poleski}, R. and {Wyrzykowski}, {\L}. and {Ulaczyk}, K. and {Koz{\l}owski}, S. and {Mr{\'o}z}, P. and {Pawlak}, M.},
        title = "{OGLE-ing the Magellanic System: Photometric Metallicity from Fundamental Mode RR Lyrae Stars}",
      journal = {\actaa},
         year = 2016,
        month = sep,
       volume = {66},
       number = {3},
        pages = {269-292},
          doi = {10.48550/arXiv.1608.00013},
archivePrefix = {arXiv},
       eprint = {1608.00013},
 primaryClass = {astro-ph.SR},
       adsurl = {https://ui.adsabs.harvard.edu/abs/2016AcA....66..269S}
}

@ARTICLE{braga2016,
       author = {{Braga}, V.~F. and {Stetson}, P.~B. and {Bono}, G. and {Dall'Ora}, M. and {Ferraro}, I. and {Fiorentino}, G. and {Freyhammer}, L.~M. and {Iannicola}, G. and {Marengo}, M. and {Neeley}, J. and {Valenti}, E. and {Buonanno}, R. and {Calamida}, A. and {Castellani}, M. and {da Silva}, R. and {Degl'Innocenti}, S. and {Di Cecco}, A. and {Fabrizio}, M. and {Freedman}, W.~L. and {Giuffrida}, G. and {Lub}, J. and {Madore}, B.~F. and {Marconi}, M. and {Marinoni}, S. and {Matsunaga}, N. and {Monelli}, M. and {Persson}, S.~E. and {Piersimoni}, A.~M. and {Pietrinferni}, A. and {Prada-Moroni}, P. and {Pulone}, L. and {Stellingwerf}, R. and {Tognelli}, E. and {Walker}, A.~R.},
        title = "{On the RR Lyrae Stars in Globulars. IV. {\ensuremath{\omega}} Centauri Optical UBVRI Photometry}",
      journal = {\aj},
         year = 2016,
        month = dec,
       volume = {152},
       number = {6},
          eid = {170},
        pages = {170},
          doi = {10.3847/0004-6256/152/6/170},
archivePrefix = {arXiv},
       eprint = {1609.04916},
 primaryClass = {astro-ph.SR},
       adsurl = {https://ui.adsabs.harvard.edu/abs/2016AJ....152..170B}
}

@ARTICLE{ngeow2016,
       author = {{Ngeow}, Chow-Choong and {Yu}, Po-Chieh and {Bellm}, Eric and {Yang}, Ting-Chang and {Chang}, Chan-Kao and {Miller}, Adam and {Laher}, Russ and {Surace}, Jason and {Ip}, Wing-Huen},
        title = "{The Palomar Transient Factory and RR Lyrae: The Metallicity-Light Curve Relation Based on ab-type RR Lyrae in the Kepler Field}",
      journal = {\apjs},
         year = 2016,
        month = dec,
       volume = {227},
       number = {2},
          eid = {30},
        pages = {30},
          doi = {10.3847/1538-4365/227/2/30},
archivePrefix = {arXiv},
       eprint = {1612.03366},
 primaryClass = {astro-ph.SR},
       adsurl = {https://ui.adsabs.harvard.edu/abs/2016ApJS..227...30N}
}

@ARTICLE{monson2017,
       author = {{Monson}, Andrew J. and {Beaton}, Rachael L. and {Scowcroft}, Victoria and {Freedman}, Wendy L. and {Madore}, Barry F. and {Rich}, Jeffrey A. and {Seibert}, Mark and {Kollmeier}, Juna A. and {Clementini}, Gisella},
        title = "{Standard Galactic Field RR Lyrae. I. Optical to Mid-infrared Phased Photometry}",
      journal = {\aj},
         year = 2017,
        month = mar,
       volume = {153},
       number = {3},
          eid = {96},
        pages = {96},
          doi = {10.3847/1538-3881/153/3/96},
archivePrefix = {arXiv},
       eprint = {1703.01520},
 primaryClass = {astro-ph.SR},
       adsurl = {https://ui.adsabs.harvard.edu/abs/2017AJ....153...96M}
}

@ARTICLE{neeley2017,
       author = {{Neeley}, Jillian R. and {Marengo}, Massimo and {Bono}, Giuseppe and {Braga}, Vittorio F. and {Dall'Ora}, Massimo and {Magurno}, Davide and {Marconi}, Marcella and {Trueba}, Nicolas and {Tognelli}, Emanuele and {Prada Moroni}, Pier G. and {Beaton}, Rachael L. and {Freedman}, Wendy L. and {Madore}, Barry F. and {Monson}, Andrew J. and {Scowcroft}, Victoria and {Seibert}, Mark and {Stetson}, Peter B.},
        title = "{On a New Theoretical Framework for RR Lyrae Stars. II. Mid-infrared Period-Luminosity-Metallicity Relations}",
      journal = {\apj},
         year = 2017,
        month = jun,
       volume = {841},
       number = {2},
          eid = {84},
        pages = {84},
          doi = {10.3847/1538-4357/aa713d},
archivePrefix = {arXiv},
       eprint = {1705.01970},
 primaryClass = {astro-ph.SR},
       adsurl = {https://ui.adsabs.harvard.edu/abs/2017ApJ...841...84N}
}

@ARTICLE{vivas2017,
       author = {{Vivas}, A. Katherina and {Saha}, Abhijit and {Olsen}, Knut and {Blum}, Robert and {Olszewski}, Edward W. and {Claver}, Jennifer and {Valdes}, Francisco and {Axelrod}, Tim and {Kaleida}, Catherine and {Kunder}, Andrea and {Narayan}, Gautham and {Matheson}, Thomas and {Walker}, Alistair},
        title = "{Absolute Magnitudes and Colors of RR Lyrae Stars in DECam Passbands from Photometry of the Globular Cluster M5}",
      journal = {\aj},
         year = 2017,
        month = sep,
       volume = {154},
       number = {3},
          eid = {85},
        pages = {85},
          doi = {10.3847/1538-3881/aa7fed},
archivePrefix = {arXiv},
       eprint = {1707.03954},
 primaryClass = {astro-ph.SR},
       adsurl = {https://ui.adsabs.harvard.edu/abs/2017AJ....154...85V}
}

@ARTICLE{braga2018,
       author = {{Braga}, V.~F. and {Stetson}, P.~B. and {Bono}, G. and {Dall'Ora}, M. and {Ferraro}, I. and {Fiorentino}, G. and {Iannicola}, G. and {Marconi}, M. and {Marengo}, M. and {Monson}, A.~J. and {Neeley}, J. and {Persson}, S.~E. and {Beaton}, R.~L. and {Buonanno}, R. and {Calamida}, A. and {Castellani}, M. and {Di Carlo}, E. and {Fabrizio}, M. and {Freedman}, W.~L. and {Inno}, L. and {Madore}, B.~F. and {Magurno}, D. and {Marchetti}, E. and {Marinoni}, S. and {Marrese}, P. and {Matsunaga}, N. and {Minniti}, D. and {Monelli}, M. and {Nonino}, M. and {Piersimoni}, A.~M. and {Pietrinferni}, A. and {Prada-Moroni}, P. and {Pulone}, L. and {Stellingwerf}, R. and {Tognelli}, E. and {Walker}, A.~R. and {Valenti}, E. and {Zoccali}, M.},
        title = "{On the RR Lyrae Stars in Globulars. V. The Complete Near-infrared (JHK $_{ s }$) Census of {\ensuremath{\omega}} Centauri RR Lyrae Variables}",
      journal = {\aj},
         year = 2018,
        month = mar,
       volume = {155},
       number = {3},
          eid = {137},
        pages = {137},
          doi = {10.3847/1538-3881/aaadab},
archivePrefix = {arXiv},
       eprint = {1802.03578},
 primaryClass = {astro-ph.SR},
       adsurl = {https://ui.adsabs.harvard.edu/abs/2018AJ....155..137B}
}

@ARTICLE{jurcsik2018,
       author = {{Jurcsik}, J. and {Hajdu}, G. and {D{\'e}k{\'a}ny}, I. and {Nuspl}, J. and {Catelan}, M. and {Grebel}, E.~K.},
        title = "{Blazhko modulation in the infrared}",
      journal = {\mnras},
         year = 2018,
        month = apr,
       volume = {475},
       number = {3},
        pages = {4208-4222},
          doi = {10.1093/mnras/sty112},
archivePrefix = {arXiv},
       eprint = {1801.03436},
 primaryClass = {astro-ph.SR},
       adsurl = {https://ui.adsabs.harvard.edu/abs/2018MNRAS.475.4208J}
}

@ARTICLE{green2018,
       author = {{Green}, Gregory M. and {Schlafly}, Edward F. and {Finkbeiner}, Douglas and {Rix}, Hans-Walter and {Martin}, Nicolas and {Burgett}, William and {Draper}, Peter W. and {Flewelling}, Heather and {Hodapp}, Klaus and {Kaiser}, Nicholas and {Kudritzki}, Rolf-Peter and {Magnier}, Eugene A. and {Metcalfe}, Nigel and {Tonry}, John L. and {Wainscoat}, Richard and {Waters}, Christopher},
        title = "{Galactic reddening in 3D from stellar photometry - an improved map}",
      journal = {\mnras},
         year = 2018,
        month = jul,
       volume = {478},
       number = {1},
        pages = {651-666},
          doi = {10.1093/mnras/sty1008},
archivePrefix = {arXiv},
       eprint = {1801.03555},
 primaryClass = {astro-ph.GA},
       adsurl = {https://ui.adsabs.harvard.edu/abs/2018MNRAS.478..651G}
}

@ARTICLE{jayasinghe2018,
       author = {{Jayasinghe}, T. and {Kochanek}, C.~S. and {Stanek}, K.~Z. and {Shappee}, B.~J. and {Holoien}, T.~W.-S. and {Thompson}, Todd A. and {Prieto}, J.~L. and {Dong}, Subo and {Pawlak}, M. and {Shields}, J.~V. and {Pojmanski}, G. and {Otero}, S. and {Britt}, C.~A. and {Will}, D.},
        title = "{The ASAS-SN catalogue of variable stars I: The Serendipitous Survey}",
      journal = {\mnras},
         year = 2018,
        month = jul,
       volume = {477},
       number = {3},
        pages = {3145-3163},
          doi = {10.1093/mnras/sty838},
archivePrefix = {arXiv},
       eprint = {1803.01001},
 primaryClass = {astro-ph.SR},
       adsurl = {https://ui.adsabs.harvard.edu/abs/2018MNRAS.477.3145J}
}

@ARTICLE{astropy2018,
       author = {{Astropy Collaboration} and {Price-Whelan}, A.~M. and
         {Sip{\H{o}}cz}, B.~M. and {G{\"u}nther}, H.~M. and {Lim}, P.~L. and
         {Crawford}, S.~M. and {Conseil}, S. and {Shupe}, D.~L. and
         {Craig}, M.~W. and {Dencheva}, N. and {Ginsburg}, A. and {Vand
        erPlas}, J.~T. and {Bradley}, L.~D. and {P{\'e}rez-Su{\'a}rez}, D. and
         {de Val-Borro}, M. and {Aldcroft}, T.~L. and {Cruz}, K.~L. and
         {Robitaille}, T.~P. and {Tollerud}, E.~J. and {Ardelean}, C. and
         {Babej}, T. and {Bach}, Y.~P. and {Bachetti}, M. and {Bakanov}, A.~V. and
         {Bamford}, S.~P. and {Barentsen}, G. and {Barmby}, P. and
         {Baumbach}, A. and {Berry}, K.~L. and {Biscani}, F. and {Boquien}, M. and
         {Bostroem}, K.~A. and {Bouma}, L.~G. and {Brammer}, G.~B. and
         {Bray}, E.~M. and {Breytenbach}, H. and {Buddelmeijer}, H. and
         {Burke}, D.~J. and {Calderone}, G. and {Cano Rodr{\'\i}guez}, J.~L. and
         {Cara}, M. and {Cardoso}, J.~V.~M. and {Cheedella}, S. and {Copin}, Y. and
         {Corrales}, L. and {Crichton}, D. and {D'Avella}, D. and {Deil}, C. and
         {Depagne}, {\'E}. and {Dietrich}, J.~P. and {Donath}, A. and
         {Droettboom}, M. and {Earl}, N. and {Erben}, T. and {Fabbro}, S. and
         {Ferreira}, L.~A. and {Finethy}, T. and {Fox}, R.~T. and
         {Garrison}, L.~H. and {Gibbons}, S.~L.~J. and {Goldstein}, D.~A. and
         {Gommers}, R. and {Greco}, J.~P. and {Greenfield}, P. and
         {Groener}, A.~M. and {Grollier}, F. and {Hagen}, A. and {Hirst}, P. and
         {Homeier}, D. and {Horton}, A.~J. and {Hosseinzadeh}, G. and {Hu}, L. and
         {Hunkeler}, J.~S. and {Ivezi{\'c}}, {\v{Z}}. and {Jain}, A. and
         {Jenness}, T. and {Kanarek}, G. and {Kendrew}, S. and {Kern}, N.~S. and
         {Kerzendorf}, W.~E. and {Khvalko}, A. and {King}, J. and {Kirkby}, D. and
         {Kulkarni}, A.~M. and {Kumar}, A. and {Lee}, A. and {Lenz}, D. and
         {Littlefair}, S.~P. and {Ma}, Z. and {Macleod}, D.~M. and
         {Mastropietro}, M. and {McCully}, C. and {Montagnac}, S. and
         {Morris}, B.~M. and {Mueller}, M. and {Mumford}, S.~J. and {Muna}, D. and
         {Murphy}, N.~A. and {Nelson}, S. and {Nguyen}, G.~H. and
         {Ninan}, J.~P. and {N{\"o}the}, M. and {Ogaz}, S. and {Oh}, S. and
         {Parejko}, J.~K. and {Parley}, N. and {Pascual}, S. and {Patil}, R. and
         {Patil}, A.~A. and {Plunkett}, A.~L. and {Prochaska}, J.~X. and
         {Rastogi}, T. and {Reddy Janga}, V. and {Sabater}, J. and
         {Sakurikar}, P. and {Seifert}, M. and {Sherbert}, L.~E. and
         {Sherwood-Taylor}, H. and {Shih}, A.~Y. and {Sick}, J. and
         {Silbiger}, M.~T. and {Singanamalla}, S. and {Singer}, L.~P. and
         {Sladen}, P.~H. and {Sooley}, K.~A. and {Sornarajah}, S. and
         {Streicher}, O. and {Teuben}, P. and {Thomas}, S.~W. and
         {Tremblay}, G.~R. and {Turner}, J.~E.~H. and {Terr{\'o}n}, V. and
         {van Kerkwijk}, M.~H. and {de la Vega}, A. and {Watkins}, L.~L. and
         {Weaver}, B.~A. and {Whitmore}, J.~B. and {Woillez}, J. and
         {Zabalza}, V. and {Astropy Contributors}},
        title = "{The Astropy Project: Building an Open-science Project and Status of the v2.0 Core Package}",
      journal = {\aj},
         year = 2018,
        month = sep,
       volume = {156},
       number = {3},
          eid = {123},
        pages = {123},
          doi = {10.3847/1538-3881/aabc4f},
archivePrefix = {arXiv},
       eprint = {1801.02634},
 primaryClass = {astro-ph.IM},
       adsurl = {https://ui.adsabs.harvard.edu/abs/2018AJ....156..123A}
}

@ARTICLE{magurno2018,
       author = {{Magurno}, D. and {Sneden}, C. and {Braga}, V.~F. and {Bono}, G. and {Mateo}, M. and {Persson}, S.~E. and {Dall'Ora}, M. and {Marengo}, M. and {Monelli}, M. and {Neeley}, J.~R.},
        title = "{Chemical Compositions of Field and Globular Cluster RR Lyrae Stars. I. NGC 3201}",
      journal = {\apj},
         year = 2018,
        month = sep,
       volume = {864},
       number = {1},
          eid = {57},
        pages = {57},
          doi = {10.3847/1538-4357/aad4a3},
archivePrefix = {arXiv},
       eprint = {1807.06681},
 primaryClass = {astro-ph.SR},
       adsurl = {https://ui.adsabs.harvard.edu/abs/2018ApJ...864...57M}
}

@ARTICLE{muraveva2018,
       author = {{Muraveva}, Tatiana and {Delgado}, Hector E. and {Clementini}, Gisella and {Sarro}, Luis M. and {Garofalo}, Alessia},
        title = "{RR Lyrae stars as standard candles in the Gaia Data Release 2 Era}",
      journal = {\mnras},
         year = 2018,
        month = nov,
       volume = {481},
       number = {1},
        pages = {1195-1211},
          doi = {10.1093/mnras/sty2241},
archivePrefix = {arXiv},
       eprint = {1805.08742},
 primaryClass = {astro-ph.SR},
       adsurl = {https://ui.adsabs.harvard.edu/abs/2018MNRAS.481.1195M}
}

@ARTICLE{ivezic2019,
       author = {{Ivezi{\'c}}, {\v{Z}}eljko and {Kahn}, Steven M. and {Tyson}, J. Anthony and {Abel}, Bob and {Acosta}, Emily and {Allsman}, Robyn and {Alonso}, David and {AlSayyad}, Yusra and {Anderson}, Scott F. and {Andrew}, John and {Angel}, James Roger P. and {Angeli}, George Z. and {Ansari}, Reza and {Antilogus}, Pierre and {Araujo}, Constanza and {Armstrong}, Robert and {Arndt}, Kirk T. and {Astier}, Pierre and {Aubourg}, {\'E}ric and {Auza}, Nicole and {Axelrod}, Tim S. and {Bard}, Deborah J. and {Barr}, Jeff D. and {Barrau}, Aurelian and {Bartlett}, James G. and {Bauer}, Amanda E. and {Bauman}, Brian J. and {Baumont}, Sylvain and {Bechtol}, Ellen and {Bechtol}, Keith and {Becker}, Andrew C. and {Becla}, Jacek and {Beldica}, Cristina and {Bellavia}, Steve and {Bianco}, Federica B. and {Biswas}, Rahul and {Blanc}, Guillaume and {Blazek}, Jonathan and {Blandford}, Roger D. and {Bloom}, Josh S. and {Bogart}, Joanne and {Bond}, Tim W. and {Booth}, Michael T. and {Borgland}, Anders W. and {Borne}, Kirk and {Bosch}, James F. and {Boutigny}, Dominique and {Brackett}, Craig A. and {Bradshaw}, Andrew and {Brandt}, William Nielsen and {Brown}, Michael E. and {Bullock}, James S. and {Burchat}, Patricia and {Burke}, David L. and {Cagnoli}, Gianpietro and {Calabrese}, Daniel and {Callahan}, Shawn and {Callen}, Alice L. and {Carlin}, Jeffrey L. and {Carlson}, Erin L. and {Chandrasekharan}, Srinivasan and {Charles-Emerson}, Glenaver and {Chesley}, Steve and {Cheu}, Elliott C. and {Chiang}, Hsin-Fang and {Chiang}, James and {Chirino}, Carol and {Chow}, Derek and {Ciardi}, David R. and {Claver}, Charles F. and {Cohen-Tanugi}, Johann and {Cockrum}, Joseph J. and {Coles}, Rebecca and {Connolly}, Andrew J. and {Cook}, Kem H. and {Cooray}, Asantha and {Covey}, Kevin R. and {Cribbs}, Chris and {Cui}, Wei and {Cutri}, Roc and {Daly}, Philip N. and {Daniel}, Scott F. and {Daruich}, Felipe and {Daubard}, Guillaume and {Daues}, Greg and {Dawson}, William and {Delgado}, Francisco and {Dellapenna}, Alfred and {de Peyster}, Robert and {de Val-Borro}, Miguel and {Digel}, Seth W. and {Doherty}, Peter and {Dubois}, Richard and {Dubois-Felsmann}, Gregory P. and {Durech}, Josef and {Economou}, Frossie and {Eifler}, Tim and {Eracleous}, Michael and {Emmons}, Benjamin L. and {Fausti Neto}, Angelo and {Ferguson}, Henry and {Figueroa}, Enrique and {Fisher-Levine}, Merlin and {Focke}, Warren and {Foss}, Michael D. and {Frank}, James and {Freemon}, Michael D. and {Gangler}, Emmanuel and {Gawiser}, Eric and {Geary}, John C. and {Gee}, Perry and {Geha}, Marla and {Gessner}, Charles J.~B. and {Gibson}, Robert R. and {Gilmore}, D. Kirk and {Glanzman}, Thomas and {Glick}, William and {Goldina}, Tatiana and {Goldstein}, Daniel A. and {Goodenow}, Iain and {Graham}, Melissa L. and {Gressler}, William J. and {Gris}, Philippe and {Guy}, Leanne P. and {Guyonnet}, Augustin and {Haller}, Gunther and {Harris}, Ron and {Hascall}, Patrick A. and {Haupt}, Justine and {Hernandez}, Fabio and {Herrmann}, Sven and {Hileman}, Edward and {Hoblitt}, Joshua and {Hodgson}, John A. and {Hogan}, Craig and {Howard}, James D. and {Huang}, Dajun and {Huffer}, Michael E. and {Ingraham}, Patrick and {Innes}, Walter R. and {Jacoby}, Suzanne H. and {Jain}, Bhuvnesh and {Jammes}, Fabrice and {Jee}, M. James and {Jenness}, Tim and {Jernigan}, Garrett and {Jevremovi{\'c}}, Darko and {Johns}, Kenneth and {Johnson}, Anthony S. and {Johnson}, Margaret W.~G. and {Jones}, R. Lynne and {Juramy-Gilles}, Claire and {Juri{\'c}}, Mario and {Kalirai}, Jason S. and {Kallivayalil}, Nitya J. and {Kalmbach}, Bryce and {Kantor}, Jeffrey P. and {Karst}, Pierre and {Kasliwal}, Mansi M. and {Kelly}, Heather and {Kessler}, Richard and {Kinnison}, Veronica and {Kirkby}, David and {Knox}, Lloyd and {Kotov}, Ivan V. and {Krabbendam}, Victor L. and {Krughoff}, K. Simon and {Kub{\'a}nek}, Petr and {Kuczewski}, John and {Kulkarni}, Shri and {Ku}, John and {Kurita}, Nadine R. and {Lage}, Craig S. and {Lambert}, Ron and {Lange}, Travis and {Langton}, J. Brian and {Le Guillou}, Laurent and {Levine}, Deborah and {Liang}, Ming and {Lim}, Kian-Tat and {Lintott}, Chris J. and {Long}, Kevin E. and {Lopez}, Margaux and {Lotz}, Paul J. and {Lupton}, Robert H. and {Lust}, Nate B. and {MacArthur}, Lauren A. and {Mahabal}, Ashish and {Mandelbaum}, Rachel and {Markiewicz}, Thomas W. and {Marsh}, Darren S. and {Marshall}, Philip J. and {Marshall}, Stuart and {May}, Morgan and {McKercher}, Robert and {McQueen}, Michelle and {Meyers}, Joshua and {Migliore}, Myriam and {Miller}, Michelle and {Mills}, David J.},
        title = "{LSST: From Science Drivers to Reference Design and Anticipated Data Products}",
      journal = {\apj},
         year = 2019,
        month = mar,
       volume = {873},
       number = {2},
          eid = {111},
        pages = {111},
          doi = {10.3847/1538-4357/ab042c},
archivePrefix = {arXiv},
       eprint = {0805.2366},
 primaryClass = {astro-ph},
       adsurl = {https://ui.adsabs.harvard.edu/abs/2019ApJ...873..111I}
}

@ARTICLE{saha2019,
       author = {{Saha}, Abhijit and {Vivas}, A. Katherina and {Olszewski}, Edward W. and {Smith}, Verne and {Olsen}, Knut and {Blum}, Robert and {Valdes}, Francisco and {Claver}, Jenna and {Calamida}, Annalisa and {Walker}, Alistair R. and {Matheson}, Thomas and {Narayan}, Gautham and {Soraisam}, Monika and {Cunha}, Katia and {Axelrod}, T. and {Bloom}, Joshua S. and {Cenko}, S. Bradley and {Frye}, Brenda and {Juric}, Mario and {Kaleida}, Catherine and {Kunder}, Andrea and {Miller}, Adam and {Nidever}, David and {Ridgway}, Stephen},
        title = "{Mapping the Interstellar Reddening and Extinction toward Baade{\textquoteright}s Window Using Minimum Light Colors of ab-type RR Lyrae Stars: Revelations from the De-reddened Color-Magnitude Diagrams}",
      journal = {\apj},
         year = 2019,
        month = mar,
       volume = {874},
       number = {1},
          eid = {30},
        pages = {30},
          doi = {10.3847/1538-4357/ab07ba},
archivePrefix = {arXiv},
       eprint = {1902.05637},
 primaryClass = {astro-ph.GA},
       adsurl = {https://ui.adsabs.harvard.edu/abs/2019ApJ...874...30S}
}

@ARTICLE{braga2019a,
       author = {{Braga}, V.~F. and {Stetson}, P.~B. and {Bono}, G. and {Dall'Ora}, M. and {Ferraro}, I. and {Fiorentino}, G. and {Iannicola}, G. and {Inno}, L. and {Marengo}, M. and {Neeley}, J. and {Beaton}, R.~L. and {Buonanno}, R. and {Calamida}, A. and {Contreras Ramos}, R. and {Chaboyer}, B. and {Fabrizio}, M. and {Freedman}, W.~L. and {Gilligan}, C.~K. and {Johnston}, K.~V. and {Lub}, J. and {Madore}, B.~F. and {Magurno}, D. and {Marconi}, M. and {Marinoni}, S. and {Marrese}, P.~M. and {Mateo}, M. and {Matsunaga}, N. and {Minniti}, D. and {Monson}, A.~J. and {Monelli}, M. and {Nonino}, M. and {Persson}, S.~E. and {Pietrinferni}, A. and {Sneden}, C. and {Storm}, J. and {Walker}, A.~R. and {Valenti}, E. and {Zoccali}, M.},
        title = "{New near-infrared JHK$_{s}$ light-curve templates for RR Lyrae variables}",
      journal = {\aap},
         year = 2019,
        month = may,
       volume = {625},
          eid = {A1},
        pages = {A1},
          doi = {10.1051/0004-6361/201834893},
archivePrefix = {arXiv},
       eprint = {1812.06372},
 primaryClass = {astro-ph.SR},
       adsurl = {https://ui.adsabs.harvard.edu/abs/2019A&A...625A...1B}
}

@ARTICLE{magurno2019,
       author = {{Magurno}, D. and {Sneden}, C. and {Bono}, G. and {Braga}, V.~F. and {Mateo}, M. and {Persson}, S.~E. and {Preston}, G. and {Th{\'e}venin}, F. and {da Silva}, R. and {Dall'Ora}, M. and {Fabrizio}, M. and {Ferraro}, I. and {Fiorentino}, G. and {Iannicola}, G. and {Inno}, L. and {Marengo}, M. and {Marinoni}, S. and {Marrese}, P.~M. and {Mart{\'\i}nez-V{\'a}zquez}, C.~E. and {Matsunaga}, N. and {Monelli}, M. and {Neeley}, J.~R. and {Nonino}, M. and {Walker}, A.~R.},
        title = "{Chemical Compositions of Field and Globular Cluster RR Lyrae Stars. II. {\ensuremath{\omega}} Centauri}",
      journal = {\apj},
         year = 2019,
        month = aug,
       volume = {881},
       number = {2},
          eid = {104},
        pages = {104},
          doi = {10.3847/1538-4357/ab2e76},
archivePrefix = {arXiv},
       eprint = {1906.08550},
 primaryClass = {astro-ph.SR},
       adsurl = {https://ui.adsabs.harvard.edu/abs/2019ApJ...881..104M}
}

@ARTICLE{fabrizio2019,
       author = {{Fabrizio}, M. and {Bono}, G. and {Braga}, V.~F. and {Magurno}, D. and {Marinoni}, S. and {Marrese}, P.~M. and {Ferraro}, I. and {Fiorentino}, G. and {Giuffrida}, G. and {Iannicola}, G. and {Monelli}, M. and {Altavilla}, G. and {Chaboyer}, B. and {Dall'Ora}, M. and {Gilligan}, C.~K. and {Layden}, A. and {Marengo}, M. and {Nonino}, M. and {Preston}, G.~W. and {Sesar}, B. and {Sneden}, C. and {Valenti}, E. and {Th{\'e}venin}, F. and {Zoccali}, E.},
        title = "{On the Use of Field RR Lyrae as Galactic Probes. I. The Oosterhoff Dichotomy Based on Fundamental Variables}",
      journal = {\apj},
         year = 2019,
        month = sep,
       volume = {882},
       number = {2},
          eid = {169},
        pages = {169},
          doi = {10.3847/1538-4357/ab3977},
archivePrefix = {arXiv},
       eprint = {1908.02064},
 primaryClass = {astro-ph.SR},
       adsurl = {https://ui.adsabs.harvard.edu/abs/2019ApJ...882..169F}
}

@ARTICLE{neeley2019,
       author = {{Neeley}, Jillian R. and {Marengo}, Massimo and {Freedman}, Wendy L. and {Madore}, Barry F. and {Beaton}, Rachael L. and {Hatt}, Dylan and {Hoyt}, Taylor and {Monson}, Andrew J. and {Rich}, Jeffrey A. and {Sarajedini}, Ata and {Seibert}, Mark and {Scowcroft}, Victoria},
        title = "{Standard Galactic field RR Lyrae II: a Gaia DR2 calibration of the period-Wesenheit-metallicity relation}",
      journal = {\mnras},
         year = 2019,
        month = dec,
       volume = {490},
       number = {3},
        pages = {4254-4270},
          doi = {10.1093/mnras/stz2814},
archivePrefix = {arXiv},
       eprint = {1910.01773},
 primaryClass = {astro-ph.SR},
       adsurl = {https://ui.adsabs.harvard.edu/abs/2019MNRAS.490.4254N}
}

@ARTICLE{soszynski2019,
       author = {{Soszy{\'n}ski}, I. and {Udalski}, A. and {Wrona}, M. and {Szyma{\'n}ski}, M.~K. and {Pietrukowicz}, P. and {Skowron}, J. and {Skowron}, D. and {Poleski}, R. and {Koz{\l}owski}, S. and {Mr{\'o}z}, P. and {Ulaczyk}, K. and {Rybicki}, K. and {Iwanek}, P. and {Gromadzki}, M.},
        title = "{Over 78 000 RR Lyrae Stars in the Galactic Bulge and Disk from the OGLE Survey}",
      journal = {\actaa},
         year = 2019,
        month = dec,
       volume = {69},
       number = {4},
        pages = {321-337},
          doi = {10.32023/0001-5237/69.4.2},
archivePrefix = {arXiv},
       eprint = {2001.00025},
 primaryClass = {astro-ph.SR},
       adsurl = {https://ui.adsabs.harvard.edu/abs/2019AcA....69..321S}
}

@ARTICLE{scipy2020,
       author = {{Virtanen}, Pauli and {Gommers}, Ralf and {Oliphant}, Travis E. and {Haberland}, Matt and {Reddy}, Tyler and {Cournapeau}, David and {Burovski}, Evgeni and {Peterson}, Pearu and {Weckesser}, Warren and {Bright}, Jonathan and {van der Walt}, St{\'e}fan J. and {Brett}, Matthew and {Wilson}, Joshua and {Millman}, K. Jarrod and {Mayorov}, Nikolay and {Nelson}, Andrew R.~J. and {Jones}, Eric and {Kern}, Robert and {Larson}, Eric and {Carey}, C.~J. and {Polat}, {\.I}lhan and {Feng}, Yu and {Moore}, Eric W. and {VanderPlas}, Jake and {Laxalde}, Denis and {Perktold}, Josef and {Cimrman}, Robert and {Henriksen}, Ian and {Quintero}, E.~A. and {Harris}, Charles R. and {Archibald}, Anne M. and {Ribeiro}, Ant{\^o}nio H. and {Pedregosa}, Fabian and {van Mulbregt}, Paul and {SciPy 1. 0 Contributors}},
        title = "{SciPy 1.0: fundamental algorithms for scientific computing in Python}",
      journal = {Nature Medicine},
         year = 2020,
        month = feb,
       volume = {17},
        pages = {261-272},
          doi = {10.1038/s41592-019-0686-2},
archivePrefix = {arXiv},
       eprint = {1907.10121},
 primaryClass = {cs.MS},
       adsurl = {https://ui.adsabs.harvard.edu/abs/2020NatMe..17..261V}
}

@ARTICLE{zinn2020,
       author = {{Zinn}, R. and {Chen}, X. and {Layden}, A.~C. and {Casetti-Dinescu}, D.~I.},
        title = "{Local RR Lyrae stars: native and alien}",
      journal = {\mnras},
         year = 2020,
        month = feb,
       volume = {492},
       number = {2},
        pages = {2161-2176},
          doi = {10.1093/mnras/stz3580},
archivePrefix = {arXiv},
       eprint = {1912.07686},
 primaryClass = {astro-ph.GA},
       adsurl = {https://ui.adsabs.harvard.edu/abs/2020MNRAS.492.2161Z}
}

@ARTICLE{numpy2020,
 title         = {Array programming with {NumPy}},
 author        = {Charles R. Harris and K. Jarrod Millman and St{\'{e}}fan J.
                 van der Walt and Ralf Gommers and Pauli Virtanen and David
                 Cournapeau and Eric Wieser and Julian Taylor and Sebastian
                 Berg and Nathaniel J. Smith and Robert Kern and Matti Picus
                 and Stephan Hoyer and Marten H. van Kerkwijk and Matthew
                 Brett and Allan Haldane and Jaime Fern{\'{a}}ndez del
                 R{\'{i}}o and Mark Wiebe and Pearu Peterson and Pierre
                 G{\'{e}}rard-Marchant and Kevin Sheppard and Tyler Reddy and
                 Warren Weckesser and Hameer Abbasi and Christoph Gohlke and
                 Travis E. Oliphant},
 year          = {2020},
 month         = sep,
 journal       = {Nature},
 volume        = {585},
 number        = {7825},
 pages         = {357--362},
 doi           = {10.1038/s41586-020-2649-2},
 publisher     = {Springer Science and Business Media {LLC}},
 url           = {https://doi.org/10.1038/s41586-020-2649-2}
}

@ARTICLE{bhardwaj2020,
       author = {{Bhardwaj}, Anupam and {Rejkuba}, Marina and {de Grijs}, Richard and {Herczeg}, Gregory J. and {Singh}, Harinder P. and {Kanbur}, Shashi and {Ngeow}, Chow-Choong},
        title = "{Near-infrared Census of RR Lyrae Variables in the Messier 3 Globular Cluster and the Period-Luminosity Relations}",
      journal = {\aj},
         year = 2020,
        month = nov,
       volume = {160},
       number = {5},
          eid = {220},
        pages = {220},
          doi = {10.3847/1538-3881/abb3f9},
archivePrefix = {arXiv},
       eprint = {2008.11745},
 primaryClass = {astro-ph.SR},
       adsurl = {https://ui.adsabs.harvard.edu/abs/2020AJ....160..220B}
}

@ARTICLE{braga2020,
       author = {{Braga}, V.~F. and {Bono}, G. and {Fiorentino}, G. and {Stetson}, P.~B. and {Dall'Ora}, M. and {Salaris}, M. and {da Silva}, R. and {Fabrizio}, M. and {Marinoni}, S. and {Marrese}, M.~P. and {Mateo}, M. and {Matsunaga}, N. and {Monelli}, M. and {Wallerstein}, G.},
        title = "{Separation between RR Lyrae and type II Cepheids and their importance for a distance determination: the case of omega Cen}",
      journal = {\aap},
         year = 2020,
        month = dec,
       volume = {644},
          eid = {A95},
        pages = {A95},
          doi = {10.1051/0004-6361/202039145},
archivePrefix = {arXiv},
       eprint = {2010.06368},
 primaryClass = {astro-ph.SR},
       adsurl = {https://ui.adsabs.harvard.edu/abs/2020A&A...644A..95B}
}

@ARTICLE{crestani2021a,
       author = {{Crestani}, J. and {Fabrizio}, M. and {Braga}, V.~F. and {Sneden}, C. and {Preston}, G. and {Ferraro}, I. and {Iannicola}, G. and {Bono}, G. and {Alves-Brito}, A. and {Nonino}, M. and {D'Orazi}, V. and {Inno}, L. and {Monelli}, M. and {Storm}, J. and {Altavilla}, G. and {Chaboyer}, B. and {Dall'Ora}, M. and {Fiorentino}, G. and {Gilligan}, C. and {Grebel}, E.~K. and {Lala}, H. and {Lemasle}, B. and {Marengo}, M. and {Marinoni}, S. and {Marrese}, P.~M. and {Mart{\'\i}nez-V{\'a}zquez}, C.~E. and {Matsunaga}, N. and {Mullen}, J.~P. and {Neeley}, J. and {Prudil}, Z. and {da Silva}, R. and {Stetson}, P.~B. and {Th{\'e}venin}, F. and {Valenti}, E. and {Walker}, A. and {Zoccali}, M.},
        title = "{On the Use of Field RR Lyrae as Galactic Probes. II. A New {\ensuremath{\Delta}}S Calibration to Estimate Their Metallicity}",
      journal = {\apj},
         year = 2021,
        month = feb,
       volume = {908},
       number = {1},
          eid = {20},
        pages = {20},
          doi = {10.3847/1538-4357/abd183},
archivePrefix = {arXiv},
       eprint = {2012.02284},
 primaryClass = {astro-ph.SR},
       adsurl = {https://ui.adsabs.harvard.edu/abs/2021ApJ...908...20C}
}

@ARTICLE{riess2021a,
       author = {{Riess}, Adam G. and {Casertano}, Stefano and {Yuan}, Wenlong and {Bowers}, J. Bradley and {Macri}, Lucas and {Zinn}, Joel C. and {Scolnic}, Dan},
        title = "{Cosmic Distances Calibrated to 1\% Precision with Gaia EDR3 Parallaxes and Hubble Space Telescope Photometry of 75 Milky Way Cepheids Confirm Tension with {\ensuremath{\Lambda}}CDM}",
      journal = {\apjl},
         year = 2021,
        month = feb,
       volume = {908},
       number = {1},
          eid = {L6},
        pages = {L6},
          doi = {10.3847/2041-8213/abdbaf},
archivePrefix = {arXiv},
       eprint = {2012.08534},
 primaryClass = {astro-ph.CO},
       adsurl = {https://ui.adsabs.harvard.edu/abs/2021ApJ...908L...6R}
}

@ARTICLE{iorio2021,
       author = {{Iorio}, Giuliano and {Belokurov}, Vasily},
        title = "{Chemo-kinematics of the Gaia RR Lyrae: the halo and the disc}",
      journal = {\mnras},
         year = 2021,
        month = apr,
       volume = {502},
       number = {4},
        pages = {5686-5710},
          doi = {10.1093/mnras/stab005},
archivePrefix = {arXiv},
       eprint = {2008.02280},
 primaryClass = {astro-ph.GA},
       adsurl = {https://ui.adsabs.harvard.edu/abs/2021MNRAS.502.5686I}
}

@ARTICLE{mullen2021,
       author = {{Mullen}, Joseph P. and {Marengo}, Massimo and {Mart{\'\i}nez-V{\'a}zquez}, Clara E. and {Neeley}, Jillian R. and {Bono}, Giuseppe and {Dall'Ora}, Massimo and {Chaboyer}, Brian and {Th{\'e}venin}, Fr{\'e}d{\'e}ric and {Braga}, Vittorio F. and {Crestani}, Juliana and {Fabrizio}, Michele and {Fiorentino}, Giuliana and {Gilligan}, Christina K. and {Monelli}, Matteo and {Stetson}, Peter B.},
        title = "{Metallicity of Galactic RR Lyrae from Optical and Infrared Light Curves. I. Period-Fourier-Metallicity Relations for Fundamental-mode RR Lyrae}",
      journal = {\apj},
         year = 2021,
        month = may,
       volume = {912},
       number = {2},
          eid = {144},
        pages = {144},
          doi = {10.3847/1538-4357/abefd4},
archivePrefix = {arXiv},
       eprint = {2103.09372},
 primaryClass = {astro-ph.SR},
       adsurl = {https://ui.adsabs.harvard.edu/abs/2021ApJ...912..144M}
}

@ARTICLE{lindegren2021,
       author = {{Lindegren}, L. and {Bastian}, U. and {Biermann}, M. and {Bombrun}, A. and {de Torres}, A. and {Gerlach}, E. and {Geyer}, R. and {Hern{\'a}ndez}, J. and {Hilger}, T. and {Hobbs}, D. and {Klioner}, S.~A. and {Lammers}, U. and {McMillan}, P.~J. and {Ramos-Lerate}, M. and {Steidelm{\"u}ller}, H. and {Stephenson}, C.~A. and {van Leeuwen}, F.},
        title = "{Gaia Early Data Release 3. Parallax bias versus magnitude, colour, and position}",
      journal = {\aap},
         year = 2021,
        month = may,
       volume = {649},
          eid = {A4},
        pages = {A4},
          doi = {10.1051/0004-6361/202039653},
archivePrefix = {arXiv},
       eprint = {2012.01742},
 primaryClass = {astro-ph.IM},
       adsurl = {https://ui.adsabs.harvard.edu/abs/2021A&A...649A...4L}
}

@ARTICLE{crestani2021b,
       author = {{Crestani}, J. and {Braga}, V.~F. and {Fabrizio}, M. and {Bono}, G. and {Sneden}, C. and {Preston}, G. and {Ferraro}, I. and {Iannicola}, G. and {Nonino}, M. and {Fiorentino}, G. and {Th{\'e}venin}, F. and {Lemasle}, B. and {Prudil}, Z. and {Alves-Brito}, A. and {Altavilla}, G. and {Chaboyer}, B. and {Dall'Ora}, M. and {D'Orazi}, V. and {Gilligan}, C. and {Grebel}, E.~K. and {Koch-Hansen}, A.~J. and {Lala}, H. and {Marengo}, M. and {Marinoni}, S. and {Marrese}, P.~M. and {Mart{\'\i}nez-V{\'a}zquez}, C. and {Matsunaga}, N. and {Monelli}, M. and {Mullen}, J.~P. and {Neeley}, J. and {da Silva}, R. and {Stetson}, P.~B. and {Salaris}, M. and {Storm}, J. and {Valenti}, E. and {Zoccali}, M.},
        title = "{On the Use of Field RR Lyrae as Galactic Probes. III. The {\ensuremath{\alpha}}-element Abundances}",
      journal = {\apj},
         year = 2021,
        month = jun,
       volume = {914},
       number = {1},
          eid = {10},
        pages = {10},
          doi = {10.3847/1538-4357/abfa23},
archivePrefix = {arXiv},
       eprint = {2104.08113},
 primaryClass = {astro-ph.GA},
       adsurl = {https://ui.adsabs.harvard.edu/abs/2021ApJ...914...10C}
}

@ARTICLE{gilligan2021,
       author = {{Gilligan}, Christina K. and {Chaboyer}, Brian and {Marengo}, Massimo and {Mullen}, Joseph P. and {Bono}, Giuseppe and {Braga}, Vittorio F. and {Crestani}, Juliana and {Dall'Ora}, Massimo and {Fiorentino}, Giuliana and {Monelli}, Matteo and {Neeley}, Jill R. and {Fabrizio}, Michele and {Mart{\'\i}nez-V{\'a}zquez}, Clara E. and {Th{\'e}venin}, Fr{\'e}d{\'e}ric and {Sneden}, Christopher},
        title = "{Metallicities from high-resolution spectra of 49 RR Lyrae variables}",
      journal = {\mnras},
         year = 2021,
        month = jun,
       volume = {503},
       number = {4},
        pages = {4719-4733},
          doi = {10.1093/mnras/stab857},
archivePrefix = {arXiv},
       eprint = {2103.11012},
 primaryClass = {astro-ph.SR},
       adsurl = {https://ui.adsabs.harvard.edu/abs/2021MNRAS.503.4719G}
}

@ARTICLE{buder2021,
       author = {{Buder}, Sven and {Sharma}, Sanjib and {Kos}, Janez and {Amarsi}, Anish M. and {Nordlander}, Thomas and {Lind}, Karin and {Martell}, Sarah L. and {Asplund}, Martin and {Bland-Hawthorn}, Joss and {Casey}, Andrew R. and {de Silva}, Gayandhi M. and {D'Orazi}, Valentina and {Freeman}, Ken C. and {Hayden}, Michael R. and {Lewis}, Geraint F. and {Lin}, Jane and {Schlesinger}, Katharine J. and {Simpson}, Jeffrey D. and {Stello}, Dennis and {Zucker}, Daniel B. and {Zwitter}, Toma{\v{z}} and {Beeson}, Kevin L. and {Buck}, Tobias and {Casagrande}, Luca and {Clark}, Jake T. and {{\v{C}}otar}, Klemen and {da Costa}, Gary S. and {de Grijs}, Richard and {Feuillet}, Diane and {Horner}, Jonathan and {Kafle}, Prajwal R. and {Khanna}, Shourya and {Kobayashi}, Chiaki and {Liu}, Fan and {Montet}, Benjamin T. and {Nandakumar}, Govind and {Nataf}, David M. and {Ness}, Melissa K. and {Spina}, Lorenzo and {Tepper-Garc{\'\i}a}, Thor and {Ting}, Yuan-Sen and {Traven}, Gregor and {Vogrin{\v{c}}i{\v{c}}}, Rok and {Wittenmyer}, Robert A. and {Wyse}, Rosemary F.~G. and {{\v{Z}}erjal}, Maru{\v{s}}a and {Galah Collaboration}},
        title = "{The GALAH+ survey: Third data release}",
      journal = {\mnras},
         year = 2021,
        month = sep,
       volume = {506},
       number = {1},
        pages = {150-201},
          doi = {10.1093/mnras/stab1242},
archivePrefix = {arXiv},
       eprint = {2011.02505},
 primaryClass = {astro-ph.GA},
       adsurl = {https://ui.adsabs.harvard.edu/abs/2021MNRAS.506..150B}
}

@ARTICLE{dekany2021,
       author = {{D{\'e}k{\'a}ny}, Istv{\'a}n and {Grebel}, Eva K. and {Pojma{\'n}ski}, Grzegorz},
        title = "{Metallicity Estimation of RR Lyrae Stars From Their I-Band Light Curves}",
      journal = {\apj},
         year = 2021,
        month = oct,
       volume = {920},
       number = {1},
          eid = {33},
        pages = {33},
          doi = {10.3847/1538-4357/ac106f},
archivePrefix = {arXiv},
       eprint = {2107.05983},
 primaryClass = {astro-ph.SR},
       adsurl = {https://ui.adsabs.harvard.edu/abs/2021ApJ...920...33D}
}

@ARTICLE{fabrizio2021,
       author = {{Fabrizio}, M. and {Braga}, V.~F. and {Crestani}, J. and {Bono}, G. and {Ferraro}, I. and {Fiorentino}, G. and {Iannicola}, G. and {Preston}, G.~W. and {Sneden}, C. and {Th{\'e}venin}, F. and {Altavilla}, G. and {Chaboyer}, B. and {Dall'Ora}, M. and {da Silva}, R. and {Grebel}, E.~K. and {Gilligan}, C.~K. and {Lala}, H. and {Lemasle}, B. and {Magurno}, D. and {Marengo}, M. and {Marinoni}, S. and {Marrese}, P.~M. and {Mart{\'\i}nez-V{\'a}zquez}, C.~E. and {Matsunaga}, N. and {Monelli}, M. and {Mullen}, J.~P. and {Neeley}, J. and {Nonino}, M. and {Prudil}, Z. and {Salaris}, M. and {Stetson}, P.~B. and {Valenti}, E. and {Zoccali}, M.},
        title = "{On the Use of Field RR Lyrae As Galactic Probes: IV. New Insights Into and Around the Oosterhoff Dichotomy}",
      journal = {\apj},
         year = 2021,
        month = oct,
       volume = {919},
       number = {2},
          eid = {118},
        pages = {118},
          doi = {10.3847/1538-4357/ac1115},
archivePrefix = {arXiv},
       eprint = {2107.00919},
 primaryClass = {astro-ph.SR},
       adsurl = {https://ui.adsabs.harvard.edu/abs/2021ApJ...919..118F}
}

@ARTICLE{bhardwaj2021,
       author = {{Bhardwaj}, Anupam and {Rejkuba}, Marina and {Sloan}, G.~C. and {Marconi}, Marcella and {Yang}, Soung-Chul},
        title = "{Optical and Near-infrared Pulsation Properties of RR Lyrae and Population II Cepheid Variables in the Messier 15 Globular Cluster}",
      journal = {\apj},
         year = 2021,
        month = nov,
       volume = {922},
       number = {1},
          eid = {20},
        pages = {20},
          doi = {10.3847/1538-4357/ac214d},
archivePrefix = {arXiv},
       eprint = {2108.11388},
 primaryClass = {astro-ph.SR},
       adsurl = {https://ui.adsabs.harvard.edu/abs/2021ApJ...922...20B}
}

@ARTICLE{garofalo2022,
       author = {{Garofalo}, A. and {Delgado}, H.~E. and {Sarro}, L.~M. and {Clementini}, G. and {Muraveva}, T. and {Marconi}, M. and {Ripepi}, V.},
        title = "{New LZ and PW(Z) relations of RR Lyrae stars calibrated with Gaia EDR3 parallaxes}",
      journal = {\mnras},
         year = 2022,
        month = jun,
       volume = {513},
       number = {1},
        pages = {788-806},
          doi = {10.1093/mnras/stac735},
archivePrefix = {arXiv},
       eprint = {2203.07435},
 primaryClass = {astro-ph.SR},
       adsurl = {https://ui.adsabs.harvard.edu/abs/2022MNRAS.513..788G}
}

@ARTICLE{mullen2022,
       author = {{Mullen}, Joseph P. and {Marengo}, Massimo and {Mart{\'\i}nez-V{\'a}zquez}, Clara E. and {Bono}, Giuseppe and {Braga}, Vittorio F. and {Chaboyer}, Brian and {Crestani}, Juliana and {Dall'Ora}, Massimo and {Fabrizio}, Michele and {Fiorentino}, Giuliana and {Monelli}, Matteo and {Neeley}, Jillian R. and {Stetson}, Peter B. and {Th{\'e}venin}, Fr{\'e}d{\'e}ric},
        title = "{Metallicity of Galactic RR Lyrae from Optical and Infrared Light Curves. II. Period-Fourier-Metallicity Relations for First Overtone RR Lyrae}",
      journal = {\apj},
         year = 2022,
        month = jun,
       volume = {931},
       number = {2},
          eid = {131},
        pages = {131},
          doi = {10.3847/1538-4357/ac67ee},
archivePrefix = {arXiv},
       eprint = {2204.07627},
 primaryClass = {astro-ph.SR},
       adsurl = {https://ui.adsabs.harvard.edu/abs/2022ApJ...931..131M}
}

@ARTICLE{marconi2022,
       author = {{Marconi}, Marcella and {Molinaro}, Roberto and {Dall'Ora}, Massimo and {Ripepi}, Vincenzo and {Musella}, Ilaria and {Bono}, Giuseppe and {Braga}, Vittorio and {Di Criscienzo}, Marcella and {Fiorentino}, Giuliana and {Leccia}, Silvio and {Monelli}, Matteo},
        title = "{New Theoretical Period-Luminosity-Metallicity Relations for RR Lyrae in the Rubin-LSST Filters}",
      journal = {\apj},
         year = 2022,
        month = jul,
       volume = {934},
       number = {1},
          eid = {29},
        pages = {29},
          doi = {10.3847/1538-4357/ac78ee},
archivePrefix = {arXiv},
       eprint = {2206.06470},
 primaryClass = {astro-ph.SR},
       adsurl = {https://ui.adsabs.harvard.edu/abs/2022ApJ...934...29M}
}

@software{pandas2022,
       author = {{The pandas development Team}},
        title = "{pandas-dev/pandas: Pandas}",
         year = 2025,
        month = jul,
          eid = {10.5281/zenodo.3509134},
          doi = {10.5281/zenodo.3509134},
      version = {v2.3.1},
    publisher = {Zenodo},
       adsurl = {https://ui.adsabs.harvard.edu/abs/2022zndo...3509134T}
}

@ARTICLE{dekany2022,
       author = {{D{\'e}k{\'a}ny}, Istv{\'a}n and {Grebel}, Eva K.},
        title = "{Photometric Metallicity Prediction of Fundamental-mode RR Lyrae Stars in the Gaia Optical and K $_{ s }$ Infrared Wave Bands by Deep Learning}",
      journal = {\apjs},
         year = 2022,
        month = aug,
       volume = {261},
       number = {2},
          eid = {33},
        pages = {33},
          doi = {10.3847/1538-4365/ac74ba},
       adsurl = {https://ui.adsabs.harvard.edu/abs/2022ApJS..261...33D}
}

@ARTICLE{astropy2022,
       author = {{Astropy Collaboration} and {Price-Whelan}, Adrian M. and {Lim}, Pey Lian and {Earl}, Nicholas and {Starkman}, Nathaniel and {Bradley}, Larry and {Shupe}, David L. and {Patil}, Aarya A. and {Corrales}, Lia and {Brasseur}, C.~E. and {N{"o}the}, Maximilian and {Donath}, Axel and {Tollerud}, Erik and {Morris}, Brett M. and {Ginsburg}, Adam and {Vaher}, Eero and {Weaver}, Benjamin A. and {Tocknell}, James and {Jamieson}, William and {van Kerkwijk}, Marten H. and {Robitaille}, Thomas P. and {Merry}, Bruce and {Bachetti}, Matteo and {G{"u}nther}, H. Moritz and {Aldcroft}, Thomas L. and {Alvarado-Montes}, Jaime A. and {Archibald}, Anne M. and {B{'o}di}, Attila and {Bapat}, Shreyas and {Barentsen}, Geert and {Baz{'a}n}, Juanjo and {Biswas}, Manish and {Boquien}, M{'e}d{'e}ric and {Burke}, D.~J. and {Cara}, Daria and {Cara}, Mihai and {Conroy}, Kyle E. and {Conseil}, Simon and {Craig}, Matthew W. and {Cross}, Robert M. and {Cruz}, Kelle L. and {D'Eugenio}, Francesco and {Dencheva}, Nadia and {Devillepoix}, Hadrien A.~R. and {Dietrich}, J{"o}rg P. and {Eigenbrot}, Arthur Davis and {Erben}, Thomas and {Ferreira}, Leonardo and {Foreman-Mackey}, Daniel and {Fox}, Ryan and {Freij}, Nabil and {Garg}, Suyog and {Geda}, Robel and {Glattly}, Lauren and {Gondhalekar}, Yash and {Gordon}, Karl D. and {Grant}, David and {Greenfield}, Perry and {Groener}, Austen M. and {Guest}, Steve and {Gurovich}, Sebastian and {Handberg}, Rasmus and {Hart}, Akeem and {Hatfield-Dodds}, Zac and {Homeier}, Derek and {Hosseinzadeh}, Griffin and {Jenness}, Tim and {Jones}, Craig K. and {Joseph}, Prajwel and {Kalmbach}, J. Bryce and {Karamehmetoglu}, Emir and {Ka{l}uszy{'n}ski}, Miko{l}aj and {Kelley}, Michael S.~P. and {Kern}, Nicholas and {Kerzendorf}, Wolfgang E. and {Koch}, Eric W. and {Kulumani}, Shankar and {Lee}, Antony and {Ly}, Chun and {Ma}, Zhiyuan and {MacBride}, Conor and {Maljaars}, Jakob M. and {Muna}, Demitri and {Murphy}, N.~A. and {Norman}, Henrik and {O'Steen}, Richard and {Oman}, Kyle A. and {Pacifici}, Camilla and {Pascual}, Sergio and {Pascual-Granado}, J. and {Patil}, Rohit R. and {Perren}, Gabriel I. and {Pickering}, Timothy E. and {Rastogi}, Tanuj and {Roulston}, Benjamin R. and {Ryan}, Daniel F. and {Rykoff}, Eli S. and {Sabater}, Jose and {Sakurikar}, Parikshit and {Salgado}, Jes{'u}s and {Sanghi}, Aniket and {Saunders}, Nicholas and {Savchenko}, Volodymyr and {Schwardt}, Ludwig and {Seifert-Eckert}, Michael and {Shih}, Albert Y. and {Jain}, Anany Shrey and {Shukla}, Gyanendra and {Sick}, Jonathan and {Simpson}, Chris and {Singanamalla}, Sudheesh and {Singer}, Leo P. and {Singhal}, Jaladh and {Sinha}, Manodeep and {Sip{H{o}}cz}, Brigitta M. and {Spitler}, Lee R. and {Stansby}, David and {Streicher}, Ole and {{{S}}umak}, Jani and {Swinbank}, John D. and {Taranu}, Dan S. and {Tewary}, Nikita and {Tremblay}, Grant R. and {Val-Borro}, Miguel de and {Van Kooten}, Samuel J. and {Vasovi{'c}}, Zlatan and {Verma}, Shresth and {de Miranda Cardoso}, Jos{'e} Vin{'i}cius and {Williams}, Peter K.~G. and {Wilson}, Tom J. and {Winkel}, Benjamin and {Wood-Vasey}, W.~M. and {Xue}, Rui and {Yoachim}, Peter and {Zhang}, Chen and {Zonca}, Andrea and {Astropy Project Contributors}},
        title = "{The Astropy Project: Sustaining and Growing a Community-oriented Open-source Project and the Latest Major Release (v5.0) of the Core Package}",
      journal = {\apj},
         year = 2022,
        month = aug,
       volume = {935},
       number = {2},
          eid = {167},
        pages = {167},
          doi = {10.3847/1538-4357/ac7c74},
archivePrefix = {arXiv},
       eprint = {2206.14220},
 primaryClass = {astro-ph.IM},
       adsurl = {https://ui.adsabs.harvard.edu/abs/2022ApJ...935..167A}
}

@ARTICLE{bhardwaj2023,
       author = {{Bhardwaj}, Anupam and {Marconi}, Marcella and {Rejkuba}, Marina and {de Grijs}, Richard and {Singh}, Harinder P. and {Braga}, Vittorio F. and {Kanbur}, Shashi and {Ngeow}, Chow-Choong and {Ripepi}, Vincenzo and {Bono}, Giuseppe and {De Somma}, Giulia and {Dall'Ora}, Massimo},
        title = "{Precise Empirical Determination of Metallicity Dependence of Near-infrared Period-Luminosity Relations for RR Lyrae Variables}",
      journal = {\apjl},
         year = 2023,
        month = feb,
       volume = {944},
       number = {2},
          eid = {L51},
        pages = {L51},
          doi = {10.3847/2041-8213/acba7f},
archivePrefix = {arXiv},
       eprint = {2302.05455},
 primaryClass = {astro-ph.SR},
       adsurl = {https://ui.adsabs.harvard.edu/abs/2023ApJ...944L..51B}
}

@ARTICLE{mullen2023,
       author = {{Mullen}, Joseph P. and {Marengo}, Massimo and {Mart{\'\i}nez-V{\'a}zquez}, Clara E. and {Chaboyer}, Brian and {Bono}, Giuseppe and {Braga}, Vittorio F. and {Dall'Ora}, Massimo and {D'Orazi}, Valentina and {Fabrizio}, Michele and {Monelli}, Matteo and {Th{\'e}venin}, Fr{\'e}d{\'e}ric},
        title = "{RR Lyrae Mid-infrared Period-Luminosity-Metallicity and Period-Wesenheit-Metallicity Relations Based on Gaia DR3 Parallaxes}",
      journal = {\apj},
         year = 2023,
        month = mar,
       volume = {945},
       number = {1},
          eid = {83},
        pages = {83},
          doi = {10.3847/1538-4357/acb20a},
archivePrefix = {arXiv},
       eprint = {2301.03777},
 primaryClass = {astro-ph.GA},
       adsurl = {https://ui.adsabs.harvard.edu/abs/2023ApJ...945...83M}
}

@ARTICLE{cruzreyes2023,
       author = {{Cruz Reyes}, Mauricio and {Anderson}, Richard I.},
        title = "{A 0.9\% calibration of the Galactic Cepheid luminosity scale based on Gaia DR3 data of open clusters and Cepheids}",
      journal = {\aap},
         year = 2023,
        month = apr,
       volume = {672},
          eid = {A85},
        pages = {A85},
          doi = {10.1051/0004-6361/202244775},
archivePrefix = {arXiv},
       eprint = {2208.09403},
 primaryClass = {astro-ph.GA},
       adsurl = {https://ui.adsabs.harvard.edu/abs/2023A&A...672A..85C}
}

@ARTICLE{eyer2023,
       author = {{Eyer}, L. and {Audard}, M. and {Holl}, B. and {Rimoldini}, L. and {Carnerero}, M.~I. and {Clementini}, G. and {De Ridder}, J. and {Distefano}, E. and {Evans}, D.~W. and {Gavras}, P. and {Gomel}, R. and {Lebzelter}, T. and {Marton}, G. and {Mowlavi}, N. and {Panahi}, A. and {Ripepi}, V. and {Wyrzykowski}, {\L}. and {Nienartowicz}, K. and {Jevardat de Fombelle}, G. and {Lecoeur-Taibi}, I. and {Rohrbasser}, L. and {Riello}, M. and {Garc{\'\i}a-Lario}, P. and {Lanzafame}, A.~C. and {Mazeh}, T. and {Raiteri}, C.~M. and {Zucker}, S. and {{\'A}brah{\'a}m}, P. and {Aerts}, C. and {Aguado}, J.~J. and {Anderson}, R.~I. and {Bashi}, D. and {Binnenfeld}, A. and {Faigler}, S. and {Garofalo}, A. and {Karbevska}, L. and {K{\'o}sp{\'a}l}, {\'A}. and {Kruszy{\'n}ska}, K. and {Kun}, M. and {Lanza}, A.~F. and {Leccia}, S. and {Marconi}, M. and {Messina}, S. and {Molinaro}, R. and {Moln{\'a}r}, L. and {Muraveva}, T. and {Musella}, I. and {Nagy}, Z. and {Pagano}, I. and {Palaversa}, L. and {Plachy}, E. and {Pr{\v{s}}a}, A. and {Rybicki}, K.~A. and {Shahaf}, S. and {Szabados}, L. and {Szegedi-Elek}, E. and {Trabucchi}, M. and {Barblan}, F. and {Grenon}, M. and {Roelens}, M. and {S{\"u}veges}, M.},
        title = "{Gaia Data Release 3. Summary of the variability processing and analysis}",
      journal = {\aap},
         year = 2023,
        month = jun,
       volume = {674},
          eid = {A13},
        pages = {A13},
          doi = {10.1051/0004-6361/202244242},
archivePrefix = {arXiv},
       eprint = {2206.06416},
 primaryClass = {astro-ph.SR},
       adsurl = {https://ui.adsabs.harvard.edu/abs/2023A&A...674A..13E}
}

@ARTICLE{clementini2023,
       author = {{Clementini}, G. and {Ripepi}, V. and {Garofalo}, A. and {Molinaro}, R. and {Muraveva}, T. and {Leccia}, S. and {Rimoldini}, L. and {Holl}, B. and {Jevardat de Fombelle}, G. and {Sartoretti}, P. and {Marchal}, O. and {Audard}, M. and {Nienartowicz}, K. and {Andrae}, R. and {Marconi}, M. and {Szabados}, L. and {Evans}, D.~W. and {Lecoeur-Taibi}, I. and {Mowlavi}, N. and {Musella}, I. and {Eyer}, L.},
        title = "{Gaia Data Release 3. Specific processing and validation of all-sky RR Lyrae and Cepheid stars: The RR Lyrae sample}",
      journal = {\aap},
         year = 2023,
        month = jun,
       volume = {674},
          eid = {A18},
        pages = {A18},
          doi = {10.1051/0004-6361/202243964},
archivePrefix = {arXiv},
       eprint = {2206.06278},
 primaryClass = {astro-ph.SR},
       adsurl = {https://ui.adsabs.harvard.edu/abs/2023A&A...674A..18C}
}

@ARTICLE{zgirski2023,
       author = {{Zgirski}, Bart{\l}omiej and {Pietrzy{\'n}ski}, Grzegorz and {G{\'o}rski}, Marek and {Gieren}, Wolfgang and {Wielg{\'o}rski}, Piotr and {Karczmarek}, Paulina and {Hajdu}, Gergely and {Lewis}, Megan and {Chini}, Rolf and {Graczyk}, Dariusz and {Ka{\l}uszy{\'n}ski}, Miko{\l}aj and {Narloch}, Weronika and {Pilecki}, Bogumi{\l} and {Garc{\'\i}a}, Gonzalo Rojas and {Suchomska}, Ksenia and {Taormina}, M{\'o}nica},
        title = "{New Near-infrared Period-Luminosity-Metallicity Relations for Galactic RR Lyrae Stars Based on Gaia EDR3 Parallaxes}",
      journal = {\apj},
         year = 2023,
        month = jul,
       volume = {951},
       number = {2},
          eid = {114},
        pages = {114},
          doi = {10.3847/1538-4357/acd63a},
archivePrefix = {arXiv},
       eprint = {2305.09414},
 primaryClass = {astro-ph.SR},
       adsurl = {https://ui.adsabs.harvard.edu/abs/2023ApJ...951..114Z}
}

@ARTICLE{zhang2023,
       author = {{Zhang}, Xiangyu and {Green}, Gregory M. and {Rix}, Hans-Walter},
        title = "{Parameters of 220 million stars from Gaia BP/RP spectra}",
      journal = {\mnras},
         year = 2023,
        month = sep,
       volume = {524},
       number = {2},
        pages = {1855-1884},
          doi = {10.1093/mnras/stad1941},
archivePrefix = {arXiv},
       eprint = {2303.03420},
 primaryClass = {astro-ph.SR},
       adsurl = {https://ui.adsabs.harvard.edu/abs/2023MNRAS.524.1855Z}
}

@ARTICLE{khan2023,
       author = {{Khan}, S. and {Anderson}, R.~I. and {Miglio}, A. and {Mosser}, B. and {Elsworth}, Y.~P.},
        title = "{Investigating Gaia EDR3 parallax systematics using asteroseismology of cool giant stars observed by Kepler, K2, and TESS. II. Deciphering Gaia parallax systematics using red clump stars}",
      journal = {\aap},
         year = 2023,
        month = dec,
       volume = {680},
          eid = {A105},
        pages = {A105},
          doi = {10.1051/0004-6361/202347919},
archivePrefix = {arXiv},
       eprint = {2310.03654},
 primaryClass = {astro-ph.SR},
       adsurl = {https://ui.adsabs.harvard.edu/abs/2023A&A...680A.105K}
}

@INPROCEEDINGS{groenewegen2024,
       author = {{Groenewegen}, M.~A.~T.},
        title = "{Primary Period-Luminosity-Relation Calibrators in the Milky Way: Cepheids and RR Lyrae Physical basis, Calibration, and Applications}",
    booktitle = {IAU Symposium},
         year = 2024,
       editor = {{de Grijs}, Richard and {Whitelock}, Patricia A. and {Catelan}, M{\'a}rcio},
       series = {IAU Symposium},
       volume = {376},
        month = jan,
        pages = {128-149},
          doi = {10.1017/S1743921323002995},
archivePrefix = {arXiv},
       eprint = {2307.03033},
 primaryClass = {astro-ph.GA},
       adsurl = {https://ui.adsabs.harvard.edu/abs/2024IAUS..376..128G}
}

@ARTICLE{edenhofer2024,
       author = {{Edenhofer}, Gordian and {Zucker}, Catherine and {Frank}, Philipp and {Saydjari}, Andrew K. and {Speagle}, Joshua S. and {Finkbeiner}, Douglas and {En{\ss}lin}, Torsten A.},
        title = "{A parsec-scale Galactic 3D dust map out to 1.25 kpc from the Sun}",
      journal = {\aap},
         year = 2024,
        month = may,
       volume = {685},
          eid = {A82},
        pages = {A82},
          doi = {10.1051/0004-6361/202347628},
archivePrefix = {arXiv},
       eprint = {2308.01295},
 primaryClass = {astro-ph.GA},
       adsurl = {https://ui.adsabs.harvard.edu/abs/2024A&A...685A..82E}
}

@ARTICLE{dorazi2024,
       author = {{D'Orazi}, Valentina and {Storm}, Nicholas and {Casey}, Andrew R. and {Braga}, Vittorio F. and {Zocchi}, Alice and {Bono}, Giuseppe and {Fabrizio}, Michele and {Sneden}, Christopher and {Massari}, Davide and {Giribaldi}, Riano E. and {Bergemann}, Maria and {Campbell}, Simon W. and {Casagrande}, Luca and {de Grijs}, Richard and {De Silva}, Gayandhi and {Lugaro}, Maria and {Zucker}, Daniel B. and {Bragaglia}, Angela and {Feuillet}, Diane and {Fiorentino}, Giuliana and {Chaboyer}, Brian and {Dall'Ora}, Massimo and {Marengo}, Massimo and {Mart{\'\i}nez-V{\'a}zquez}, Clara E. and {Matsunaga}, Noriyuki and {Monelli}, Matteo and {Mullen}, Joseph P. and {Nataf}, David and {Tantalo}, Maria and {Thevenin}, Frederic and {Vitello}, Fabio R. and {Kudritzki}, Rolf-Peter and {Bland-Hawthorn}, Joss and {Buder}, Sven and {Freeman}, Ken and {Kos}, Janez and {Lewis}, Geraint F. and {Lind}, Karin and {Martell}, Sarah and {Sharma}, Sanjib and {Stello}, Dennis and {Zwitter}, Toma{\v{z}}},
        title = "{The GALAH survey: tracing the Milky Way's formation and evolution through RR Lyrae stars}",
      journal = {\mnras},
         year = 2024,
        month = jun,
       volume = {531},
       number = {1},
        pages = {137-162},
          doi = {10.1093/mnras/stae1149},
archivePrefix = {arXiv},
       eprint = {2405.04580},
 primaryClass = {astro-ph.GA},
       adsurl = {https://ui.adsabs.harvard.edu/abs/2024MNRAS.531..137D}
}

@ARTICLE{braga2024,
       author = {{Braga}, V.~F. and {Monelli}, M. and {Dall'Ora}, M. and {Mullen}, J.~P. and {Molinaro}, R. and {Marconi}, M. and {Szab{\'o}}, R. and {Gallart}, C.},
        title = "{On the use of field RR Lyrae as Galactic probes: VII. Light curve templates in the LSST photometric system}",
      journal = {\aap},
         year = 2024,
        month = sep,
       volume = {689},
          eid = {A349},
        pages = {A349},
          doi = {10.1051/0004-6361/202450971},
archivePrefix = {arXiv},
       eprint = {2407.20813},
 primaryClass = {astro-ph.SR},
       adsurl = {https://ui.adsabs.harvard.edu/abs/2024A&A...689A.349B}
}

@ARTICLE{muraveva2025,
       author = {{Muraveva}, Tatiana and {Giannetti}, Andrea and {Clementini}, Gisella and {Garofalo}, Alessia and {Monti}, Lorenzo},
        title = "{Metallicity of RR Lyrae stars from the Gaia Data Release 3 catalogue computed with Machine Learning algorithms}",
      journal = {\mnras},
         year = 2025,
        month = jan,
       volume = {536},
       number = {3},
        pages = {2749-2769},
          doi = {10.1093/mnras/stae2679},
archivePrefix = {arXiv},
       eprint = {2407.05815},
 primaryClass = {astro-ph.SR},
       adsurl = {https://ui.adsabs.harvard.edu/abs/2025MNRAS.536.2749M}
}

@ARTICLE{dorazi2025a,
       author = {{D'Orazi}, V. and {Braga}, V. and {Bono}, G. and {Fabrizio}, M. and {Fiorentino}, G. and {Storm}, N. and {Pietrinferni}, A. and {Sneden}, C. and {S{\'a}nchez-Benavente}, M. and {Monelli}, M. and {Sestito}, F. and {J{\"o}nsson}, H. and {Buder}, S. and {Bobrick}, A. and {Iorio}, G. and {Matsunaga}, N. and {Marconi}, M. and {Marengo}, M. and {Mart{\'\i}nez-V{\'a}zquez}, C.~E. and {Mullen}, J. and {Takayama}, M. and {Testa}, V. and {Cusano}, F. and {Crestani}, J.},
        title = "{The elderly among the oldest: new evidence for extremely metal-poor RR Lyrae stars}",
      journal = {\aap},
         year = 2025,
        month = feb,
       volume = {694},
          eid = {A158},
        pages = {A158},
          doi = {10.1051/0004-6361/202453202},
archivePrefix = {arXiv},
       eprint = {2501.05807},
 primaryClass = {astro-ph.SR},
       adsurl = {https://ui.adsabs.harvard.edu/abs/2025A&A...694A.158D}
}

@ARTICLE{bono2026,
       author = {{Bono}, G. and {Braga}, V.~F. and {Fabrizio}, M. and {Tantalo}, M. and {Baeza-Villagra}, K. and {Crestani}, J. and {D'Orazi}, V. and {Dall'Ora}, M. and {Di Criscienzo}, M. and {Fiorentino}, G. and {Gholami}, M. and {Marengo}, M. and {Mart{\'\i}nez-V{\'a}zquez}, C.~E. and {Monelli}, M. and {Mullen}, J.~P. and {Nunnari}, A. and {Pipwala}, V.~D. and {Prudil}, Z. and {Sneden}, C. and {Altavilla}, G. and {Bergemann}, M. and {B{\"o}cek Topcu}, G. and {Buonanno}, R. and {Calamida}, A. and {Carretta}, E. and {Ceci}, G. and {Chaboyer}, B. and {Correnti}, M. and {da Silva}, R. and {Ferraro}, I. and {G{\'o}mez}, F.~A. and {Iannicola}, G. and {Kudritzki}, R.-P. and {Kunder}, A. and {Kwak}, S. and {Marconi}, M. and {Marinoni}, S. and {Matsunaga}, N. and {Matteucci}, F. and {Monachesi}, A. and {Musella}, I. and {Navarro Ovando}, M.~G. and {Preston}, G.~W. and {Ripepi}, V. and {Salaris}, M. and {S{\'a}nchez-Benavente}, M. and {Spitoni}, E. and {Stetson}, P.~B. and {Th{\'e}venin}, F. and {Thompson}, I.~B. and {Tissera}, P.~B. and {Tsujimoto}, T. and {Valenti}, E. and {Vivas}, A.~K. and {Walker}, A.~R. and {Zoccali}, M. and {Zocchi}, A.},
        title = "{On the Use of Field RR Lyrae as Galactic Probes. VIII. Early Formation of the Galactic Spheroid}",
      journal = {\apj},
         year = 2026,
        month = feb,
       volume = {998},
       number = {1},
          eid = {86},
        pages = {86},
          doi = {10.3847/1538-4357/ae2c60},
archivePrefix = {arXiv},
       eprint = {2601.16523},
 primaryClass = {astro-ph.GA},
       adsurl = {https://ui.adsabs.harvard.edu/abs/2026ApJ...998...86B}
}

@ARTICLE{bono1997b,
   author = {{Bono}, G. and {Caputo}, F. and {Castellani}, V. and {Marconi}, M.
	},
    title = "{Nonlinear investigation of the pulsational properties of RR Lyrae variables}",
  journal = {\aaps},
   eprint = {astro-ph/9606030},
     year = 1997,
    month = feb,
   volume = 121,
    pages = {327-342},
      doi = {10.1051/aas:1997289},
   adsurl = {http://adsabs.harvard.edu/abs/1997A%26AS..121..327B}
}

\begin{appendix}
\nolinenumbers
\section{Unified linear fit procedure} \label{sec:linear_fit}

We summarise here the approach taken to compute the linear correlations presented in the main text, following \citet{hogg2010}. We model every linear relation with $K$ independent variables as
\begin{equation} \label{eq:linear_fit}
y = \alpha + \boldsymbol{\beta} \cdot \mathbf{x}=\alpha + \sum^{K}_{k=1}{\beta_k x_k},
\end{equation} 
where $\boldsymbol{\beta}=\left(\beta_1,\ldots,\beta_k\right)$ is the vector of slopes and $\alpha$ is the intercept. For $K=1$ this is a straight line in a 2D space, and for $K=2$ a plane in a 3D space. The independent variable measurements $\mathbf{x}_i = (x_{1,i},\ldots,x_{K,i})$ and the dependent variable $y_i$ carry Gaussian uncertainties $(\sigma_{x_1,i},\ldots,\sigma_{x_K,i})$ and $\sigma_{y,i}$, assumed independent.
Instead of employing an ordinary least-squares algorithm, we take a Markov Chain Monte Carlo (MCMC) approach. This allows for an effective exploration of the parameter space of the model, providing the full posterior distribution of each of these parameters. To implement a likelihood function we, therefore, calculate the orthogonal residual between the line and each point $(\mathbf{x_i},y_i)$ as

\begin{equation} \label{eq:perp_distance}
d_{i}=\frac{\boldsymbol{\beta} \cdot \mathbf{x_i} + \alpha - y_i}{\sqrt{\lvert \beta \rvert^{2}+1}},
\end{equation} 

with the resulting projected error $\sigma_i$
\begin{equation} \label{eq:sigma_perp}
\sigma_{i}^{2}=\frac{\sum^{K}_{k=1}{\beta_{k}^{2} \sigma_{x,i}^{2}} + \sigma_{y,i}^{2}}{{\lvert \beta \rvert^{2}+1}}.
\end{equation} 

Including an additional term to take into account the possible intrinsic scatter of the relation, $\sigma_{int}$, and adding it in quadrature to the measurement contribution, the log-likelihood function of this fit is defined as
\begin{dmath} \label{eq:ll_sigmaint}
\ln\mathcal{L}_{i}\left(\boldsymbol{\beta},\alpha,\sigma_{int}\right)\propto
-\frac{1}{2}\left[\frac{d_{i}^2}{\sigma_{i}^2+\sigma_{int}^2}+\ln\left({\sigma_{i}^2+\sigma_{int}^2}\right)\right]
\end{dmath} 

To reduce sensitivity to possible outliers, we included a mixture model in the final log-likelihood, i.e, an inlier Gaussian with a variance $\sigma_{i}^2+\sigma_{int}^2$, and a wider, outlier Gaussian with variance $\sigma_{i}^2+\sigma_{out}^2$, each of them appropriately weighted by the fraction of outliers $f_{out}$. The log-likelihood, then, for each of these components is given, respectively, by

\begin{equation} \label{eq:ll_inliers}
\ln\mathcal{L}_{in,i}\left(\boldsymbol{\beta},\alpha,\sigma_{int},f_{out}\right)\propto
\ln\left(1-f_{out}\right)+\ln\mathcal{L}_{i}\left(\boldsymbol{\beta},\alpha,\sigma_{int}\right),
\end{equation} 
\begin{dmath} \label{eq:ll_outliers}
\ln\mathcal{L}_{out,i}\left(\boldsymbol{\beta},\alpha,\sigma_{out},f_{out}\right)\propto
\ln\left(f_{out}\right)+\ln\mathcal{L}_{i}\left(\boldsymbol{\beta},\alpha,\sigma_{out}\right),
\end{dmath} 

with the complete log-likelihood for the fit becoming
\begin{dmath} \label{eq:ll_total}
\ln\mathcal{L}_{total}\left(\boldsymbol{\beta},\alpha,\sigma_{int},\sigma_{out},f_{out}\right)=
\sum_{i}\ln\left(\mathcal{L}_{in,i}+\mathcal{L}_{out,i}\right)-\frac{1}{2}\sum_{i}\ln\left(1+\lvert \beta \rvert^2\right).
\end{dmath} 

For the priors, we adopted 
\begin{equation}
p\left(\boldsymbol{\beta},\alpha\right)\propto\left(1+\lvert\boldsymbol{\beta}\rvert^{2}\right)^{-(K+1)/2},
\end{equation}

for the fit parameters, equivalent to imposing a uniform prior on $\alpha$ and an isotropic prior on the orientation of the fitted model in the $(K+1)$-dimensional variable space. For $K=1$ (a line) this reduces to a uniform prior on $\theta$, with $\beta=\tan(\theta)$, while for $K=2$ (a plane) it corresponds to a uniform distribution over solid angle for the plane's unit normal vector, that is, an orientation that is spherically isotropic. We also adopted uniform priors in $\ln{\sigma}$ (with $\ln{\sigma_{int}}\in\left[-5,1\right]$ and $\ln{\sigma_{out}}\in\left[-0.5,1.5\right]$); and for $f_{out}$ we impose a Beta distribution $B(x,y)$ with $x=1$ and $y=30$ as a prior (with $f_{out}\in\left[0,0.3\right]$). Uniformity in $\ln{\sigma}$ corresponds to a Jeffreys prior for scale parameters, $p\left(\sigma\right)\propto1/\sigma$. 

To prevent the outlier component from collapsing onto the inlier component, the
prior is further multiplied by a smooth lower limit on the ratio of their widths,
\begin{equation}
p_{\mathrm{sep}}\left(\sigma_{\mathrm{out}},\sigma_{\mathrm{int}}\right)=
\left[1+\exp\left(-\frac{\ln\left(\sigma_{\mathrm{out}}/\sigma_{\mathrm{int}}\right)-\ln3}{s}\right)\right]^{-1},
\end{equation}
with $s=0.25$. This factor equals $1/2$ at $\sigma_{\mathrm{out}}=3\,\sigma_{\mathrm{int}}$, falls to $\simeq0.01$ for $\sigma_{\mathrm{out}}=\sigma_{\mathrm{int}}$, and tends to unity for $\sigma_{\mathrm{out}}\gtrsim5\,\sigma_{\mathrm{int}}$, so that it disfavours outlier widths comparable to the intrinsic scatter without restricting how broad the outlier component can be.

Every fit is performed on standardised variables,
\begin{equation}
v'=\frac{v-\overline{v}}{\sigma_v},\,
\Delta v'=\frac{\Delta v}{\sigma_v}
\end{equation}
with $\overline{v}$ being the mean value of the variable itself, $\sigma_v$ its standard deviation, and $\Delta v$ its uncertainty. This way, we improve numerical stability, reduce parameter correlations in the posterior, and make the behaviour of the priors more predictable; final posterior samples are transformed back to the original units after sampling.

To carry out the sampling, we employed the \texttt{emcee} Python package \citep{foreman-mackey2013}, based on the algorithm proposed by \cite{goodman2010}). Walkers are initialised near an ordinary least-squares (OLS) solution computed on the standardised data. The non-linear parameters are initialised as $\ln{\sigma_{int}}=-2$, $\ln{\sigma_{out}}=0.5$, $f_{out}=0.05$, with a small random perturbation per walker. The MCMC is run with 128 walkers for $10{,}000$ steps. The autocorrelation time $\tau$ is estimated for each parameter and the first $10\tau$ steps are discarded as burn-in; the remaining samples are used for parameter estimates and credible intervals.

This joint treatment offers a specific advantage over simpler alternatives. Ordinary or weighted least squares treats one variable as error free, which is inappropriate for relations where both axes carry comparable, often large, uncertainty. The classical bisector avoids this asymmetry but ignores individual measurement uncertainties entirely. More sophisticated alternatives such as orthogonal distance regression (ODR) and the BCES estimator \citep{akritas1996} incorporate errors on both axes, but neither jointly estimates an intrinsic scatter term alongside the slope and intercept, nor includes an outlier model. Our approach estimates $\sigma_{int}$ (Eq.~\ref{eq:ll_sigmaint}) and the outlier fraction (Eqs.~\ref{eq:ll_outliers}-\ref{eq:ll_total}) jointly with $(\alpha, \beta)$ within a single likelihood, capturing their full joint posterior.

\section{Tables of RRL sample properties} \label{sec:table_properties}

\begin{table*}[t]
\caption{Properties of the RR~Lyrae sample.}
\label{table:rrl_sample}
\centering
\small
\renewcommand{\arraystretch}{1.3}
{\setlength{\tabcolsep}{4pt}
\begin{tabular}{l c r c c c c c c c}
\hline\hline
Name & Type$^a$ & \multicolumn{1}{c}{Gaia ID} & RA & Dec.\@ & $\varpi$ & Period & $P_{\rm Gaia} - P$ & [Fe/H] & $E(B{-}V)^b$ \\
 & & & (h) & ($^{\circ}$) & (mas) & (d) & ($10^{-6}$\,d) & & (mag) \\
\hline
\hline
RU Scl & RRab$^*$ & 2336550174250087936 & 0:02:48.17 & $-24:56:43.36$ & $1.280$ $\pm$ 0.032 & $0.493356$ & $-3.8$ & $-1.45$ $\pm$ 0.06 & $0.022$ $\pm$ 0.004 \\
UU Cet & RRab & 2414817603803476992 & 0:04:05.16 & $-16:59:51.62$ & $0.533$ $\pm$ 0.019 & $0.606075$ & $+4.6$ & $-1.72$ $\pm$ 0.05 & $0.031$ $\pm$ 0.005 \\
SW And & RRab$^*$ & 2857456211775108608 & 0:23:43.08 & $+29:24:03.32$ & $1.995$ $\pm$ 0.028 & $0.442259$ & $+4.2$ & $-0.18$ $\pm$ 0.06 & $0.038$ $\pm$ 0.005 \\
RX Cet & RRab$^*$ & 2373827054405627904 & 0:33:38.25 & $-15:29:15.75$ & $0.765$ $\pm$ 0.030 & $0.573687$ & $+7.5$ & $-1.53$ $\pm$ 0.04 & $0.028$ $\pm$ 0.006 \\
RU Psc & RRc & 294072906063827072 & 1:14:26.15 & $+24:24:55.71$ & $1.278$ $\pm$ 0.029 & $0.390348$ & $-31.8$ & $-1.90$ & $0.057$ $\pm$ 0.007 \\
XX And & RRab & 370067649378653440 & 1:17:27.49 & $+38:57:01.52$ & $0.890$ $\pm$ 0.020 & $0.722759$ & \nodata & $-1.97$ $\pm$ 0.07 & $0.038$ $\pm$ 0.006 \\
RR Cet & RRab & 2558296724402139136 & 1:32:08.19 & $+1:20:29.52$ & $1.642$ $\pm$ 0.021 & $0.553030$ & $-3.0$ & $-1.53$ $\pm$ 0.06 & $0.019$ $\pm$ 0.004 \\
SV Scl & RRc & 5022411786734717952 & 1:44:59.66 & $-30:03:33.96$ & $0.704$ $\pm$ 0.027 & $0.377355$ & $+0.8$ & $-1.45$ & $0.025$ $\pm$ 0.006 \\
AO Peg & RRab & 1786827307055763968 & 21:27:03.47 & $+18:35:57.27$ & $0.396$ $\pm$ 0.027 & $0.547249$ & $-7.5$ & $-1.55$ $\pm$ 0.08 & $0.161^\dagger$ $\pm$ 0.057 \\
\hline
\multicolumn{10}{l}{The complete table will be available in the online version.}
\end{tabular}}
\tablefoot{When no error is available for a measurement, only the measurement itself appears in the corresponding column.}
\tablefoottext{a}{Blazhko variables are marked with an asterisk.}
\tablefoottext{b}{Stars whose $E(B{-}V)$ values are derived from the procedure described in Section~\ref{subsec:colour_min} are marked with a dagger.}
\end{table*}

\begin{table*}[th]
\caption{Mean magnitudes of the RR~Lyrae sample.}
\label{table:magnitude}
\centering
\resizebox{\textwidth}{!}{%
\small
\renewcommand{\arraystretch}{1.3}
{\setlength{\tabcolsep}{4pt}
\begin{tabular}{l c c c c c c c c c c c}
\hline\hline
Name & $U$ & $B$ & $V$ & $R$ & $I$ & $J$ & $H$ & $K_s$ & $G$ & $BP$ & $RP$ \\
\hline
\hline
RU Scl & $10.346$ $\pm$ 0.004 & $10.547$ $\pm$ 0.002 & $10.225$ $\pm$ 0.002 & $9.986$ $\pm$ 0.001 & $9.797$ $\pm$ 0.001 & $9.464$ $\pm$ 0.006 & $9.230$ $\pm$ 0.006 & $9.224$ $\pm$ 0.006 & $10.180$ $\pm$ 0.012 & $10.355$ $\pm$ 0.016 & $9.854$ $\pm$ 0.008 \\
UU Cet & $12.303$ $\pm$ 0.003 & $12.461$ $\pm$ 0.001 & $12.105$ $\pm$ 0.001 & $11.764$ $\pm$ 0.001 & $11.526$ $\pm$ 0.001 & $11.120$ $\pm$ 0.008 & $10.851$ $\pm$ 0.006 & $10.805$ $\pm$ 0.010 & $11.938$ $\pm$ 0.003 & $12.160$ $\pm$ 0.003 & $11.563$ $\pm$ 0.002 \\
SW And & $10.130$ $\pm$ 0.001 & $10.075$ $\pm$ 0.001 & $9.681$ $\pm$ 0.001 & $9.441$ $\pm$ 0.001 & $9.148$ $\pm$ 0.001 & $8.685$ $\pm$ 0.207 & $8.590$ $\pm$ 0.039 & \nodata & $9.588$ $\pm$ 0.003 & $9.809$ $\pm$ 0.004 & $9.213$ $\pm$ 0.002 \\
RX Cet & $11.707$ $\pm$ 0.009 & $11.840$ $\pm$ 0.008 & $11.396$ $\pm$ 0.005 & $11.186$ $\pm$ 0.004 & $10.885$ $\pm$ 0.003 & $10.507$ $\pm$ 0.007 & $10.271$ $\pm$ 0.005 & $10.253$ $\pm$ 0.008 & $11.328$ $\pm$ 0.009 & $11.539$ $\pm$ 0.011 & $10.964$ $\pm$ 0.007 \\
RU Psc & $10.831$ $\pm$ 0.003 & $10.383$ $\pm$ 0.001 & $10.159$ $\pm$ 0.001 & $9.944$ $\pm$ 0.001 & $9.739$ $\pm$ 0.001 & $9.361$ $\pm$ 0.005 & $9.151$ $\pm$ 0.004 & $9.124$ $\pm$ 0.007 & $10.079$ $\pm$ 0.007 & $10.249$ $\pm$ 0.009 & $9.767$ $\pm$ 0.008 \\
XX And & $10.937$ $\pm$ 0.002 & $11.044$ $\pm$ 0.001 & $10.681$ $\pm$ 0.001 & $10.440$ $\pm$ 0.001 & $10.166$ $\pm$ 0.001 & \nodata & \nodata & \nodata & $10.577$ $\pm$ 0.005 & $10.783$ $\pm$ 0.003 & $10.204$ $\pm$ 0.003 \\
RR Cet & $9.956$ $\pm$ 0.003 & $10.046$ $\pm$ 0.002 & $9.717$ $\pm$ 0.001 & $9.500$ $\pm$ 0.001 & $9.198$ $\pm$ 0.001 & $8.857$ $\pm$ 0.010 & $8.614$ $\pm$ 0.010 & $8.568$ $\pm$ 0.022 & $9.613$ $\pm$ 0.004 & $9.811$ $\pm$ 0.003 & $9.255$ $\pm$ 0.002 \\
SV Scl & \nodata & \nodata & \nodata & \nodata & \nodata & $10.730$ $\pm$ 0.003 & $10.547$ $\pm$ 0.003 & $10.527$ $\pm$ 0.004 & $11.304$ $\pm$ 0.002 & $11.435$ $\pm$ 0.002 & $11.049$ $\pm$ 0.001 \\
AO Peg & \nodata & \nodata & \nodata & \nodata & \nodata & $11.939$ $\pm$ 0.009 & $11.637$ $\pm$ 0.006 & $11.583$ $\pm$ 0.008 & $12.775$ $\pm$ 0.007 & $12.994$ $\pm$ 0.008 & $12.389$ $\pm$ 0.003 \\
\hline
\multicolumn{10}{l}{The complete table will be available in the online version.}
\end{tabular}}}
\end{table*}

\begin{table*}[t]
\caption{Luminosity amplitudes of the RR~Lyrae sample.}
\label{table:amplitude}
\centering
\resizebox{\textwidth}{!}{%
\small
\renewcommand{\arraystretch}{1.3}
{\setlength{\tabcolsep}{4pt}
\begin{tabular}{l c c c c c c c c c c c}
\hline\hline
Name & $U$ & $B$ & $V$ & $R$ & $I$ & $J$ & $H$ & $K_s$ & $G$ & $BP$ & $RP$ \\
\hline
\hline
RU Scl & $1.700$ $\pm$ 0.009 & $1.645$ $\pm$ 0.011 & $1.289$ $\pm$ 0.007 & $1.060$ $\pm$ 0.005 & $0.854$ $\pm$ 0.006 & $0.439$ $\pm$ 0.022 & $0.306$ $\pm$ 0.023 & $0.326$ $\pm$ 0.025 & $0.925$ $\pm$ 0.091 & $1.080$ $\pm$ 0.074 & $0.701$ $\pm$ 0.054 \\
UU Cet & $0.909$ $\pm$ 0.018 & $0.824$ $\pm$ 0.006 & $0.644$ $\pm$ 0.005 & $0.521$ $\pm$ 0.006 & $0.422$ $\pm$ 0.003 & $0.316$ $\pm$ 0.028 & $0.275$ $\pm$ 0.023 & $0.284$ $\pm$ 0.035 & $0.559$ $\pm$ 0.014 & $0.677$ $\pm$ 0.011 & $0.427$ $\pm$ 0.006 \\
SW And & $1.383$ $\pm$ 0.008 & $1.284$ $\pm$ 0.006 & $0.955$ $\pm$ 0.005 & $0.760$ $\pm$ 0.005 & $0.608$ $\pm$ 0.004 & $0.694$ $\pm$ 0.349 & $0.209$ $\pm$ 0.155 & \nodata & $0.808$ $\pm$ 0.015 & $1.003$ $\pm$ 0.028 & $0.600$ $\pm$ 0.009 \\
RX Cet & $1.210$ $\pm$ 0.046 & $1.018$ $\pm$ 0.057 & $0.872$ $\pm$ 0.029 & $0.733$ $\pm$ 0.026 & $0.577$ $\pm$ 0.021 & $0.281$ $\pm$ 0.024 & $0.236$ $\pm$ 0.017 & $0.263$ $\pm$ 0.026 & $0.670$ $\pm$ 0.027 & $0.785$ $\pm$ 0.043 & $0.503$ $\pm$ 0.025 \\
RU Psc & $0.621$ $\pm$ 0.015 & $0.582$ $\pm$ 0.005 & $0.450$ $\pm$ 0.004 & $0.355$ $\pm$ 0.003 & $0.284$ $\pm$ 0.002 & $0.183$ $\pm$ 0.014 & $0.095$ $\pm$ 0.011 & $0.131$ $\pm$ 0.017 & $0.377$ $\pm$ 0.015 & $0.470$ $\pm$ 0.016 & $0.275$ $\pm$ 0.012 \\
XX And & $1.242$ $\pm$ 0.006 & $1.262$ $\pm$ 0.004 & $0.995$ $\pm$ 0.003 & $0.798$ $\pm$ 0.004 & $0.642$ $\pm$ 0.003 & \nodata & \nodata & \nodata & $0.888$ $\pm$ 0.019 & $1.046$ $\pm$ 0.008 & $0.669$ $\pm$ 0.007 \\
RR Cet & $1.217$ $\pm$ 0.013 & $1.177$ $\pm$ 0.009 & $0.955$ $\pm$ 0.008 & $0.770$ $\pm$ 0.005 & $0.620$ $\pm$ 0.005 & $0.376$ $\pm$ 0.032 & $0.280$ $\pm$ 0.027 & $0.395$ $\pm$ 0.057 & $0.836$ $\pm$ 0.029 & $0.992$ $\pm$ 0.022 & $0.610$ $\pm$ 0.012 \\
SV Scl & \nodata & \nodata & \nodata & \nodata & \nodata & $0.204$ $\pm$ 0.010 & $0.133$ $\pm$ 0.008 & $0.123$ $\pm$ 0.010 & $0.452$ $\pm$ 0.008 & $0.540$ $\pm$ 0.008 & $0.336$ $\pm$ 0.005 \\
AO Peg & \nodata & \nodata & \nodata & \nodata & \nodata & $0.401$ $\pm$ 0.022 & $0.312$ $\pm$ 0.017 & $0.316$ $\pm$ 0.017 & $0.847$ $\pm$ 0.024 & $1.018$ $\pm$ 0.017 & $0.648$ $\pm$ 0.013 \\
\hline
\multicolumn{10}{l}{The complete table will be available in the online version.}
\end{tabular}}}
\end{table*}

\begin{table*}[t]
\caption{Fourier parameter $\phi_{31}$ of the RR~Lyrae sample.}
\label{table:phi31}
\centering
\resizebox{\textwidth}{!}{%
\small
\renewcommand{\arraystretch}{1.3}
{\setlength{\tabcolsep}{4pt}
\begin{tabular}{l c c c c c c c c c c c}
\hline\hline
Name & $U$ & $B$ & $V$ & $R$ & $I$ & $J$ & $H$ & $K_s$ & $G$ & $BP$ & $RP$ \\
\hline
\hline
RU Scl & $4.637$ $\pm$ 0.040 & $4.782$ $\pm$ 0.031 & $4.925$ $\pm$ 0.031 & $5.089$ $\pm$ 0.027 & $5.413$ $\pm$ 0.037 & $6.270$ $\pm$ 0.302 & $7.744$ $\pm$ 0.357 & $7.631$ $\pm$ 0.560 & $5.049$ $\pm$ 0.129 & $4.828$ $\pm$ 0.150 & $5.427$ $\pm$ 0.110 \\
UU Cet & $5.182$ $\pm$ 0.095 & $4.864$ $\pm$ 0.023 & $5.227$ $\pm$ 0.023 & $5.528$ $\pm$ 0.025 & $6.053$ $\pm$ 0.026 & $7.861$ $\pm$ 0.263 & $8.500$ $\pm$ 0.376 & $8.816$ $\pm$ 0.512 & $5.501$ $\pm$ 0.067 & $5.301$ $\pm$ 0.066 & $6.032$ $\pm$ 0.054 \\
SW And & $5.245$ $\pm$ 0.017 & $5.272$ $\pm$ 0.015 & $5.532$ $\pm$ 0.017 & $5.732$ $\pm$ 0.020 & $6.046$ $\pm$ 0.018 & $6.638$ $\pm$ 1.130 & $6.671$ $\pm$ 1.396 & \nodata & $5.582$ $\pm$ 0.052 & $5.402$ $\pm$ 0.056 & $6.049$ $\pm$ 0.048 \\
RX Cet & $4.804$ $\pm$ 0.082 & $5.003$ $\pm$ 0.092 & $5.157$ $\pm$ 0.065 & $5.313$ $\pm$ 0.052 & $5.677$ $\pm$ 0.058 & $6.809$ $\pm$ 0.486 & $8.270$ $\pm$ 0.482 & $9.386$ $\pm$ 0.840 & $5.680$ $\pm$ 0.306 & $5.220$ $\pm$ 0.350 & $5.901$ $\pm$ 0.271 \\
RU Psc & $4.378$ $\pm$ 0.122 & $4.077$ $\pm$ 0.049 & $3.944$ $\pm$ 0.051 & $3.906$ $\pm$ 0.062 & $4.131$ $\pm$ 0.075 & $4.975$ $\pm$ 0.888 & $5.741$ $\pm$ 2.103 & $2.115$ $\pm$ 0.839 & $4.093$ $\pm$ 0.689 & $3.591$ $\pm$ 1.103 & $4.123$ $\pm$ 0.783 \\
XX And & $5.123$ $\pm$ 0.020 & $5.075$ $\pm$ 0.017 & $5.346$ $\pm$ 0.016 & $5.532$ $\pm$ 0.019 & $5.895$ $\pm$ 0.023 & \nodata & \nodata & \nodata & $5.466$ $\pm$ 0.112 & $5.307$ $\pm$ 0.106 & $5.887$ $\pm$ 0.120 \\
RR Cet & $4.895$ $\pm$ 0.035 & $4.863$ $\pm$ 0.030 & $5.088$ $\pm$ 0.023 & $5.300$ $\pm$ 0.019 & $5.693$ $\pm$ 0.023 & $7.327$ $\pm$ 0.461 & $8.297$ $\pm$ 0.581 & $8.925$ $\pm$ 0.555 & $5.223$ $\pm$ 0.060 & $5.046$ $\pm$ 0.057 & $5.733$ $\pm$ 0.057 \\
SV Scl & \nodata & \nodata & \nodata & \nodata & \nodata & $4.251$ $\pm$ 0.672 & $5.781$ $\pm$ 1.077 & $4.743$ $\pm$ 0.564 & $3.325$ $\pm$ 0.173 & $3.487$ $\pm$ 0.149 & $3.473$ $\pm$ 0.174 \\
AO Peg & \nodata & \nodata & \nodata & \nodata & \nodata & $6.628$ $\pm$ 0.390 & $7.728$ $\pm$ 0.396 & $7.406$ $\pm$ 0.566 & $5.154$ $\pm$ 0.102 & $4.968$ $\pm$ 0.114 & $5.551$ $\pm$ 0.084 \\
\hline
\multicolumn{10}{l}{The complete table will be available in the online version.}
\end{tabular}}}
\end{table*}

\section{Tables of results} \label{sec:table_results}

\begin{table*}[tbp]
\caption{Relation between Fourier parameter $\phi_{31}$ in different filters and $V$ filter.}
\label{table:phi31_relation}
\centering
\renewcommand{\arraystretch}{1.5}
\begin{tabular}{l ccccc ccccc}
\hline\hline
Filter & \multicolumn{5}{c}{RRab} & \multicolumn{5}{c}{RRc} \\
\cmidrule(lr){2-6} \cmidrule(lr){7-11}
 & $\phi_{31,0}^{s,V}$ & $\phi_{31,0}^{s,\lambda}$ & $\beta$ & $\alpha$ & $\sigma$ & $\phi_{31,0}^{c,V}$ & $\phi_{31,0}^{c,\lambda}$ & $\beta$ & $\alpha$ & $\sigma$\\
\hline
$U$ & $5.202$ & $4.999$ & $1.113^{+0.048}_{-0.051}$ & $0.004^{+0.010}_{-0.010}$ & $0.035^{+0.009}_{-0.008}$ & $3.334$ & $3.232$ & $0.801^{+0.089}_{-0.091}$ & $-0.154^{+0.080}_{-0.072}$ & $0.138^{+0.084}_{-0.061}$ \\
$B$ & $5.202$ & $4.997$ & $1.167^{+0.044}_{-0.043}$ & $-0.003^{+0.010}_{-0.009}$ & $0.030^{+0.010}_{-0.010}$ & $3.334$ & $3.411$ & $0.946^{+0.062}_{-0.055}$ & $0.022^{+0.044}_{-0.044}$ & $0.075^{+0.047}_{-0.033}$ \\
$R$ & $5.202$ & $5.428$ & $0.873^{+0.022}_{-0.023}$ & $0.001^{+0.007}_{-0.007}$ & $0.024^{+0.006}_{-0.006}$ & $3.334$ & $3.461$ & $0.993^{+0.056}_{-0.063}$ & $0.092^{+0.042}_{-0.042}$ & $0.056^{+0.034}_{-0.026}$ \\
$I$ & $5.202$ & $5.767$ & $0.735^{+0.031}_{-0.032}$ & $-0.000^{+0.011}_{-0.011}$ & $0.060^{+0.008}_{-0.007}$ & $3.334$ & $3.570$ & $0.956^{+0.057}_{-0.064}$ & $0.052^{+0.045}_{-0.045}$ & $0.061^{+0.044}_{-0.030}$ \\
$J$ & $5.155$ & $6.614$ & $0.362^{+0.080}_{-0.075}$ & $-0.010^{+0.040}_{-0.039}$ & $0.145^{+0.043}_{-0.049}$ & $3.467$ & $4.124$ & \nodata & \nodata & \nodata \\
$H$ & $5.171$ & $7.931$ & $0.240^{+0.089}_{-0.080}$ & $0.002^{+0.048}_{-0.047}$ & $0.200^{+0.052}_{-0.067}$ & $3.467$ & $3.560$ & \nodata & \nodata & \nodata \\
$K_s$ & $5.168$ & $7.791$ & $0.194^{+0.096}_{-0.081}$ & $0.002^{+0.052}_{-0.053}$ & $0.221^{+0.053}_{-0.054}$ & $3.322$ & $3.647$ & \nodata & \nodata & \nodata \\
$G$ & $5.212$ & $5.376$ & $0.884^{+0.053}_{-0.054}$ & $0.028^{+0.014}_{-0.015}$ & $0.042^{+0.014}_{-0.013}$ & $3.204$ & $3.248$ & $0.77^{+0.13}_{-0.11}$ & $-0.037^{+0.095}_{-0.089}$ & $0.100^{+0.086}_{-0.054}$ \\
$BP$ & $5.208$ & $5.176$ & $1.012^{+0.057}_{-0.058}$ & $0.031^{+0.014}_{-0.014}$ & $0.027^{+0.013}_{-0.012}$ & $3.204$ & $3.237$ & $0.68^{+0.11}_{-0.10}$ & $-0.069^{+0.087}_{-0.081}$ & $0.093^{+0.085}_{-0.050}$ \\
$RP$ & $5.191$ & $5.711$ & $0.681^{+0.049}_{-0.049}$ & $0.003^{+0.018}_{-0.018}$ & $0.077^{+0.015}_{-0.013}$ & $3.204$ & $3.377$ & $0.68^{+0.12}_{-0.10}$ & $-0.020^{+0.095}_{-0.093}$ & $0.094^{+0.088}_{-0.052}$ \\
\hline
\end{tabular}
\tablefoot{Results on the table are presented following the equation $\phi_{31}^{V}-\phi_{31,0}^{V}=\alpha+\beta \left(\phi_{31}^{\lambda}-\phi_{31,0}^{\lambda} \right)$. Pivots for the $y$ variable ($\phi_{31,0}^{V}$) vary among fits because, even if the $y$ variable is always the same, the number of stars in common between the two filters varies with the pair of filters selected.

The superindex $^s$ or $^c$ indicate whether $\phi_{31}$ was obtained with the sinus form or the cosinus form of the Fourier series.}
\end{table*}

\begin{table*}[t]
\caption{Amplitude ratio as a function of period relations.}
\label{table:ampl_ratio_relation}
\centering
\renewcommand{\arraystretch}{1.5}
\begin{tabular}{l l ccccc cc}
\hline\hline
Reference filter & Filter & \multicolumn{5}{c}{RRab} & \multicolumn{2}{c}{RRc} \\
\cmidrule(lr){3-7} \cmidrule(lr){8-9}
 & & $r_{0}^{\lambda}$ & $L_0$ & $\beta$ & $\alpha$ & $\sigma$ & $\overline{r^{\lambda}}$ & $\sigma$ \\
\hline
\multirow{10}{*}{$V$}
& $U$ & $1.341$ & $-0.277$ & $-0.67^{+0.12}_{-0.12}$ & $0.004^{+0.010}_{-0.010}$ & $0.055^{+0.009}_{-0.008}$ & $1.213$ & $0.051$ \\
& $B$ & $1.307$ & $-0.277$ & $-0.177^{+0.050}_{-0.050}$ & $-0.002^{+0.004}_{-0.004}$ & $0.023^{+0.004}_{-0.004}$ & $1.282$ & $0.022$ \\
& $R$ & $0.807$ & $-0.277$ & $0.038^{+0.022}_{-0.021}$ & $-0.001^{+0.002}_{-0.002}$ & $0.007^{+0.002}_{-0.002}$ & $0.801$ & $0.009$ \\
& $I$ & $0.650$ & $-0.277$ & $0.039^{+0.020}_{-0.019}$ & $-0.001^{+0.002}_{-0.002}$ & $0.007^{+0.002}_{-0.002}$ & $0.622$ & $0.013$ \\
& $J$ & $0.415$ & $-0.280$ & $0.329^{+0.071}_{-0.071}$ & $-0.004^{+0.005}_{-0.005}$ & $0.020^{+0.006}_{-0.006}$ & $0.373$ & $0.020$ \\
& $H$ & $0.300$ & $-0.280$ & $0.453^{+0.075}_{-0.073}$ & $-0.004^{+0.005}_{-0.005}$ & $0.022^{+0.005}_{-0.005}$ & $0.224$ & $0.032$ \\
& $K_s$ & $0.315$ & $-0.279$ & $0.41^{+0.10}_{-0.10}$ & $-0.007^{+0.008}_{-0.008}$ & $0.031^{+0.007}_{-0.006}$ & $0.243$ & $0.036$ \\
& $G$ & $0.866$ & $-0.277$ & $0.084^{+0.057}_{-0.059}$ & $0.005^{+0.005}_{-0.005}$ & $0.019^{+0.005}_{-0.005}$ & $0.866$ & $0.032$ \\
& $BP$ & $1.048$ & $-0.277$ & $-0.056^{+0.063}_{-0.062}$ & $0.010^{+0.005}_{-0.006}$ & $0.019^{+0.007}_{-0.008}$ & $1.025$ & $0.032$ \\
& $RP$ & $0.665$ & $-0.277$ & $0.094^{+0.043}_{-0.045}$ & $0.000^{+0.004}_{-0.004}$ & $0.017^{+0.005}_{-0.005}$ & $0.624$ & $0.028$ \\
\hline
\multirow{10}{*}{$G$}
& $U$ & $1.557$ & $-0.277$ & $-0.83^{+0.19}_{-0.18}$ & $0.000^{+0.015}_{-0.015}$ & $0.069^{+0.015}_{-0.012}$ & $1.405$ & $0.072$ \\
& $B$ & $1.518$ & $-0.277$ & $-0.34^{+0.11}_{-0.11}$ & $-0.015^{+0.008}_{-0.008}$ & $0.036^{+0.010}_{-0.010}$ & $1.462$ & $0.054$ \\
& $V$ & $1.162$ & $-0.277$ & $-0.105^{+0.078}_{-0.076}$ & $-0.016^{+0.006}_{-0.006}$ & $0.025^{+0.007}_{-0.006}$ & $1.151$ & $0.043$ \\
& $R$ & $0.938$ & $-0.277$ & $-0.020^{+0.074}_{-0.073}$ & $-0.013^{+0.006}_{-0.006}$ & $0.029^{+0.006}_{-0.006}$ & $0.918$ & $0.036$ \\
& $I$ & $0.755$ & $-0.277$ & $-0.010^{+0.057}_{-0.054}$ & $-0.010^{+0.004}_{-0.004}$ & $0.018^{+0.005}_{-0.004}$ & $0.710$ & $0.024$ \\
& $J$ & $0.502$ & $-0.277$ & $0.371^{+0.064}_{-0.063}$ & $-0.013^{+0.005}_{-0.005}$ & $0.032^{+0.005}_{-0.005}$ & $0.427$ & $0.029$ \\
& $H$ & $0.367$ & $-0.277$ & $0.521^{+0.069}_{-0.068}$ & $-0.007^{+0.006}_{-0.006}$ & $0.039^{+0.005}_{-0.004}$ & $0.271$ & $0.039$ \\
& $K_s$ & $0.369$ & $-0.275$ & $0.404^{+0.072}_{-0.070}$ & $-0.013^{+0.006}_{-0.006}$ & $0.032^{+0.005}_{-0.005}$ & $0.282$ & $0.044$ \\
& $BP$ & $1.217$ & $-0.276$ & $-0.119^{+0.051}_{-0.051}$ & $-0.006^{+0.004}_{-0.004}$ & $0.006^{+0.004}_{-0.003}$ & $1.192$ & $0.016$ \\
& $RP$ & $0.774$ & $-0.276$ & $0.042^{+0.032}_{-0.032}$ & $-0.011^{+0.003}_{-0.003}$ & $0.004^{+0.003}_{-0.002}$ & $0.733$ & $0.014$ \\
\midrule
\hline
\end{tabular}
\tablefoot{Results on the table are presented following the equation $r^{\lambda}-r_{0}^{\lambda}=\alpha+\beta \left( \log_{10}(P)-L_0 \right)$, where $r^{\lambda}$ is the ratio between the filter in the second column and the reference filter in the first column, $r^{\lambda}_0=\overline{r^{\lambda}}$, and $L_0=\overline{\log_{10}(P)}$. Pivots for the $x$ variable ($L_0$) varies among fits because, even if the $x$ variable is always the same, the number of stars in common between the two filters varies with the pair of filters selected.}
\end{table*}

\begin{table*}[t]
\caption{Colour at minimum relations as a function of pulsation period and metallicity.}
\label{table:colour_min_relation}
\centering
\resizebox{\textwidth}{!}{%
\renewcommand{\arraystretch}{1.35}
\begin{tabular}{c cccccccc ccccc}
\hline\hline
Colour & \multicolumn{8}{c}{RRab} & \multicolumn{5}{c}{RRc} \\
\cmidrule(lr){2-9} \cmidrule(lr){10-14}
 & $C^{min}_0$ & $L_0$ & $\feh_0$ & $\alpha$ & $\beta$ & $\gamma$ & $\sigma_{\rm int}$ & $\Delta$BIC & $C^{min}_0$ & $L_0$ & $\alpha$ & $\beta$ & $\sigma$ \\
\hline
\multirow{2}{*}{$U$$-$$B$} & \multirow{2}{*}{$-0.061$} & \multirow{2}{*}{$-0.279$} & \multirow{2}{*}{$-1.291$} & $-0.002^{+0.010}_{-0.010}$ & $0.41^{+0.18}_{-0.18}$ & $0.124^{+0.019}_{-0.020}$ & $0.060^{+0.010}_{-0.009}$ & \multirow{2}{*}{$28.6$} & \multirow{2}{*}{$-0.079$} & \multirow{2}{*}{$-0.489$} & \multirow{2}{*}{$-0.021^{+0.022}_{-0.021}$} & \multirow{2}{*}{$-0.03^{+0.40}_{-0.40}$} & \multirow{2}{*}{$0.068^{+0.018}_{-0.014}$} \\
 &  &  &  & $-0.000^{+0.012}_{-0.012}$ & $-0.37^{+0.16}_{-0.16}$ & \nodata & $0.087^{+0.011}_{-0.010}$ &  &  &  &  &  &  \\
\hline
\multirow{2}{*}{$U$$-$$V$} & \multirow{2}{*}{$0.361$} & \multirow{2}{*}{$-0.279$} & \multirow{2}{*}{$-1.291$} & $-0.004^{+0.010}_{-0.010}$ & $0.69^{+0.18}_{-0.18}$ & $0.178^{+0.020}_{-0.020}$ & $0.054^{+0.011}_{-0.009}$ & \multirow{2}{*}{$50.5$} & \multirow{2}{*}{$0.170$} & \multirow{2}{*}{$-0.489$} & \multirow{2}{*}{$-0.026^{+0.018}_{-0.017}$} & \multirow{2}{*}{$0.33^{+0.33}_{-0.34}$} & \multirow{2}{*}{$0.053^{+0.016}_{-0.012}$} \\
 &  &  &  & $-0.001^{+0.015}_{-0.015}$ & $-0.44^{+0.19}_{-0.19}$ & \nodata & $0.105^{+0.013}_{-0.012}$ &  &  &  &  &  &  \\
\hline
\multirow{2}{*}{$U$$-$$R$} & \multirow{2}{*}{$0.658$} & \multirow{2}{*}{$-0.279$} & \multirow{2}{*}{$-1.291$} & $-0.004^{+0.010}_{-0.010}$ & $0.81^{+0.19}_{-0.19}$ & $0.186^{+0.021}_{-0.021}$ & $0.054^{+0.010}_{-0.008}$ & \multirow{2}{*}{$51.8$} & \multirow{2}{*}{$0.362$} & \multirow{2}{*}{$-0.489$} & \multirow{2}{*}{$-0.023^{+0.017}_{-0.017}$} & \multirow{2}{*}{$0.54^{+0.32}_{-0.33}$} & \multirow{2}{*}{$0.049^{+0.016}_{-0.012}$} \\
 &  &  &  & $-0.001^{+0.015}_{-0.015}$ & $-0.39^{+0.19}_{-0.19}$ & \nodata & $0.111^{+0.014}_{-0.012}$ &  &  &  &  &  &  \\
\hline
\multirow{2}{*}{$U$$-$$I$} & \multirow{2}{*}{$0.942$} & \multirow{2}{*}{$-0.279$} & \multirow{2}{*}{$-1.291$} & $-0.001^{+0.011}_{-0.011}$ & $1.02^{+0.21}_{-0.21}$ & $0.191^{+0.023}_{-0.023}$ & $0.057^{+0.011}_{-0.009}$ & \multirow{2}{*}{$47.2$} & \multirow{2}{*}{$0.573$} & \multirow{2}{*}{$-0.489$} & \multirow{2}{*}{$-0.019^{+0.017}_{-0.016}$} & \multirow{2}{*}{$0.71^{+0.32}_{-0.32}$} & \multirow{2}{*}{$0.044^{+0.015}_{-0.011}$} \\
 &  &  &  & $-0.000^{+0.016}_{-0.016}$ & $-0.22^{+0.20}_{-0.20}$ & \nodata & $0.125^{+0.013}_{-0.012}$ &  &  &  &  &  &  \\
\hline
\multirow{2}{*}{$U$$-$$J$} & \multirow{2}{*}{$1.296$} & \multirow{2}{*}{$-0.278$} & \multirow{2}{*}{$-1.368$} & $-0.000^{+0.018}_{-0.018}$ & $1.42^{+0.33}_{-0.33}$ & $0.210^{+0.036}_{-0.037}$ & $0.059^{+0.016}_{-0.012}$ & \multirow{2}{*}{$23.4$} & \multirow{2}{*}{$0.888$} & \multirow{2}{*}{$-0.493$} & \multirow{2}{*}{$-0.024^{+0.049}_{-0.041}$} & \multirow{2}{*}{$0.47^{+0.87}_{-0.78}$} & \multirow{2}{*}{$0.082^{+0.044}_{-0.026}$} \\
 &  &  &  & $0.000^{+0.025}_{-0.024}$ & $0.18^{+0.32}_{-0.33}$ & \nodata & $0.144^{+0.021}_{-0.020}$ &  &  &  &  &  &  \\
\hline
\multirow{2}{*}{$U$$-$$H$} & \multirow{2}{*}{$1.576$} & \multirow{2}{*}{$-0.278$} & \multirow{2}{*}{$-1.368$} & $0.002^{+0.019}_{-0.019}$ & $1.51^{+0.34}_{-0.34}$ & $0.200^{+0.037}_{-0.038}$ & $0.059^{+0.016}_{-0.012}$ & \multirow{2}{*}{$20.1$} & \multirow{2}{*}{$1.081$} & \multirow{2}{*}{$-0.493$} & \multirow{2}{*}{$-0.030^{+0.045}_{-0.036}$} & \multirow{2}{*}{$0.77^{+0.81}_{-0.71}$} & \multirow{2}{*}{$0.066^{+0.038}_{-0.023}$} \\
 &  &  &  & $0.002^{+0.025}_{-0.025}$ & $0.32^{+0.32}_{-0.32}$ & \nodata & $0.141^{+0.023}_{-0.021}$ &  &  &  &  &  &  \\
\hline
\multirow{2}{*}{$U$$-$$K_s$} & \multirow{2}{*}{$1.616$} & \multirow{2}{*}{$-0.277$} & \multirow{2}{*}{$-1.394$} & $-0.003^{+0.020}_{-0.020}$ & $1.53^{+0.38}_{-0.38}$ & $0.196^{+0.044}_{-0.044}$ & $0.062^{+0.018}_{-0.013}$ & \multirow{2}{*}{$15.4$} & \multirow{2}{*}{$1.099$} & \multirow{2}{*}{$-0.498$} & \multirow{2}{*}{$-0.019^{+0.052}_{-0.046}$} & \multirow{2}{*}{$1.05^{+0.93}_{-0.89}$} & \multirow{2}{*}{$0.074^{+0.044}_{-0.026}$} \\
 &  &  &  & $0.001^{+0.025}_{-0.025}$ & $0.35^{+0.32}_{-0.32}$ & \nodata & $0.138^{+0.022}_{-0.021}$ &  &  &  &  &  &  \\
\hline
\multirow{2}{*}{$B$$-$$V$} & \multirow{2}{*}{$0.422$} & \multirow{2}{*}{$-0.279$} & \multirow{2}{*}{$-1.291$} & $-0.004^{+0.004}_{-0.004}$ & $0.293^{+0.081}_{-0.081}$ & $0.049^{+0.009}_{-0.009}$ & $0.019^{+0.004}_{-0.004}$ & \multirow{2}{*}{$24.2$} & \multirow{2}{*}{$0.249$} & \multirow{2}{*}{$-0.489$} & \multirow{2}{*}{$-0.015^{+0.014}_{-0.013}$} & \multirow{2}{*}{$0.26^{+0.22}_{-0.23}$} & \multirow{2}{*}{$0.034^{+0.014}_{-0.013}$} \\
 &  &  &  & $-0.002^{+0.005}_{-0.005}$ & $-0.031^{+0.065}_{-0.064}$ & \nodata & $0.028^{+0.005}_{-0.005}$ &  &  &  &  &  &  \\
\hline
\multirow{2}{*}{$B$$-$$R$} & \multirow{2}{*}{$0.719$} & \multirow{2}{*}{$-0.279$} & \multirow{2}{*}{$-1.291$} & $-0.004^{+0.005}_{-0.005}$ & $0.368^{+0.094}_{-0.093}$ & $0.053^{+0.011}_{-0.010}$ & $0.025^{+0.005}_{-0.004}$ & \multirow{2}{*}{$22.5$} & \multirow{2}{*}{$0.441$} & \multirow{2}{*}{$-0.489$} & \multirow{2}{*}{$-0.010^{+0.014}_{-0.013}$} & \multirow{2}{*}{$0.45^{+0.24}_{-0.25}$} & \multirow{2}{*}{$0.036^{+0.014}_{-0.013}$} \\
 &  &  &  & $-0.003^{+0.006}_{-0.006}$ & $0.024^{+0.071}_{-0.072}$ & \nodata & $0.035^{+0.005}_{-0.005}$ &  &  &  &  &  &  \\
\hline
\multirow{2}{*}{$B$$-$$I$} & \multirow{2}{*}{$1.003$} & \multirow{2}{*}{$-0.279$} & \multirow{2}{*}{$-1.291$} & $-0.002^{+0.005}_{-0.005}$ & $0.610^{+0.091}_{-0.093}$ & $0.063^{+0.010}_{-0.010}$ & $0.024^{+0.004}_{-0.004}$ & \multirow{2}{*}{$31.2$} & \multirow{2}{*}{$0.652$} & \multirow{2}{*}{$-0.489$} & \multirow{2}{*}{$-0.008^{+0.016}_{-0.016}$} & \multirow{2}{*}{$0.52^{+0.31}_{-0.30}$} & \multirow{2}{*}{$0.051^{+0.015}_{-0.014}$} \\
 &  &  &  & $-0.001^{+0.006}_{-0.006}$ & $0.196^{+0.075}_{-0.077}$ & \nodata & $0.039^{+0.006}_{-0.005}$ &  &  &  &  &  &  \\
\hline
\multirow{2}{*}{$B$$-$$J$} & \multirow{2}{*}{$1.372$} & \multirow{2}{*}{$-0.278$} & \multirow{2}{*}{$-1.368$} & $0.003^{+0.008}_{-0.008}$ & $0.94^{+0.14}_{-0.14}$ & $0.073^{+0.016}_{-0.016}$ & $0.027^{+0.006}_{-0.005}$ & \multirow{2}{*}{$14.9$} & \multirow{2}{*}{$0.896$} & \multirow{2}{*}{$-0.493$} & \multirow{2}{*}{$0.005^{+0.016}_{-0.017}$} & \multirow{2}{*}{$-0.00^{+0.27}_{-0.28}$} & \multirow{2}{*}{$0.026^{+0.016}_{-0.012}$} \\
 &  &  &  & $0.004^{+0.010}_{-0.009}$ & $0.50^{+0.12}_{-0.12}$ & \nodata & $0.043^{+0.009}_{-0.008}$ &  &  &  &  &  &  \\
\hline
\multirow{2}{*}{$B$$-$$H$} & \multirow{2}{*}{$1.652$} & \multirow{2}{*}{$-0.278$} & \multirow{2}{*}{$-1.368$} & $0.005^{+0.008}_{-0.007}$ & $1.08^{+0.13}_{-0.14}$ & $0.066^{+0.015}_{-0.016}$ & $0.023^{+0.006}_{-0.004}$ & \multirow{2}{*}{$13.9$} & \multirow{2}{*}{$1.088$} & \multirow{2}{*}{$-0.493$} & \multirow{2}{*}{$0.002^{+0.018}_{-0.018}$} & \multirow{2}{*}{$0.29^{+0.31}_{-0.31}$} & \multirow{2}{*}{$0.031^{+0.018}_{-0.013}$} \\
 &  &  &  & $0.005^{+0.009}_{-0.009}$ & $0.69^{+0.12}_{-0.12}$ & \nodata & $0.037^{+0.008}_{-0.007}$ &  &  &  &  &  &  \\
\hline
\multirow{2}{*}{$B$$-$$K_s$} & \multirow{2}{*}{$1.690$} & \multirow{2}{*}{$-0.277$} & \multirow{2}{*}{$-1.394$} & $0.003^{+0.010}_{-0.009}$ & $1.13^{+0.17}_{-0.18}$ & $0.072^{+0.021}_{-0.021}$ & $0.030^{+0.007}_{-0.006}$ & \multirow{2}{*}{$8.2$} & \multirow{2}{*}{$1.102$} & \multirow{2}{*}{$-0.498$} & \multirow{2}{*}{$0.008^{+0.018}_{-0.019}$} & \multirow{2}{*}{$0.18^{+0.31}_{-0.30}$} & \multirow{2}{*}{$0.030^{+0.018}_{-0.012}$} \\
 &  &  &  & $0.005^{+0.011}_{-0.011}$ & $0.70^{+0.14}_{-0.14}$ & \nodata & $0.045^{+0.009}_{-0.008}$ &  &  &  &  &  &  \\
\hline
$V$$-$$R$ & $0.297$ & $-0.279$ & \nodata & $-0.002^{+0.004}_{-0.004}$ & $0.062^{+0.052}_{-0.052}$ & \nodata & $0.023^{+0.004}_{-0.004}$ & $-3.5$ & $0.192$ & $-0.489$ & $0.005^{+0.007}_{-0.007}$ & $0.20^{+0.13}_{-0.13}$ & $0.009^{+0.005}_{-0.003}$ \\
\hline
$V$$-$$I$ & $0.581$ & $-0.279$ & \nodata & $-0.000^{+0.004}_{-0.004}$ & $0.237^{+0.055}_{-0.054}$ & \nodata & $0.026^{+0.004}_{-0.003}$ & $-2.1$ & $0.403$ & $-0.489$ & $0.002^{+0.010}_{-0.010}$ & $0.26^{+0.19}_{-0.19}$ & $0.032^{+0.009}_{-0.008}$ \\
\hline
$V$$-$$J$ & $0.953$ & $-0.278$ & \nodata & $0.004^{+0.007}_{-0.007}$ & $0.567^{+0.099}_{-0.100}$ & \nodata & $0.034^{+0.006}_{-0.005}$ & $1.7$ & $0.666$ & $-0.493$ & $0.006^{+0.014}_{-0.014}$ & $0.03^{+0.23}_{-0.23}$ & $0.026^{+0.013}_{-0.011}$ \\
\hline
$V$$-$$H$ & $1.233$ & $-0.278$ & \nodata & $0.005^{+0.007}_{-0.007}$ & $0.753^{+0.095}_{-0.098}$ & \nodata & $0.029^{+0.006}_{-0.005}$ & $-0.6$ & $0.859$ & $-0.493$ & $0.004^{+0.014}_{-0.014}$ & $0.34^{+0.24}_{-0.24}$ & $0.023^{+0.014}_{-0.009}$ \\
\hline
$V$$-$$K_s$ & $1.271$ & $-0.277$ & \nodata & $0.004^{+0.009}_{-0.009}$ & $0.77^{+0.11}_{-0.12}$ & \nodata & $0.036^{+0.007}_{-0.006}$ & $-0.2$ & $0.874$ & $-0.498$ & $0.006^{+0.015}_{-0.016}$ & $0.24^{+0.26}_{-0.25}$ & $0.026^{+0.014}_{-0.010}$ \\
\hline
$R$$-$$I$ & $0.284$ & $-0.279$ & \nodata & $0.002^{+0.003}_{-0.003}$ & $0.181^{+0.042}_{-0.042}$ & \nodata & $0.018^{+0.003}_{-0.003}$ & $-2.4$ & $0.211$ & $-0.489$ & $-0.001^{+0.008}_{-0.008}$ & $0.07^{+0.14}_{-0.14}$ & $0.025^{+0.007}_{-0.007}$ \\
\hline
$R$$-$$J$ & $0.656$ & $-0.278$ & \nodata & $0.004^{+0.007}_{-0.006}$ & $0.557^{+0.086}_{-0.087}$ & \nodata & $0.029^{+0.006}_{-0.005}$ & $-2.9$ & $0.477$ & $-0.493$ & $0.002^{+0.010}_{-0.010}$ & $-0.10^{+0.17}_{-0.17}$ & $0.014^{+0.010}_{-0.006}$ \\
\hline
$R$$-$$H$ & $0.936$ & $-0.278$ & \nodata & $0.005^{+0.006}_{-0.006}$ & $0.746^{+0.085}_{-0.087}$ & \nodata & $0.026^{+0.005}_{-0.005}$ & $-3.7$ & $0.670$ & $-0.493$ & $-0.002^{+0.010}_{-0.010}$ & $0.18^{+0.17}_{-0.17}$ & $0.012^{+0.009}_{-0.005}$ \\
\hline
$R$$-$$K_s$ & $0.974$ & $-0.277$ & \nodata & $0.002^{+0.008}_{-0.008}$ & $0.77^{+0.10}_{-0.10}$ & \nodata & $0.032^{+0.006}_{-0.005}$ & $-3.4$ & $0.684$ & $-0.498$ & $0.005^{+0.011}_{-0.011}$ & $0.05^{+0.19}_{-0.19}$ & $0.017^{+0.011}_{-0.007}$ \\
\hline
$I$$-$$J$ & $0.373$ & $-0.278$ & \nodata & $0.003^{+0.006}_{-0.006}$ & $0.252^{+0.074}_{-0.075}$ & \nodata & $0.028^{+0.006}_{-0.006}$ & $-2.7$ & $0.285$ & $-0.493$ & $0.002^{+0.014}_{-0.014}$ & $-0.11^{+0.23}_{-0.23}$ & $0.034^{+0.011}_{-0.009}$ \\
\hline
$I$$-$$H$ & $0.653$ & $-0.278$ & \nodata & $0.002^{+0.006}_{-0.006}$ & $0.439^{+0.074}_{-0.076}$ & \nodata & $0.027^{+0.005}_{-0.005}$ & $-3.9$ & $0.478$ & $-0.493$ & $0.001^{+0.012}_{-0.012}$ & $0.22^{+0.20}_{-0.21}$ & $0.028^{+0.010}_{-0.008}$ \\
\hline
$I$$-$$K_s$ & $0.688$ & $-0.277$ & \nodata & $0.002^{+0.007}_{-0.007}$ & $0.463^{+0.084}_{-0.085}$ & \nodata & $0.029^{+0.007}_{-0.006}$ & $-3.1$ & $0.494$ & $-0.498$ & $0.004^{+0.016}_{-0.017}$ & $0.19^{+0.27}_{-0.26}$ & $0.038^{+0.013}_{-0.011}$ \\
\hline
$J$$-$$H$ & $0.281$ & $-0.283$ & \nodata & $-0.002^{+0.002}_{-0.002}$ & $0.196^{+0.031}_{-0.032}$ & \nodata & $0.013^{+0.003}_{-0.002}$ & $-1.1$ & $0.187$ & $-0.481$ & $0.001^{+0.007}_{-0.007}$ & $0.35^{+0.12}_{-0.12}$ & $0.019^{+0.008}_{-0.006}$ \\
\hline
$J$$-$$K_s$ & $0.327$ & $-0.281$ & \nodata & $0.000^{+0.003}_{-0.003}$ & $0.189^{+0.037}_{-0.037}$ & \nodata & $0.014^{+0.003}_{-0.003}$ & $-4.4$ & $0.203$ & $-0.483$ & $0.005^{+0.006}_{-0.006}$ & $0.29^{+0.10}_{-0.11}$ & $0.014^{+0.006}_{-0.004}$ \\
\hline
$H$$-$$K_s$ & $0.045$ & $-0.281$ & \nodata & $0.002^{+0.002}_{-0.002}$ & $-0.008^{+0.035}_{-0.035}$ & \nodata & $0.014^{+0.003}_{-0.003}$ & $-3.0$ & $0.020$ & $-0.483$ & $0.000^{+0.007}_{-0.008}$ & $-0.03^{+0.12}_{-0.11}$ & $0.021^{+0.008}_{-0.006}$ \\
\hline
\hline
\end{tabular}}
\tablefoot{Dependence with metallicity only stated where statistically significant (as explained in Section~\ref{subsec:colour_min}). Results on the table are presented following the equation $C^{min}-C^{min}_{0}=\alpha+\beta \left( \log_{10}(P)-L_0 \right)+\gamma \left( \feh-\feh_0 \right)$, where $C^{min}$ is the minimum colour as defined in Section~\ref{subsec:colour_min}, $C^{min}_{0}=\overline{C^{min}}$, $L_0=\overline{\log_{10}(P)}$, and $\feh_0 = \overline{\feh}$. Pivots for the independent variables ($L_0$, $\feh_0$) vary among fits because the number of stars in common between the two filters varies with the pair of filters selected. The column $\Delta$BIC presents the difference in the BIC between the fit with one independent variable (period) or two (period and $\feh$).}
\end{table*}

\section{Comparison with literature for $\phi_{31}$ transformations} \label{sec:phi31_comparison}

We compare here the relation derived in Section~\ref{subsec:fourier_relation} for the transformation of the $\phi_{31}$ Fourier parameters between the $I$ and $V$ filters. This transformation is particularly useful for two reasons: i) the recent calibrations of the $\feh$-period-$\phi_{31}$ relation in the $V$ band provided by \citet{mullen2021,mullen2022} with a large sample of RRL stars with spectroscopic $\feh$; and ii) surveys containing large number of RRL either in the $V$ or $I$ bands, or in both \citep{jayasinghe2018,soszynski2019}. In Figure~\ref{plot:phi31_comparison}, we compare the $\phi_{31}^{I}-\phi_{31}^{V}$ relation derived in the present work (see Section~\ref{subsec:fourier_relation}) with the analogous relations from three different works in the literature: \citet[DF99]{dorfi1999}, based on light curves obtained from theoretical pulsational models; \citet[DS10]{deb2010}, based on a small sample of 29 RRab and 16 RRc from the GC M3; and \citet[S16]{skowron2016}, based on a large sample of Milky Way bulge and Large Magellanic Cloud RRL from the OGLE-IV project. The first two relations were computed using the cosine form of the Fourier series for RRab and RRc stars, so we transformed our RRab $\phi^{s,I}_{31}$ parameters to their cosinus form by substracting $\pi$, and then took the resulting $\phi^{c,V}$ back to the sinus form by adding $\pi$. The theoretical prediction from DF99 seems to be the one that is in better agreement with our data for RRab stars, while DS10 and S16 systematically lie below our relation (and our data). For RRc there are no significant differences between the fits.

\begin{figure*}[th]
\sidecaption
\includegraphics[width=12cm]{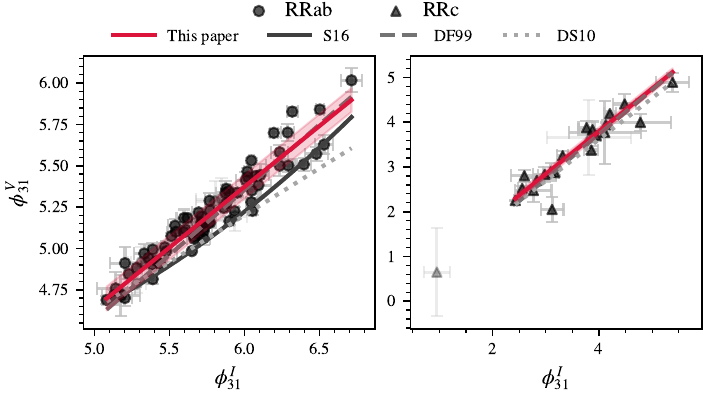}
  \caption{Left panel: Fourier parameter $\phi_{31}$ in the $I$ band versus $V$ band for RRab (circles). Grey points come from light curves not fulfilling the minimum requirements established in Section~\ref{subsec:quality_crit} and therefore were not included in the fit procedure. The red line depicts the median value of the relation obtained through the process described in Section~\ref{subsec:fourier_relation}. The dark red shaded area shows the 1$\sigma$ prediction interval for the fit, including the intrinsic scatter. The different black lines show the literature relations for DF99, DS10, and S16 (as defined in the text in Appendix~\ref{sec:phi31_comparison}).
  Right panel: same as the left, but for RRc (triangles).}
   \label{plot:phi31_comparison}
\end{figure*}

\section{Validation of colour at minimum relations} \label{sec:colour_min_validation}

We selected a data set separate from those used in deriving our period–minimum colour relations (Section~\ref{subsec:colour_min}) to use as an independent check of this work. This data set is composed of 44 GC with at least two RRL, coming from the homogeneous photometric database of P. B. Stetson\footnote{\url{https://www.canfar.net/storage/list/STETSON/homogeneous/Latest_photometry_for_targets_with_at_least_BVI}. For additional information on the database, readers may also contact Peter B. Stetson directly at peter.stetson@nrc-cnrc.gc.ca}. The data in the database were collected from ground-based telescopes using archival data from 1984 to the present. For further information about the summary of observing runs for each cluster, the bands observed, and the imagers and telescopes used, we direct the reader to the database. The details of this data set will be discussed in detail in the forthcoming Paper II. For the purpose of this validations, it is worth noting that for all the GC in the sample there is at least data in the $BVI$ filters, with some of them also containing light curves in $U$ and $R$.

We measured the mean colour at minimum light (as described in Section~\ref{subsec:colour_min}) for every filter combination for the light curves that met the quality criteria established in Section~\ref{subsec:quality_crit}. Then, applying the corresponding relation from Table~\ref{table:colour_min_relation} and the reddening law by \cite{cardelli1989}, we computed $E(B-V)$ and its uncertainty for each star in the cluster, RRab and RRc. After clipping outliers outside 3$\sigma$ around the average, we computed the weighted average and standard deviation as the measures of the GC $E(B-V)$, and we compared it with two independent estimates: i) the values reported by the \citet[H10]{harris1996, harris2010} GC catalogue, and ii) the values obtained from the \citet[SF11]{schlafly2011} 2D reddening map. We show an example of one of these comparisons in Figure~\ref{plot:colour_min_validation_VI} for the $VI$ colour. For the computation of the residuals, we limited the sample to those GC with at least five light curves in the database after the quality cuts. In Figure~\ref{plot:colour_min_validation_all} we show the result of this same procedure but applied to all the relations listed in Table~\ref{table:colour_min_relation}. For the relations requiring a GC metallicity estimate, we adopted the $\feh$ values from \citet{carretta2009}, applying a correction of $-0.08$ dex to account for the difference between their metallicity scale and that adopted in this work, following the calibration of \citet{crestani2021a}.

As shown in Fig.~\ref{plot:colour_min_validation_all}, all the relations reproduce the literature $E(B-V)$ values of the GCs in good agreement with the reference values. Relations involving NIR passbands exhibit either very small uncertainties when combined with optical filters or substantially larger uncertainties when paired with other NIR passbands. This behaviour reflects the limited number of NIR light curves available for measuring $E(B-V)$. For all relations involving blue ($UB$) passbands, the inclusion of the $\feh$ term reduces the residuals with respect to the literature $E(B-V)$ values, bringing them closer to zero.

\begin{figure*}[th]
\sidecaption
\includegraphics[width=12cm]{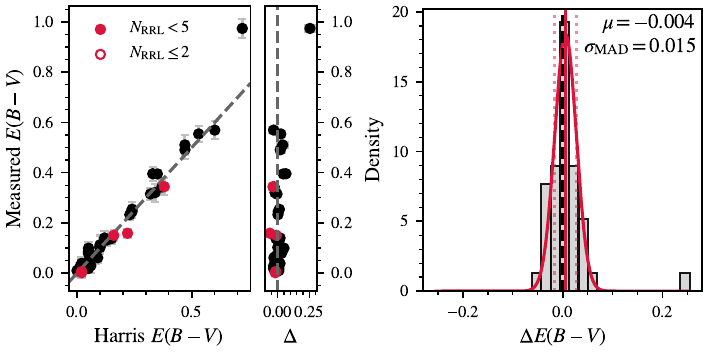}
  \caption{Left panel: $E(B-V)$ as given by H10 versus $E(B-V)$ measured from the period(-metallicity)-colour at minimum relations computed in Section~\ref{subsec:colour_min}, for a sample of GCs, in the $V-I$ colour. Error bars correspond to the dispersion on $E(B-V)$ measured from the RRL of the GC. Filled red points indicate GCs with less than five RRL, while open red symbols indicate GCs with less than two. The grey dashed line is the 1:1 relation.
  Middle panel: residual between the two measurements of $E(B-V)$ from the left panel. The grey dashed line shows zero difference, for reference.
  Right panel: histogram of the residuals, with its median value and standard deviation estimated from the median absolute deviation. GCs with less than five points were not included in the computation of these values. The red line depicts a normal distribution of the given mean and standard deviation. The black dashed line shows zero as a reference value.}
   \label{plot:colour_min_validation_VI}
\end{figure*}

\begin{figure*}[th]
\centering
\includegraphics[width=\textwidth]{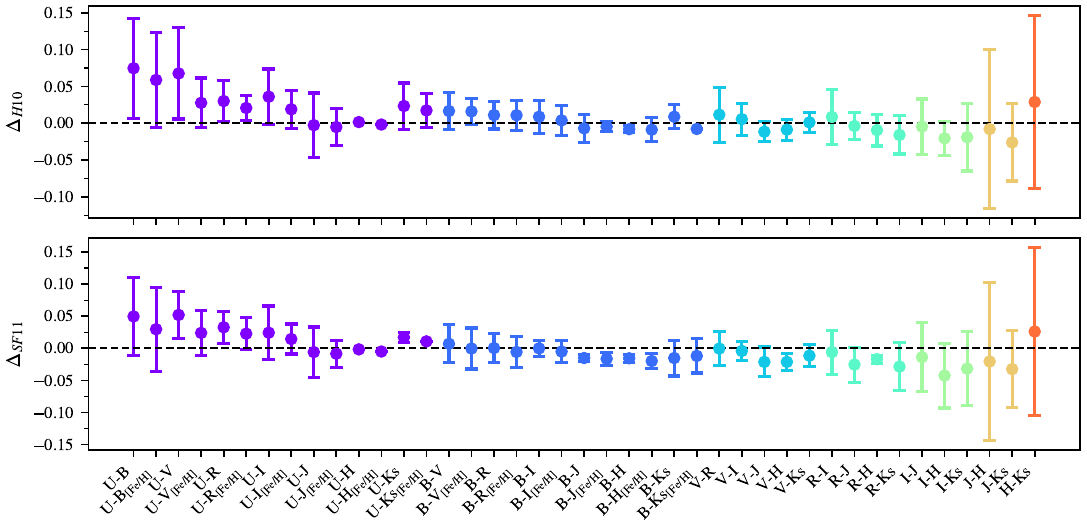}
  \caption{Top panel: median and MAD based standard deviation of the residuals between the $E(B-V)$ values derived from the relations presented in this work and the $E(B-V)$ estimates from H10 for the sample of GCs, computed separately for each relation listed in Table~\ref{table:colour_min_relation}.
  Bottom panel: same, but comparing with the $E(B-V)$ estimates from SF11.
  }
   \label{plot:colour_min_validation_all}
\end{figure*}

\end{appendix}

\end{document}